\documentclass[twocolumn, prb, amsmath,amssymb, aps, floatfix]{revtex4-2}

\usepackage{graphicx}
\usepackage{dcolumn}
\usepackage{bm}
\usepackage{dutchcal}
\usepackage[protrusion=true, expansion=true]{microtype}

\usepackage{braket}
\usepackage{bm,xcolor}
\usepackage{comment}
\usepackage{appendix}
\usepackage{amsmath}
\usepackage{amssymb}

\usepackage{siunitx}
\DeclareMathOperator{\tr}{tr}
\DeclareMathOperator{\Tr}{Tr}

\begin{document}

\title{Phase transitions induced by standard and feedback measurements in transmon arrays}

\author{Gonzalo Martín-Vázquez}
\author{Taneli Tolppanen}
\author{Matti Silveri}
\affiliation{Nano and Molecular Systems Research Unit, University of Oulu, FI-90014 Oulu, Finland}

\date{\today}

\begin{abstract}
The confluence of unitary dynamics and non-unitary measurements gives rise to intriguing and relevant phenomena, generally referred to as measurement-induced phase transitions. These transitions have been observed in quantum systems composed of trapped ions and superconducting quantum devices. However, their experimental realization demands substantial resources, primarily owing to the classical tracking of measurement outcomes, known as post-selection of trajectories. In this work, we first describe the statistical properties of an interacting transmon array which is repeatedly measured, and predict the behavior of relevant quantities in the area-law phase using a combination of the replica method and non-Hermitian perturbation theory. We show numerically that a transmon array, modeled by an attractive Bose-Hubbard model, in which local measurements of the number of bosons are probabilistically interleaved, exhibits a phase transition in the entanglement entropy properties of the ensemble of trajectories in the steady state. Furthermore, by using deterministic \textit{feedback} operations after the local number measurements, the distribution of the number of bosons measured at a single site carries information on the phase in the entanglement of individual trajectories. Interestingly, we can extract information about the phase and the phase transition from simple observables without considering an absorbing state in the feedback pattern. This implies that the \textit{feedback} measurement approach might be a viable experimental option to use simple observables to study some aspects of the entanglement phase transition in individual trajectories.
\end{abstract}

\maketitle

\section{Introduction}\label{sec:introduction}
Complex quantum systems can undergo a phase transition when subjected to quantum measurements, known as a measurement-induced phase transition~(MIPT)~\cite{Potter2022}. Described years ago within the context of quantum to classical transition~\citep{aharonov00}, its study has experienced a resurgence due to the advent of quantum technologies~\cite{li18, li19, chan19, skinner19, szyniszewski19, szyniszewski20, lunt20, tang20, bao20, jian20, choi20, gullans20, fuji20, gullans20, yang20, rossini20, ivanov20, bao21, sang21, Iaconis21, lavasani21, ippoliti21b, ippoliti21c, zabalo22, jian21b, sierant21, lu21, cote21, muller22, sharma22, liu2023}. In general, this phase transition is addressed in hybrid circuits composed of unitary evolution and non-unitary quantum measurements, which tend to increase and eliminate the entanglement between the elements of the system, respectively. In this way, two phases are defined: for infrequent measurements, the entanglement of the subsystems follows a volume-law, while for frequent measurements it follows an area-law. The relevant parameter of this phase transition is given by the measurement rate in the case of projective measurements~\citep{li18, skinner19, bao20}, the strength of weak measurements~\citep{szyniszewski19, szyniszewski20, bao20, buchhold21, doggen22} or the type of measurements applied~\citep{sang21}; and may even be achieved solely by measurements due to frustration~\citep{ippoliti21b}. Based mainly on numerical studies of random quantum circuits, it has been argued that such a phase transition should be described by a 2D non-unitary conformal field theory, which explains the universal scaling of entanglement entropy near the critical point, although a complete analytical understanding is still lacking.~\citep{li19, li21, zabalo20, zabalo22, skinner19, jian20}.

The recent development of noisy intermediate-scale quantum~\citep{preskill18} devices has also motivated the study of the~MIPT within the context of open quantum systems since the interaction of a random quantum circuit with its environment can be interpreted as a closed system continuously being measured~\citep{Potter2022, skinner19, choi20, gullans20}. Therefore, the connections between the entanglement entropy transition and other phenomena related to quantum information and communication are especially relevant. In this regard, the phase transition can be understood as a transition in the system's capability of purifying an initially mixed state~\citep{gullans20}, in the threshold of its quantum error correction properties~\citep{choi20, gullans21}, or the quantum channel capacity~\citep{choi20, kelly22}, as well as in the information that can be extracted about the initial state of the system, quantified as the Fisher information~\citep{bao20}.

\begin{figure*}
\includegraphics[width=\linewidth]{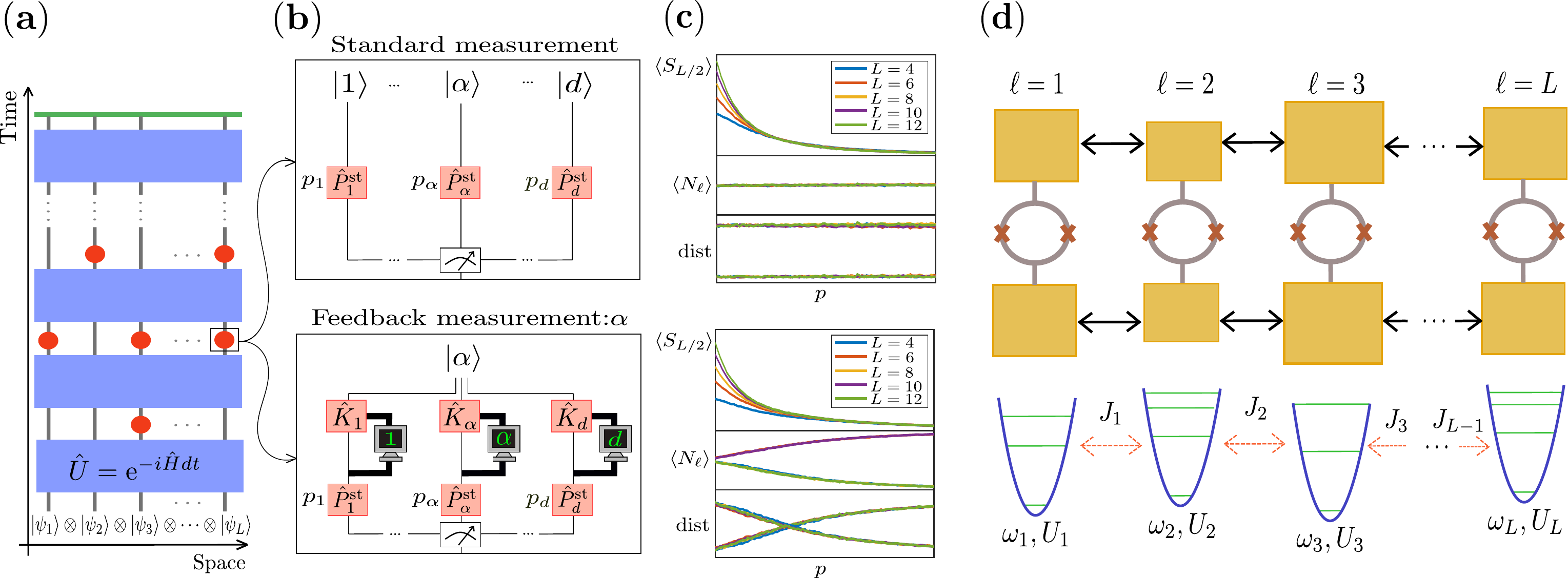}
\caption{\label{fig:circuit_scheme} (a)~Hybrid circuit of $L$ subsystems of dimension $d$ evolving under a unitary time evolution (blue rectangles) and probabilistic projective measurements (red circles) for an initial product state. The green line at the top corresponds to the steady state in which the observables are measured. (b)~The \textit{standard} measurements consists of a projective measurement $\hat{P}_n^\textnormal{st}=\ket{n}\bra{n}$ applied with a probability $p$ on each site $\ell$ whose outcome probabilities are given by Born's rule $p_n=\braket{\psi|\hat{P}_n^\textnormal{st}|\psi} $, such that $ \sum_{n=1}^{d}p_n=1$ defining $d$ different trajectories $\{\ket{1},\ket{2},...,\ket{d}  \}$. The \textit{feedback} measurement consists of a \textit{standard} measurement after which we access classically the result of the outcome, schematized by the thick wires and the monitors, based on which we apply an outcome-dependent unitary operator $\hat{K}_n=\ket{n}\bra{\alpha}+\ket{\alpha}\bra{n}$ to the system forcing the subsystem to be in the state $\ket{\alpha}$ in all the trajectories. Note that here the measurement schemes are represented for a single site~$\ell$, but the measurement is evaluated following a probability $p$ at every site and time step. (c)~Numerical results for different observables averaged over circuit iterations at long times for $d=2$. For the \textit{standard} measurement, the system undergoes a phase transition with increasing measurement probability. Here the averaged results and distributions of simple observables do not provide any relevant critical information. For the \textit{feedback} measurement, we observe a similar phase transition but, in this case, the averaged results and, especially, its distribution provide useful statistical information about the location of the entanglement phase transition in the individual quantum trajectories. (d)~Scheme of an array of $L$ interacting transmons modeled as anharmonic oscillators with on-site energies $\omega_\ell $ and anharmonicities $U_\ell $, interacting capacitively with strengths given by $J_{\ell}$.  }
\end{figure*}

The MIPT is characterized by a transition in the statistical properties of the system dynamics that can only be detected by examining individual quantum trajectories. These trajectories are pure states associated with specific measurement outcomes or trajectory-averaged quantities involving higher orders of the density matrix such as entanglement entropy or fluctuations of observables~\citep{bao21}. However, detecting these trajectories experimentally requires post-selecting all measurements to reconstruct the final state or calculate trajectory-averaged expectation values that can hinder experimental performance due to the need for multiple circuit iterations. There are different proposals to eliminate or reduce post-selection, such as the use of space-time duals of random circuits in which post-selection is only necessary for the final measurements~\citep{ippoliti21c}, the averaging of a reference ancilla that is entangled with the circuit~\citep{gullans20, dehghani2023}, or by considering swapping between the circuit and the environment instead of measurements~\citep{weinstein22}. Recently, an~MIPT in trapped ions using Clifford gates has been experimentally observed without the need for post-selection, using reference ancillae to detect the purified phase~\citep{Noel2022}; it has been argued that it is also possible to perform a similar experiment to detect the unpurified phase~\citep{yoshida21}. However, the features of superconducting devices, such as their scalability, speed, and richness of dynamics due to easy access to larger Hilbert spaces~\citep{cazalilla11, barends13, hacohen-gourgy15, roushan17, ma19, yan19, kjaergaard20, martinvazquez20}, make this platform an ideal device to study~MIPTs. Interestingly, its existence has recently been experimentally demonstrated by explicit post-selection in a superconducting quantum processor~\citep{Koh2023}. It is important to note that the~MIPT, observed in diverse systems, is a generic property of quantum trajectories in open systems, regardless of implementation details in different devices~\citep{Potter2022}.

To perform post-selection, one needs to know all the results of the measurements taken during the temporal evolution. To simplify this process, one can use a different type of measurement where the outcome is predetermined in advance, such that the probability is based solely on whether or not the measurement was conducted. This has been previously considered in systems evolving under a Bose-Hubbard Hamiltonian~\citep{tang20}, to include possible unwanted effects of projective measurements on trapped ions~\citep{Czischek2021}, or to study PT-symmetry breaking in non-Hermitian Hamiltonians~\citep{gopalakrishnan21}. It has been argued that this type of measurement instead generates a forced measurement-induced phase transition, which may belong to a different universality class than the~MIPT described above~\citep{nahum21}. Recently, several works have addressed the effect of including some feedback after each measurement, where they have demonstrated the existence of a phase transition in the averaged density matrix, which can be detected using simple linear observables that are easy to measure experimentally~\citep{buchhold22, iadecola22, odea22, ravindranath22, piotr2023, piroli2023}. However, this phase transition consists of an absorbing state phase transition (APT) and generally belongs to a different universality class than the MIPT observed in the entanglement of individual quantum trajectories~\citep{odea22}. Interestingly, including feedback corrections induces the same MIPT in individual quantum trajectories as seen in hybrid circuits without feedback~\citep{piroli2023}. Under certain conditions, the critical parameters of both transitions can coincide: in the limit of infinite local Hilbert-space dimension in Haar random and Clifford-like circuits~\citep{piroli2023}, in the limit of applying a feedback correction after each measurement~\citep{ravindranath22,odea22}, when the feedback involves long-range entangling operations~\citep{piotr2023}. Both transitions can even exhibit the same critical behavior, as is the case for free fermions governed by the essential scaling of a Berezhinskii-Kosterlitz-Thouless (BKT) transition~\citep{buchhold22}, or in the case of random circuits with long-range feedback operations with specific features where the entanglement entropy inherit the behavior of the absorbing state phase transition~\citep{piotr2023}. It has been suggested that this APT generally falls into the direct percolation (DP) universality class, and it is expected to hold in local models targeting short-range correlated states without additional symmetries~\citep{odea22}. In the presence of symmetries, the APT has also been associated with a parity-conserving universality class~\citep{ravindranath22}.

In this work, we focus on an array of transmon devices as the physical platform for implementing measurement-induced phenomena. Transmons are multilevel quantum systems, that is, qudits with $d$ levels, see Fig.~\ref{fig:circuit_scheme}(d). We compare a \textit{standard} measurement of a transmon occupation that produces one of the $d$ possible results, with a \textit{feedback} measurement that projects the system to a predetermined state, see Fig.~\ref{fig:circuit_scheme}(a-b). In the case of a \text{standard} measurement, $d$ different results can be produced after performing the measurement, thus defining branching into $d$ different trajectories. A trajectory~$n$ has a probability of occurrence~$p_n$ given by Born's rule. The state after the measurement is obtained by projecting the measured state by the operator~$\hat{P}_n^{\textnormal{st}}=\ket{n}\bra{n}$. By contrast, the \textit{feedback} measurement consists of two events: first, a \textit{standard} measurement is performed in the same way as in the previous case, and then a local unitary operator~$\hat{K}_n$, which depends on the result of the measurement $n$, is applied to the measured site. This implies that we need to have short-term classical access to the result of the measurement to apply one or another unitary operator~$\hat{K}_n $, thus forcing the system to be projected to a predetermined state and collapsing all the possible $d$ trajectories to a single one, without discarding any trajectory. In this way, the probabilities associated with Born's rule are eliminated. The measurement probability is the only relevant parameter with respect to reducing entanglement from the system. Note that throughout this paper, all measurements and their feedback are both strictly local, i.e., measuring one site involves applying a feedback unitary gate to that particular site.

Precisely, in this work, we propose that using \textit{feedback} measurements may inform us about the location of the critical parameter of the entanglement phase transition in individual trajectories based on the statistical distribution of simple local observables without the need to carry out an explicit post-selection. It is important to note that we are not studying phase transitions either in the simple observable or in the averaged density matrix; instead, we are using this information to estimate some properties of the entanglement phase transition in individual quantum trajectories. We do not propose an equivalence; rather, we claim that the entanglement phase transition in individual trajectories induced by \textit{feedback} measurements generates a pattern in the simple observable that we can use to estimate the value of the critical parameter of the entanglement phase transition in individual trajectories. In Fig.~\ref{fig:circuit_scheme}(c), we summarize the main result of the article. By measuring the boson number at one single site in the steady state, we can compare the distribution of the results with the theoretically expected distribution for area-law and volume-law phases. In the case of a \textit{standard} measurement, the distributions are the same in the two phases, and therefore, the observable does not give any critical information. On the other hand, by introducing \textit{feedback} measurements, the boson number distributions are different in different phases. Therefore, we can indirectly find the expected approximate location of the phase transition as the crossing point of the fit of the observed distributions with the theoretical ones. Note that the distribution of the number of bosons at the steady states is obtained by merely collecting the outcomes, but it is not necessary to keep any other information. Importantly, the final results obtained for transmons modeling hard-core bosons are applicable and can be extended to a wide range of scenarios, including arrays of subsystems with higher dimensions, disorder, and interactions. Both the analytical and numerical findings have a general nature, making them suitable for describing various systems.

The article is organized as follows: The attractive Bose-Hubbard model, which describes the dynamics of interacting transmons, is presented in Sec.~\ref{sec:model} and App.~\ref{appsec:trotter}, along with the procedure for creating a hybrid circuit consisting of unitary gates and non-unitary measurements. In Sec.~\ref{sec:statistics}, we present the methods. We briefly describe the replica method approach used to study the relevant statistical properties of the hybrid circuit encoded in the ground state of an effective non-Hermitian Hamiltonian. Additionally, we introduce statistical arguments to adequately describe the dispersion of simple observables, a quantity that does not require post-selection. In Sec.~\ref{sec:trajectory-averaged}, we present analytical results for the obtention of an effective Hamiltonian describing the statistics of the circuits and the averaged observables in the area-law phase of hard-core bosons for both \textit{standard} and \textit{feedback} measurements. Additionally, we propose simple observables that can indirectly help us estimate the critical value of the control parameter of the phase transition in the entanglement entropy of individual trajectories when \textit{feedback} measurements are involved. This section is extended in App.~\ref{appsec:replica_method}-\ref{appsec:averaged_observable}. In Sec.~\ref{sec:numerics} and App.~\ref{appsec:numerical_simulation}, we show numerical simulations to test the analytical predictions for hard-core bosons extended in App.~\ref{appsec:circuit_average}. Finally, Sec.~\ref{sec:discussion} is dedicated to conclusions and suggestions for future work.

\section{A repeatedly measured transmon array}
\label{sec:model}
To study an~MIPT on an array of transmons undergoing projective measurements, we create a hybrid circuit consisting of unitary gates originating from the intrinsic dynamics of the transmons and non-unitary measurements introduced externally to monitor the system, see Fig.~\ref{fig:circuit_scheme}(a). Regarding the unitary elements, the dynamics of a one-dimensional array of $L$ interacting transmons [Fig.~\ref{fig:circuit_scheme}(d)] can be described by the disordered attractive Bose-Hubbard model~\citep{hacohen-gourgy15, roushan17} with Hamiltonian
\begin{equation}
\frac{\hat{H}}{\hbar} = \sum_{\ell=1}^L \left[\omega_{\ell} \hat{n}_\ell- \frac{U_{\ell}}{2} \hat{n}_{\ell} (\hat{n}_{\ell}-1)+J_{\ell} \left(\hat{a}_{\ell}^\dagger \hat{a}^{}_{\ell+1}+\textrm{h.c.} \right)\right],
\label{Hamiltonian_BH}
\end{equation}
where $\hat{a}_\ell$ and $\hat{a}^\dagger_\ell $ are the bosonic annihilation and creation operators at site $\ell$, and $\hat{n}_\ell=\hat{a}^\dagger_\ell \hat{a}^{}_\ell $ is the corresponding number operator. Within this description, $\omega_\ell$ accounts for the on-site energy and $U_\ell$ for the attractive interaction strength at site $\ell$, to which the bosonic excitations are subject. The term $J_\ell$ refers to the hopping rate of excitations between sites $\ell$ and $\ell+1$, and $J_L$ implicitly includes the boundary of the array, i.e.~whether it has open or periodic boundary conditions. The Hamiltonian of Eq.~\eqref{Hamiltonian_BH} conserves the total number of excitations, since $[\hat{H},\hat{N}]=0 $, where $\hat{N}=\sum_{\ell=1}^{L}\hat{n}_\ell$ is the total number operator. This implies that the dynamics occur in a single sector of a fixed number of excitations when initializing the system with a definite number of excitations. 

For experimental purposes, it is convenient to take into account that the typical values of the parameters are around $\omega_\ell/2\pi \sim \SI{5}{\giga\hertz}$, $ U_\ell/2\pi \sim~\SI{200}{\mega\hertz}$, and $J_\ell/2\pi \sim~\SI{10}{\mega\hertz}$~\citep{koch2007, paik2011, arute2019}, and range within the ratios $U_\ell/J_\ell \sim 2-30 $ and $\omega_\ell/J_\ell \sim 50-1000$~\citep{hacohen-gourgy15, roushan17}. Due to manufacturing defects, the exact parameter values differ between transmons and should be understood as being taken from a certain distribution. In most of the analysis, we will consider constant values $\omega_\ell \equiv \omega $, $U_\ell \equiv U $ and $J_\ell \equiv J$, corresponding to the mean values of Gaussian distributions with variances $\sigma_\omega^2 $, $\sigma_U^2 $ and $\sigma_J^2 $. Importantly, volume-law states can also be obtained even in the presence of a certain amount of disorder~\cite{Orell2019}.

To create an analog circuit for analytic calculations corresponding to the time evolution generated by the Bose-Hubbard Hamiltonian~\eqref{Hamiltonian_BH}, we use a Suzuki-Trotter decomposition to design a unitary layer corresponding to a time step $dt$ in terms of two-site gates~\citep{Pierkarska2018, Jaschke2018, Orell2019, Sieberer19, Barbiero2020, kargi21}. Briefly, we split the Hamiltonian of Eq.~\eqref{Hamiltonian_BH} into odd and even sites $\hat{H}=\sum_{\textnormal{odd } \ell}^L \hat{H}_\ell+\sum_{\textnormal{even } \ell}^L \hat{H}_\ell$, such that we can express the unitary time evolution operator as $e^{-i\hat{H}dt/\hbar} \approx \prod_{\text{even } \ell_b}^L e^{-i\hat{H}_{\ell_b} dt/\hbar} \prod_{\text{odd } \ell_a}^L e^{-i\hat{H}_{\ell_a} dt/\hbar} $ at first order in $dt$ (more details in App.~\ref{appsec:trotter}). After each layer of gates, we artificially introduce a probabilistic measurement layer in such a way that each time step, which is an effective layer of the hybrid circuit, can be expressed as
\begin{equation}
\ket{\psi_{t+dt}}=\prod_{\ell}^L \hat M_{\ell}(p) \prod_{\text{even } \ell_b}^L e^{-i\hat{H}_{\ell_b} dt/\hbar} \prod_{\text{odd } \ell_a}^L e^{-i\hat{H}_{\ell_a} dt/\hbar}  \ket{\psi_t},
\label{Trotter_expansion_BH_measurements}
\end{equation}
where $\hat M_\ell (p)$ represent the measurement operations performed with a certain probability $p$ at each site $\ell$. Furthermore, the state $\ket{\psi_{t+dt}}$ needs to be renormalized. Note that we introduce the measurements \textit{ad hoc}, assigning them a time scale $dt$ of the order of the trotterized gates, so we assume that the probability of measuring a particular site $\ell$ depends on the time scale $dt$ and a measurement rate $\Gamma$, such that $p=\Gamma dt$. For the numerical simulations, we define the layers simply by evolving the Hamiltonian~\eqref{Hamiltonian_BH} for a time $dt$ and then performing a measurement with a probability~$p$, and finally renormalizing the state.

The type of measurement implemented is crucial. First, we consider a \textit{standard} number measurement, whose outcome $n$ after performing a measurement range from $0$ to $d-1$ with an associate probability of occurrence given by the Born's rule, the projection operators are $\hat{P}_{\ell,n}^{\text{st}}=\ket{n}\bra{n}$ and here $\ell$ is the site index. Second, we introduce a \textit{feedback} measurement with projection operators $ \hat{P}_{\ell,n}^{\text{fb}}=\ket{\alpha_\ell}\bra{n} $ for a given spatial profile $(\alpha_1, \alpha_2, ..., \alpha_L)$ that define the site-dependent outcomes after performing a measurement. This measurement can be understood so that first a \textit{standard} measurement is performed at site~$\ell$ whose result depends on Born's rule; second, we classically access this result, and, third, based on this, we perform a local operation on the site~$\ell$ to project it onto state $\ket{\alpha_\ell}$. This means formally that $ \hat{P}_{\ell,n}^{\text{fb}} \equiv \hat{K}_{\ell,n}\hat{P}_{\ell,n}^\textnormal{st}$, see Fig.~\ref{fig:circuit_scheme}(b). In both cases, the operators fulfill the measurement condition $ \sum_{n=0}^{d-1} \hat{P}_{\ell,n} ^\dagger \hat{P}_{\ell,n} = \hat{I} $. Interestingly the \text{standard} measurement conserves the total number of bosons while the \textit{feedback} measurement does not. Projectors of a similar nature have been employed in studies that successfully observed an~MIPT~\citep{Czischek2021, piroli2023}. Note that here we consider a general spatial profile $(\alpha_1, \alpha_2, ..., \alpha_L)$ for the feedback measurements. Particularly, we do not consider the profiles that would lead to an absorbed state, such as $\alpha_\ell=0$ for all $l$.

\section{Statistical methods}
\label{sec:statistics}
\subsection{Replica method}
We now analyze the long-term statistical behavior of the hybrid circuits including unitary gates and probabilistically interleaved measurements. However, this analysis becomes complex due to the numerous possibilities involved, both analytically and experimentally. For the analytical study of the statistical properties of the system, we make use of the replica method, which has been used to describe simple random unitary circuits~\cite{zhou19, vasseur19, bao20, barbier21, bao21} whose statistical properties can be mapped to a classical mechanics model in which relevant statistical properties can be easily calculated, allowing studies of phase transitions in the entanglement entropy~\cite{jian21, jian21b}.

For the implementation of the replica method to the Bose-Hubbard model with interspersed measurements, we follow the work carried out by Bao \textit{et al.}~in Ref.~\cite{bao21}. In contrast to their $\mathbb{Z}_2 $ symmetry-preserving circuits, we consider circuits that conserve the total number of bosons~\citep{oshima2023}. The symmetry of the conserved total number of bosons can be broken by the presence of \textit{feedback} measurements. Since we are interested in the ensemble of trajectories, we start by labeling the states at time $t$ with a sequence of measurement outcomes $ m_{\bold{i}} $ and a set of gate parameters~$\theta_{\bold{i}}$~as:
\begin{equation}
\hat{\rho}_{m_{\bold{i}},\theta_{\bold{i}}}(t)=\left(\hat{P}^{}_{m_t} \hat{U}^{}_{\theta_t}  \cdots  \hat{P}^{}_{m_1} \hat{U}^{}_{\theta_1} \right) \hat{\rho}_0 \left(\hat{U}_{\theta_1}^{\dagger} \hat{P}_{m_1}^\dagger  \cdots \hat{U}_{\theta_t}^{\dagger} \hat{P}_{m_t}^\dagger\right),
\label{trajectories}
\end{equation}
where $\hat{\rho}_0$ is the initial state, $\hat{U}_{\theta_{\bold{i}}} $ are the set of unitary evolution gates, $ \hat{P}_{m_{\bold{i}}}$ are the string of projection operators associated with measurement outcomes $ m_\bold{i} $, and $\bold{i}$ refers to all the positions in the circuit space-time. For studying the steady-state properties of the ensemble of states $\hat{\rho}_{m_{\bold{i}},\theta_{\bold{i}}}(t)$, and its associated probabilities, we consider the dynamics of $n$ copies\----the replicas\----of the density matrices $ \ket{\ket{\rho_{m_{\bold{i}},\theta_{\bold{i}}}^{(n)}}} \equiv \hat{\rho}_{m_{\bold{i}},\theta_{\bold{i}}}^{\otimes n} $ interpreted as state vectors in the replicated Hilbert space $ \mathcal{H}^{(n)}= ( \mathcal{H} \otimes \mathcal{H}^* )^{\otimes n} $. Replicated unitary operators $\check{\mathcal{U}}_{\theta_i}^{(n)} \equiv ( \hat{U}_{\theta_i} \otimes \hat{U}_{\theta_i}^* )^{\otimes n}$, measurement projection operators $\check{\mathcal{M}}_{m_i}^{(n)} \equiv ( \hat{P}_{m_i} \otimes \hat{P}_{m_i}^\dagger )^{\otimes n} $, as well as general operators $\check{\mathcal{O}}^{(n)} \equiv ( \hat{O} \otimes \hat{I} )^{\otimes n} $ are going to be used for computing observables.

Taking into account the non-normalized averaged state of the ensemble $ \ket{\ket{ \rho^{(n)}(t)}} = \sum_{m_{\bold{i}},\theta_{\bold{i}}} \ket{\ket{ \rho_{m_{\bold{i}},\theta_{\bold{i}}}^{(n)}(t)}}=e^{-t \check{\mathcal{H}}_{\textnormal{eff}}}\ket{\ket{ \rho_0^{(n)}}}$, we can exactly map the dynamics to an imaginary time evolution generated by an effective quantum Hamiltonian $\check{\mathcal{H}}_{\textnormal{eff}}$, such that the properties of the averaged state of the ensemble at long times are encoded in its ground state. Note that from now on we will consider $\hbar=1$ to simplify the notation. Using this formalism, we can compute the trajectory-averaged $k$-moment of an observable $\hat{O}$, which is given by
\begin{align}
\overline{\left\langle O_k \right\rangle}=& \lim_{n \to 1} \frac{\braket{\braket{ \mathcal{I}^{(n)} ||  \check{\mathcal{O}}_k^{(n)} || \rho^{(n)}}}}{\braket{\braket{ \mathcal{I}^{(n)} || \rho^{(n)} }}} = \lim_{n \to 1}O_k^{(n)}, 
\label{replicated_quantities}
\end{align}
where $ \overline{\cdot} $ refers to the average over gate parameters and $\left\langle \cdot \right\rangle $ to the average over measurements outcomes and the inner product is defined by $ \braket{\braket{ \mu|| \sigma}} \equiv \tr \left( \hat{\mu}^\dagger \hat{\sigma} \right) $, for arbitrary states $ \hat{\mu} $ and  $ \hat{\sigma} $ and a reference state $\ket{\ket{\mathcal{I}^{(n)}}} $ in the replicated Hilbert space. Therefore, the quantities we are going to study for addressing phase transitions are the objects $O_k^{(n)}$ in Eq.~\eqref{replicated_quantities}, which corresponds to the exact trajectory-averaged quantum mechanical observables only in the replica limit $n \to 1$. It has been shown, at least for the von Neumann entropy, that although not being the same quantity, both share critical properties in the~MIPT~\cite{bao20}. More details on this particular implementation of the formalism can be found in App.~\ref{appsec:replica_method}~and~in~Ref.~\cite{bao21}. 

\begin{figure}
\includegraphics[width=\linewidth]{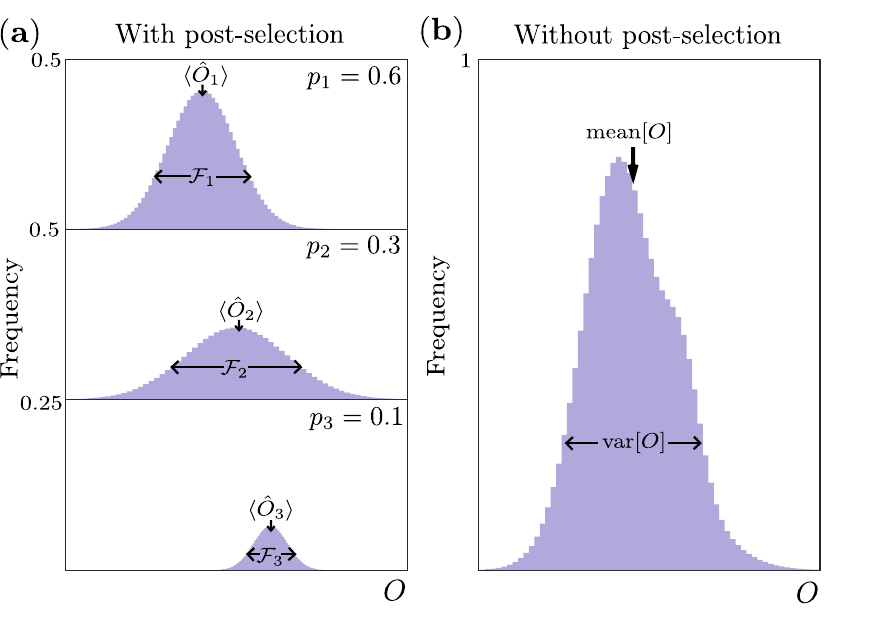}
\caption{\label{fig:postselection_nopostelection} A scheme highlighting experimental differences in computing the mean and variance of an arbitrary observable~$\hat{O}$ with and without post-selection. For the sake of simplicity, in this example, the circuit produces only three different trajectories with associated probabilities $\{p_1,p_2,p_3\}$. (a)~Post-selected histogram of the distribution of measurement outcomes. We separate each iteration of the circuit into three different subgroups of trajectories. The mean values of the distribution of the observable correspond to the expectation values $\{\braket{\hat{O}_1},\braket{ \hat{O}_2},\braket{\hat{O}_3}\}$ and the variances to the fluctuations $\{\mathcal{F}_1,\mathcal{F}_2,\mathcal{F}_3\}$ of the operator $\hat{O}$ for each trajectory type. (b)~Histogram of the measurement outcomes without post-selection: We mix all the iterations of the circuit without taking into account to which trajectory corresponds, and compute the mean value of the observable $\textnormal{mean}[O]$ and the variance $\textnormal{var}[O] \equiv \Delta O$ of the distribution of the observable among all iterations. Since the expectation value involves the first moment $k=1$ of the density matrix, the mean value among all iterations corresponds to the average over the expectation values of the trajectories $\textnormal{mean}[O] \equiv \overline{\braket{O_1}} = p_1\braket{\hat{O}_1} + p_2 \braket{\hat{O}_2} + p_3\braket{\hat{O}_3}$. However, the fluctuations of the operator involve the second moments $k=2$ of the density matrix. Therefore, the average of the fluctuations of the trajectories $\overline{\braket{\mathcal{F}}} = p_1\mathcal{F}_1 + p_2\mathcal {F}_2 + p_3\mathcal{ F}_3$ and the variance among all iterations $\Delta O$, which we name dispersion, are not generally equivalent. However, both quantities are similar under certain conditions, as shown in Eqs.~\eqref{averaged_observables_F_L2_comp} and \eqref{averaged_DeltaN}.}
\end{figure}

\subsection{Direct average of circuit realizations}
Next, we elaborate the role of the moment $k$ of Eq.~\eqref{replicated_quantities} in the post-selection of trajectories to calculate trajectory-averaged quantities. For that, it is useful to express the trajectory-averaged quantities directly in terms of the measurement probability $p$ and probability distributions of the gate parameters $p_{\theta_\bold{i}} $ and the outcomes of the interleaved measurements $p_{m_\bold{b}}(\theta) $, such that
\begin{align}
    \overline{\langle O_k \rangle}= \sum_{b=0}^M p^{b} (1-p)^{M-b}\langle O_k^b \rangle_{\textnormal{m},\theta} ,
    \label{averaged_observables}
\end{align}
where the average expectation value is 
\begin{align}
\langle O_k^b \rangle_{\substack{\textnormal{m},\theta}} =  \int d\theta_{\bold{i}}   \sum_{\bold{b}}^{\binom{M}{b}} \sum_{m_{\bold{b}}}^{d^b} p_{m_\bold{b},\theta_{\bold{i}}} \left[\tr\left(\hat{O}\hat{\rho}^{\prime}_{m_{\bold{b}},\theta_\bold{i}} \right)\right]^k,
\label{averaged_observables_distribution}
\end{align}
where $\hat{\rho}^{\prime}_{m_{\bold{b}},\theta_\bold{i}}$ is the normalized density matrix of an individual trajectory, $\hat{O}$ can be any operator, and $p_{m_{\bold{b}},\theta_{\bold{i}}} \equiv p_{\theta_\bold{i}} p_{m_\bold{b}}(\theta) $, $\binom{M}{b}=\frac{M!}{b!(M-b)!}$, $\bold{b}$ is an array with all the possible $\binom{M}{b}$ combinations of arranging $b$ measurements in the total $M$ positions of space-time. In other words, $M$ is the maximum number of measurements that is performed when $p=1$ and $\bold{b}=\bold{i}$; while no measurements are performed when $p=0$ and $\bold{b}=\emptyset \to p_{m_{\bold{b}},\theta_{\bold{i}}} = p_{\theta_\bold{i}} $. The probability distributions are normalized such that $\sum_{b=0}^M p^{b} (1-p)^{M-b} {\binom{M}{b}}=1 $, $\sum_{m}^{d^b} p_{m_\bold{b}}(\theta)=1$ and $\int d\theta_{\bold{i}} p_{\theta_\bold{i}}=1$. We have considered that $p_{m_\bold{i}}(\theta_\bold{i}) \to p_{m_\bold{b}}(\theta_\bold{i})$, because for the probability distribution we need to take into account those positions $\bold{b} $ where measurements have been performed and not all the possible positions $\bold{i}$ where a measurement could have been performed.

The trajectory-averaged first moment $k=1$ quantities, such as the number of bosons $\overline{\braket{N_1}}$ or the dispersion of the number of bosons over different circuit iterations $\Delta N$, can be obtained in a realistic experimental device by averaging all the results obtained from different experiments, i.e.~iterations, without taking into account the final states. By contrast, the trajectory-averaged second moment $k=2$ quantities, such as the second Rényi entanglement entropy (related to the von Neumann entropy $\overline{\braket{S}}$ in the replica limit) and the fluctuation of the number operator $\overline{\braket{\mathcal{F}}}$~\citep{agrawal2022, oshima2023}, require to repeat the experiment for each final state independently, post-selecting the different trajectories from the different iterations of the circuit, see Fig.~\ref{fig:postselection_nopostelection}.

The role of $k$ can be seen by simplifying Eqs.~\eqref{averaged_observables} and~\eqref{averaged_observables_distribution} to a generic expression $\overline{f}=\sum_x p_1(x) \sum_y [p_2(x,y)f(y)]^k$ where $p_1(x)$ is the probability of each trajectory and $p_2(x,y)$ is the probability of obtaining the different outcomes $f(y)$. This can be further simplified as $\overline{f}=\sum_z p(z) f(z)$ if $k=1$ and thus obtained from a general distribution, i.e.~collecting results from different experiments without taking into account the trajectories. Here $p(z)$ is the probability of obtaining an outcome $f(z)$. The simplification does not hold for a general case if $k\geq2$. Notice the subtle difference between averaging over trajectories analytically and averaging over iterations of the circuit numerically/experimentally, see Fig.~\ref{fig:postselection_nopostelection} and App.~\ref{appsec:numerical_simulation}.

We can go further and use the simple way of describing trajectory-averaged quantities in Eq.~\eqref{averaged_observables}, to study quantities related to the variance of observables in the high measurement regime. In what follows, we consider the first $k=1$ and second $k=2$ moments of observables related to the boson number without any disorder in the parameters, such that Eq.~\eqref{averaged_observables} becomes
\begin{align}
    \langle N_k \rangle= \sum_{b=0}^M p^{b} (1-p)^{M-b}  \sum_{\bold{b}}^{\binom{M}{b}} \sum_{m}^{d^b} p_{m_\bold{b}} [\tr(\hat{N}\hat{\rho}^{\prime}_{m_{\bold{b}}} )]^k,
    \label{averaged_observables_2}
\end{align}
where the usual number of bosons is given for $k=1$. We are particularly interested in the trajectory-averaged fluctuations of $\hat{N}$ for each trajectory
\begin{align}
    \langle \mathcal{F} \rangle= & \sum_{b=0}^M  p^{b}  (1-p)^{M-b}     \label{averaged_observables_F} \\
    & \times \sum_{\bold{b}}^{\binom{M}{b}} \sum_{m}^{d^b} p_{m_\bold{b}} \left \{ \tr\left(\hat{N}^2\hat{\rho}^{\prime}_{m_{\bold{b}}} \right)-\left[\tr(\hat{N}\hat{\rho}^{\prime}_{m_{\bold{b}}} )\right]^2 \right\}, \nonumber 
\end{align}
for which we calculate the variance of $\hat{N}$ for the final state of each trajectory, and then we average over all possible trajectories. We are also going to calculate the dispersion in the number of bosons with the average given by Eq.~\eqref{averaged_observables_2}
\begin{align}
    \Delta N \equiv & \langle (N^2)_1 \rangle -\langle N_1 \rangle^2.   \label{averaged_observables_DN}
\end{align}
To simplify notation, we now omit the sub-index $k=1$ when referring to the number of bosons. 

\section{Analytical results for trajectory-averaged observables}\label{sec:trajectory-averaged}

In this section, we derive analytical expressions for the effective Hamiltonian in the replica space describing the circuit statistics, as well as for the trajectory-averaged observables. By combining the replica method and the non-Hermitian perturbation theory we generate specific results for hard-core bosons and both measurement types in the area-law phase at low measurement rates. The main results can be extended to higher local subsystem dimensions too.

\subsection{Boson dynamics in an enlarged space}
We now consider $n=2$ replicas, which is the lowest number of replicas needed to capture the relevant~MIPT properties~\citep{bao20,bao21}. The transfer matrix between the state of the system at $t+dt$ and $t$ is then obtained by averaging the evolution over the distribution of unitary gates and the probabilities $p=\Gamma dt $ of applying measurement operators $\braket{\braket{ \rho_m^{(2)}(t+dt) || \prod_{\ell'}^L \braket{\check{\mathcal{M}}_{\ell'}} \prod_{\ell}^L \overline{\check{\mathcal{U}}_{\ell}} ||  \rho_m^{(2)}(t)}}$. At first order in $dt$, we have that $ \prod_{\ell'}^L \braket{\check{\mathcal{M}}_{\ell'}} \prod_{\ell}^L \overline{\check{\mathcal{U}}_{\ell}} \simeq e^{- d t \check{\mathcal{H}}_{\textnormal{eff}}}$, where the effective Hamiltonian is given by
\begin{align}
\check{\mathcal{H}}_{\textnormal{eff}}=\sum_{\ell}^L & \left \{ \Gamma_{\ell} \left(\check{\mathcal{I}}- \sum_{m}^d \check{\mathcal{P}}_{\ell,m} \right)  \right. \label{H_eff} \\
& \left.  +i  \left[\omega\check{\mathcal{n}}_{\ell}+ J \left(\check{\mathcal{a}}_\ell^\dagger \check{\mathcal{a}}^{}_{\ell+1}+\check{\mathcal{a}}^{}_\ell \check{\mathcal{a}}_{\ell+1}^\dagger \right) -\frac{U}{2} \check{\mathcal{u}}_{\ell} \right]\right\},  \nonumber
\end{align}
where each $\check{\mathcal{P}}_{\ell,m} $ derives from a measurement operator and acts on the two replicas, $\check{\mathcal{n}}_{\ell} $, the operators $\check{\mathcal{a}}_\ell $, $\check{\mathcal{a}}_\ell^\dagger $ and $\check{\mathcal{u}}_{\ell} $ derive from the unitary gates and each of them are composed of operators acting on one of the replicas, and finally $\check{\mathcal{I}}$ is the identity operator. The exact expressions can be found in App.~\ref{appsec:replica_method}, see  Eqs.~\eqref{replicated_n}-\eqref{replicated_aad}. Using the effective Hamiltonian given by Eq.~\eqref{H_eff}, we will calculate various trajectory-averaged observables, related to the first moment $k=1$ not requiring post-selection and second moment $k=2$ requiring post-selection: $S^{(2)}$ is the $2$-conditional Renyi entropy that results in the properly trajectory-averaged von Neumann entropy in the replica limit in Eq.~\eqref{observable_entropy}, and, similarly, the other trajectory-averaged quantities result, in the replica limit, in the number of bosons $N^{(2)}$ in Eq.~\eqref{observable_number} and the fluctuation of the number operator $\mathcal{F}^{(2)}$ in Eq.~\eqref{observable_fluctuation}.

\begin{figure}
\includegraphics[width=0.9\linewidth]{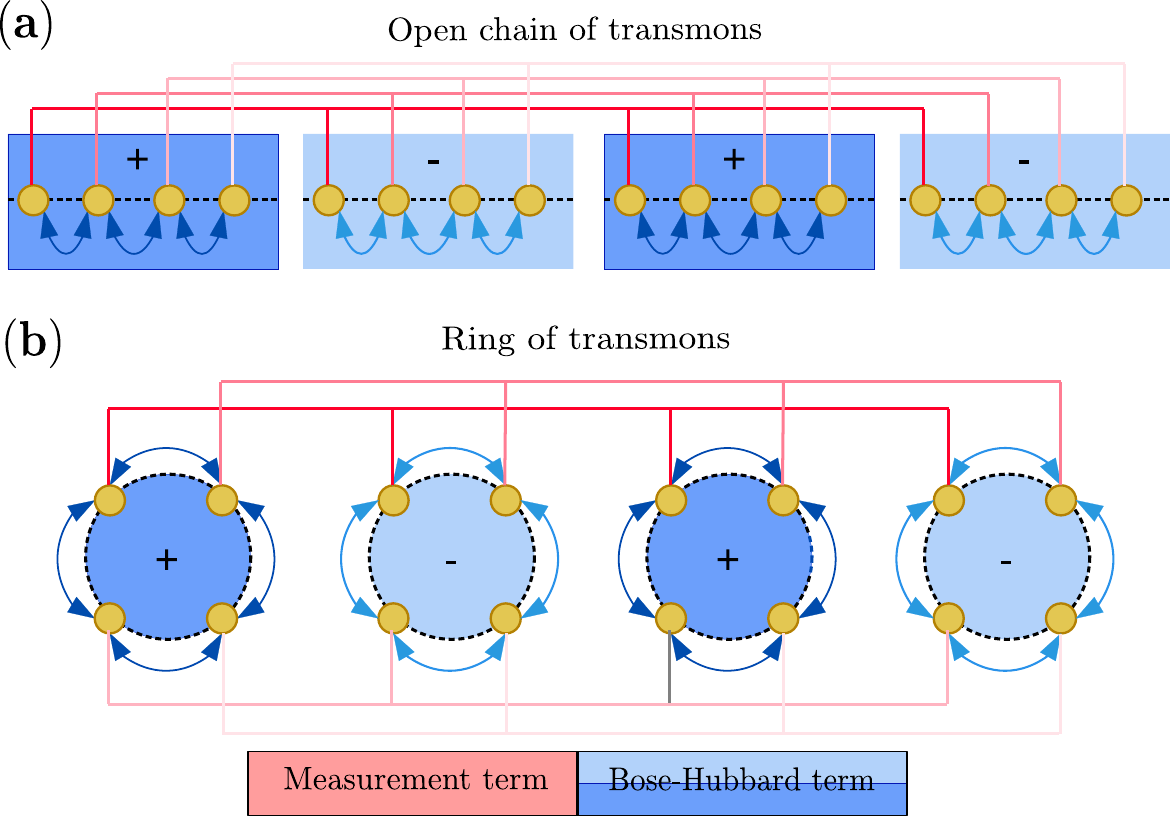}
\caption{\label{fig:enlarged_space_scheme} Scheme of the enlarged space where the effective Hamiltonian $\hat{H}_\textnormal{eff} $ is defined for a hybrid circuit of $L=4$ transmons (yellow circles) arranged (a) linearly or (b) in a ring configuration. Each rectangle represents the blocks of the replicas (dark blue for the \textit{kets} and light blue for the \textit{bras}) and the sign is given by the Eq.~\eqref{W_function}. All the terms belonging to the operator with origin in the measurements [red lines, Eq.~\eqref{HM_enlarged}] act on all the blocks of all replicas, while the terms belonging to the operator with origin in the unitary evolution of the Bose-Hubbard Hamiltonian [blue areas and arrows, Eq.~\eqref{HBH_enlarged}] act on each block independently.}
\end{figure}

The ground state of the effective Hamiltonian of Eq.~\eqref{H_eff} encapsulates the statistical information of the ensemble of trajectories of the original hybrid circuit at long times, which allows us to compute relevant trajectory-averaged quantities related to the entanglement entropy and the number of bosons. Since the effective Hamiltonian is non-Hermitian, we cannot rule out the existence of complex eigenenergies. Therefore, we define the ground state as the eigenstate with the eigenenergy having the smallest real part. Unlike previous works~\citep{bao21}, in this case, the effective Hamiltonian given by $\check{\mathcal{H}}_{\textnormal{eff}}=\check{\mathcal{H}}_\mathcal{M}+i\check{\mathcal{H}}_{\mathcal{U}} $ can be non-Hermitian. This kind of dynamics has been used before for describing other continuously measured systems~\cite{thiel20, dubey21, gopalakrishnan21}. Note that the term $\check{\mathcal{H}}_{\mathcal{U}}$ is Hermitian since it has its origin in the replicated Bose-Hubbard Hamiltonian, while the term $\check{\mathcal{H}}_{\mathcal{M}}$ has its origin in the measurements and its hermicity will depend on the type of measurements implemented in the circuit: the \textit{standard} measurement yields a Hermitian operator, while a \textit{feedback} measurement yields a non-Hermitian operator since it is real and non-symmetric.

To study the ground state of the Hamiltonian in Eq.~\eqref{H_eff} it is useful to interpret it as an effective Bose-Hubbard Hamiltonian in an enlarged space so that the bosons move in an enlarged space and have additional interacting terms arising from measurements. This implies that the $d^{4L}$-dimensional effective Hamiltonian $\check{\mathcal{H}}_{\textnormal{eff}}=\check{\mathcal{H}}_\mathcal{M}+i\check{\mathcal{H}}_{\mathcal{U}}$ constructed by the tensor product of four copies of operators describing dynamics in a $L$-sites real space becomes a $d^{4L}$-dimensional effective Hamiltonian $ \hat{H}_{\textnormal{eff}}=\hat{H}^M+i\hat{H}^{\rm BH} $ formed by operators describing the dynamics in a $4L$-sites enlarged space. Therefore, the original circuit consisting of $L$ transmons defines four different blocks in this enlarged space of $4L$ sites: $[1,L]$, $[L+1,2L]$, $[2L+1,3L]$ and $[3L+1,4L]$ (Fig.~\ref{fig:enlarged_space_scheme}). In this way, the term $\hat{H}^M$ represents an interaction between blocks at the sites $l,l+L,l+2L,l+3L$ where $l$ corresponds to the site $\ell$ of the circuit where the measurement is performed, while $\hat{H}^{\rm BH}$ is simply a Bose-Hubbard Hamiltonian of $4L$ sites, whose parameters, i.e.~$J$, $\omega$, and $U$, have different signs between contiguous blocks. Note that the local terms of the on-site energies $\omega$ and interactions $U$ cover the full $4L$ space, while the hopping terms $J$ are zero between the blocks. Therefore, we can reinterpret the terms of the effective Hamiltonian~\eqref{H_eff} in such a way that
\begin{equation}
    \hat{H}_{\textnormal{eff}}=\hat{H}^M+i\hat{H}^{\rm BH},
    \label{H_eff_enlarged}
\end{equation}
where the measurement and Bose-Hubbard terms are
\begin{align}
    \hat{H}^{\rm M} &=\Gamma \sum_{l=1}^{L} \left(\hat{I}-\sum_{n=0}^{d-1}  \hat{P}_{l,n} \hat{P}_{l+L,n} \hat{P}_{l+2L,n} \hat{P}_{l+3L,n}  \right),  \label{HM_enlarged}\\
    \hat{H}^{\rm BH} &=\sum_{l=1}^{4L} W_l\left[ \omega \hat{n}_{l}+J\left( \hat{a}_l^\dagger \hat{a}^{}_{l+1} + \text{h.c.} \right) -\frac{U}{2}\hat{n}_l (\hat{n}_l-\hat{I}) \right],
    \label{HBH_enlarged}
\end{align}
with the sign function
\begin{equation}
W_l=\begin{cases}
+1, l \in \left[1,L\right] \cup \left[2L+1,3L\right],  \\
-1, l \in \left[L+1,2L\right] \cup \left[3L+1,4L\right].
\end{cases} 
\label{W_function}
\end{equation}
Note that Eqs.~\eqref{H_eff_enlarged}-\eqref{W_function} apply to both \textit{standard} and \textit{feedback} measurements; the only difference lies in the specific projector $\hat{P}_{l,n}$ required for each case. We substitute $\hat{P}_{l,n}$ with $\hat{P}_{l,n}^{\text{st}}$ for \textit{standard} and with $\hat{P}_{l,n}^{\text{fb}}$ for \textit{feedback} measurements. For exact details about obtaining the vectorized states from operators and the averaged observables, see App.~\ref{appsec:replica_method_effective_hamiltonian_enlarged}. 

Since the effective Hamiltonian describes the statistical properties of the circuit trajectories, some of its features are expected to contain information about the phase transition.~MIPT is directly related to a spontaneous symmetry breaking, which arises because the relevant quantities,  such as entanglement entropy and fluctuations of observables, for the phase transition, can be observed only in the nonlinear moments of the density matrix, whose time evolution can be expressed by the evolution of $n \geq 2$ replicas~\citep{bao21}. In this way, the dynamic has a permutation symmetry between the $n$ replicas, that is, both between \textit{kets} and between \textit{bras}, which is preserved in the area-law phase and broken in the volume-law phase.

Thus, we can study the nature of the ground states of $\hat{H}_{\textnormal{eff}}$ to understand the spontaneous symmetry breaking in limiting cases of the measurement rate $\Gamma$~\citep{altland22}. Deep in the area-law phase, when $\Gamma \to \infty$, the term $\hat{H}^M $ dominates in the effective Hamiltonian. For the case of \textit{feedback} measurements, there is a non-degenerate ground state that is independent of the number of replicas and it preserves the permutation symmetry. It can be seen in Fig.~\ref{fig:enlarged_space_scheme} that adding replicas does not affect the degeneracy of the ground state. Note that in the case of using a \textit{standard} measurement, there are $d^L$ degenerated ground states corresponding to the possible outcomes of the measurements that are also independent of the number of replicas. When $\Gamma\to 0$, the term $\hat{H}^{\rm BH}$ predominates in the effective Hamiltonian, in which, as we see in Fig.~\ref{fig:enlarged_space_scheme}, the number of terms increases by adding replicas. However, in the effective Hamiltonian, $\hat{H}^{\rm BH}$ is purely imaginary at order $dt$, and therefore does not have a well-defined ground state. This means that all eigenstates have a zero real part and the degeneracy trivially increases by increasing the number of replicas. But at higher orders of $dt$, $\hat{H}^{\rm BH}$ does have real components~\citep{bao21}, whose terms also depend on the number of replicas, thus increasing the degeneracy of the ground states with the number of replicas, and breaking the permutation symmetry. Note that we cannot directly use the second-order term in the expression for the evolution operator in the replica space, Eq.~\eqref{circuit_integral2}, since it would be necessary to have previously performed a second-order Suzuki-Trotter decomposition, making the analytical expression considerably cumbersome. These are general arguments, to prove the existence of an MIPT, each case must be addressed individually and verified through numerical finite-size scaling analysis.

\subsection{Perturbation theory for non-Hermitian Hamiltonians}
The non-hermicity of the Hamiltonian of Eq.~\eqref{H_eff_enlarged} hinders the use of standard quantum mechanical techniques to determine its ground state, mainly due to the non-orthogonality of eigenvectors. Thus, we will follow the bi-orthogonal quantum mechanical formalism~\citep{sterheim1972, Brody_2013}, in which we obtain the eigenstates and eigenenergies for the operator $\hat{H}_{\textnormal{eff}}\ket{\phi_m}=E_m\ket{\phi_m}$ and its Hermitian conjugate $\hat{H}_{\textnormal{eff}}^\dagger \ket{{\varphi}_m}={\varepsilon}_m\ket{{\varphi}_m}$. In this way, we have that $ {\varepsilon}_n=E_n^*$ is fulfilled, and $ \{ {\varphi}_m,\phi_m \} $ forms a bi-orthogonal set such that $ \braket{{\varphi}_m|\phi_n}=\delta_{m,n}$ and $\hat{I}=\sum_m \ket{\phi_m}\bra{{\varphi}_m}$. The ground state will be defined as the eigenstate with the lowest real part of the eigenenergy $\mathfrak{Re}(E_m)$.

Since obtaining the exact analytical expression for the ground state of $\hat{H}_{\textnormal{eff}}$ is rather complicated, we will make use of the perturbation theory. We can define two regimes as a function of the measurement rate $\Gamma$: $\Gamma/J \ll 1 $ where $\hat{H}^{\rm M}$ acts as a perturbation; and $\Gamma/J \gg 1 $ where $\hat{H}^{\rm BH}$ acts as an imaginary perturbation. In this work, we will focus on the area-law phase. Briefly, we expand the eigenenergies and eigenstates in terms of the parameter $\lambda \equiv J/\Gamma \ll 1$ and solve the Schrödinger equation for $\hat{H}_{\textnormal{eff}} $ and $\hat{H}_{\textnormal{eff}}^\dagger$ (see Eqs.~\eqref{H_expanded_lambda}-\eqref{Hdagger_expanded_lambda} in App.~\ref{appsec:non-Hermitian}). Up to second order in $\lambda$, the normalized non-degenerate state $\ket{\phi_\alpha}$ is
\begin{align}
    \ket{\phi_\alpha}=&\left(1-\frac{\lambda^2}{2}\sum_{n \neq \alpha} \left\vert \frac{V_{{n}\alpha}}{(E_\alpha^{(0)}-E_n^{(0)} )} \right\vert^2 \right)\ket{\phi_\alpha^{(0)}} \nonumber \\
    &\quad +i\lambda \sum_{n \neq \alpha} \frac{V_{{n}\alpha}}{(E_\alpha^{(0)}-E_n^{(0)} )}\ket{\phi_n^{(0)}} \nonumber \\
    &\quad -\lambda^2 \sum_{n,m \neq \alpha} \frac{V_{{n}m}V_{{m}\alpha}}{(E_\alpha^{(0)}-E_n^{(0)} )(E_\alpha^{(0)}-E_m^{(0)} )}\ket{\phi_n^{(0)}} \nonumber \\
    &\quad +\lambda^2 \sum_{n \neq \alpha} \frac{V_{{\alpha}\alpha}V_{{n}\alpha}}{(E_\alpha^{(0)}-E_n^{(0)} )^2}\ket{\phi_n^{(0)}}, \label{perturbation_correction}
\end{align}
and the corresponding energy is 
\begin{equation}
   E_\alpha=E_\alpha^{(0)}+i\lambda V_{{\alpha}\alpha}-\lambda^2 \sum_{n \neq \alpha} \frac{V_{{\alpha}n}V_{{n}\alpha}}{( E_\alpha^{(0)}-E_n^{(0)} )},
    \label{perturbation_energy_correction}
\end{equation}
where the matrix elements are $V_{{a}b}=\braket{{\varphi}_a^{(0)} | \hat{H}^{\rm BH}| \phi_b^{(0)} } $, and  $ \{ {\varphi}_m^{(0)},\phi_m^{(0)} \} $  is the bi-orthogonal basis of the unperturbed Hamiltonian $\hat{H}^M$. It can be shown that up to the second order in $\lambda$, the on-site energy $\sum_l^{4L} W_l \hat{n}_l $ and interaction $\sum_l^{4L} W_l \hat{n}_l(\hat{n}_l-\hat{I}) $ terms do not play any role in obtaining the half-chain entanglement or boson number of Eqs.~\eqref{observable_entropy}-\eqref{observable_number} because their effects vanish by symmetry. We obtain the same result by considering the direct average over different circuit realizations, see App.~\ref{appsec:circuit_average}. This implies that in the high measurement regime, where $J/\Gamma \ll 1 $ (i.e.~deep in the area-law phase), we only need to focus on the hopping terms $\sum_l^{4L}W_l J(\hat{a}_l^\dagger \hat{a}^{}_{l+1} + \text{h.c.})$. 

\subsubsection{Feedback measurements}
\label{sec:hardcore}
To study the \textit{feedback} measurement, we need to establish a predetermined spatial profile for the local number of bosons $\{\alpha_1, \alpha_2, \dots, \alpha_L \} $ where $\alpha_\ell =0,1, \dots,d-1$, which will be forced by the measurement at each site in the original circuit $\ell \in [1,L]$. Although we focus on non-absorbing states, they can be achieved by setting all $\alpha_\ell$ either to zero or to $d-1$. The associated projectors will be given by $ \hat{P}_{\ell,m}^{\textnormal{fb}}=\ket{\alpha_\ell}\bra{m} $, where $ \sum_{m=0}^{d-1} \hat{P}_{\ell,m}^{\textnormal{fb}\dagger} \hat{P}_{\ell,m}^\textnormal{fb} = \hat{I} $. Since $\hat{H}^M$ is non-Hermitian because it is a real non-symmetric operator that does not conserve the total number of bosons, we cannot use the boson number basis of $\hat{N}=\sum_l^{4L} \hat{n}_l $ as the unperturbed basis, and we need to obtain the full bi-orthogonal basis explicitly. We can start by considering the bi-orthogonal basis for the unperturbed effective Hamiltonian, Eq. \eqref{HM_enlarged}, of one transmon of dimension $d$, which is given by
\begin{align}
    \ket{\phi^{(0)}}=\begin{cases}\ket{\overline{\alpha_\ell}}, & E=0, \quad  \#1 \\
    \ket{\overline{a}}-\ket{\overline{\alpha_\ell}}, & E=\Gamma,\quad  \#d-1 \\
     \ket{ijkl}, &  E=\Gamma, \quad  \#d^4-d \end{cases}\label{biortobasis1} \\
    \ket{{\varphi}^{(0)}}=\begin{cases}\sum_{x=0}^{d-1}\ket{\overline{x}}, & \tilde{E}=0, \quad \#1 \\
     \ket{\overline{b}}, & \tilde{E}=\Gamma, \quad \#d-1 \\
     \ket{mnpq}, &  \tilde{E}=\Gamma, \quad \#d^4-d \end{cases} \label{biortobasis6}
\end{align}
where $\alpha_\ell, a,b,i,j,k,l,m,n,p,q=0,1,...,d-1$ and satisfy the following conditions: $a \neq \alpha_\ell$, $b \neq \alpha_\ell$, no $i=j=k=l$, and no $m=n=p=q$. Note that we adopt the notation $\ket{\overline{n}} \equiv \ket{nnnn}$. The basis for the unperturbed effective Hamiltonian has a dimension $D=d^4$, fulfills the bi-orthonormality condition $\braket{{\varphi}_i^{(0)}|\phi_j^{(0)}}=\delta_{ij}$, and has a non-degenerate ground state. For obtaining the eigenstates for an arbitrary number of transmons $L$, we consider all possible combinations of Eqs.~\eqref{biortobasis1}-\eqref{biortobasis6}, such that
\begin{align}
\ket{\Phi_{i_1,i_2,...,i_L}^{\rm{fb}{(0)}}}=\ket{\phi_{i_i}^{(0)}}\ket{\phi_{i_2}^{(0)}} \cdots \ket{\phi_{i_L}^{(0)}}, \label{biortogonal1_states}\\
\ket{{\Theta}_{i_1,i_2,...,i_L}^{\rm{fb}(0)}}=\ket{{\varphi}_{i_i}^{(0)}}\ket{{\varphi}_{i_2}^{(0)}} \cdots \ket{{\varphi}_{i_L}^{(0)}}, \label{biortogonal2_states}
\end{align}
where $i_1,i_2,..,i_L$ are indices that run over all $d^4$ possibilities in the  Eqs.~\eqref{biortobasis1}-\eqref{biortobasis6}, $ \braket{{\Theta}_{i_1,i_2,...,i_L}^{(0)}|\Phi_{j_1,j_2,...,j_L}^{(0)}}=\delta_{i_1,j_1}\delta_{i_2,j_2}\cdots\delta_{i_L,j_L}$, and the total dimension is $D=d^{4L}$. Taking into account Eqs. \eqref{HM_enlarged}, \eqref{biortobasis1} and \eqref{biortobasis6}, we can see that the energy of the states is given by the number of times each site differs from the pattern $\alpha_\ell$, i.e. $0\leq E_{i_1,i_2,...,i_L}=\Gamma \sum_{\ell=1}^L (1- \delta_{i_\ell,1}) \leq L\Gamma $, such that the ground states have an energy $E_{1,1,...,1}={\varepsilon}_{1,1,...,1}=0$, and are given by
\begin{align}
    &\ket{\Phi_{1,1,...,1}^{\rm{fb}(0)}}=\ket{\overline{\alpha_1}}\ket{\overline{\alpha_2}} \cdots \ket{\overline{\alpha_L}}, \label{biortogonal1_ground}\\
    &\ket{{\Theta}_{1,1,...,1}^{\rm{fb}(0)}}=\sum_{x_1,x_2,...x_L=0}^{d-1} \ket{\overline{x_1}}\ket{\overline{x_2}} \cdots \ket{\overline{x_L}}, \label{biortogonal2_ground}
\end{align}
where ${\alpha_1,\alpha_2,...,\alpha_L}$ are the boson number subspace projected at each site. For obtaining the bi-orthogonal basis, we have made use of a different notation which eases the calculations, such that the composite basis for $L \geq 2$ should be understood as
\begin{align}
    \ket{i_1j_1k_1l_1}&\ket{i_2j_2k_2l_2}\cdots\ket{i_Lj_Lk_Ll_L}= \nonumber \\ &\ket{i_1i_2...i_Lj_1j_2...j_Lk_1k_2...k_Ll_1l_2...l_L},
\end{align}
where $i,j,k,l$ refer to the different blocks of the enlarged space, which arise from the \textit{kets} and \textit{bras} of the two replicas. As regards the perturbation theory, we will use the state of Eq.~\eqref{biortogonal1_ground} in Eq.~\eqref{perturbation_correction} as the non-degenerate ground state, considering inner products with states from Eq.~\eqref{biortogonal1_states} excluding the other bi-orthogonal ground state of Eq.~\eqref{biortogonal2_ground} when necessary. 

\subsubsection{Standard measurements}
In the case of \textit{standard} measurements, there are multiple degenerate ground states, specifically $d^L$ states. To simplify this degeneracy, we can consider a specific manifold determined by the definite initial state of the number of bosons, as $N$ remains constant during both unitary and non-unitary dynamics. The dimension of this manifold is given by $\binom{L+N-1}{N}$. Instead of developing a complete degenerate-perturbation theory, we focus here on a specific case and employ certain arguments, which we describe below. These arguments enable us to directly utilize Eq.~\eqref{perturbation_correction}. Since, in the rest of the paper, we are going to consider hard-core bosons and a half-filling initial state, we can further simplify the dimension to $\binom{L}{L/2}$. The energy correction up to the second order, as described in Eq.~\eqref{perturbation_energy_correction}, breaks the degeneracy for the $L=2$ cases based on the number of density walls. The minimum value corresponds to a single-density wall where all bosons are stacked on one side of the chain. The remaining degeneracy lies between symmetric and antisymmetric superpositions of bosons stacked on the left and right sides. It can be proven that these two states never intersect in subsequent perturbation orders, and the degeneracy is eliminated at an order of $4(L/2)^2$, being the symmetric state the one with the smallest correction, such that the ground state is given by
\begin{equation}
    \ket{\Phi_0^{st}}=\frac{1}{\sqrt{2}} \left( \ket{1,\dots,1,0,\dots,0}+\ket{0,\dots,0,1,\dots,1} \right).
    \label{standard_ground}
\end{equation}
These ideas have been numerically proven for up to $4$ transmons and can be extended to larger systems. Since the standard $\hat{H}^M$ is Hermitian, we can utilize the usual basis in the number of bosons.

\subsection{Observables for hard-core bosons in the area-law phase}
In this subsection, we examine the repeatedly measured transmon chain via modeling the dynamics through the hard-core bosons starting from a Néel state $\ket{\psi_{t=0}}=\ket{1010\ldots10}$. In this section, we calculate the quantities from Eqs.~\eqref{observable_entropy} and \eqref{observable_number} for the perturbed ground states of Eqs.~\eqref{biortogonal1_ground} and \eqref{standard_ground}. These quantities ultimately correspond to the trajectory-averaged observables in the proper replica limit. The final result will be presented here, while App.~\ref{appsec:averaged_observable} provides details and a didactic example on related non-physical projective measurements.

\subsubsection{Feedback measurements}
First, we consider the \textit{feedback} measurements projecting to the half-filling sector consisting of operators $\hat{P}_{\ell,m}^{\textnormal{fb}}=\ket{1}\bra{m}$ and $\hat{P}_{\ell',m}^{\textnormal{fb}}=\ket{0}\bra{m}$, for $\ell$ and $\ell'$ odd and even, respectively. In other words, we make projections to $\ket{1}$ at odd sites and $\ket{0}$ at even sites, that is, the \emph{feedback} measurements try to project the system towards the state $\ket{101\ldots10}$.

Up to the second order in $ J/\Gamma $, i.e.~deep in the area-law phase, we have that for the half of the chain $S_{L/2}^{(2)} \simeq (J/\Gamma)^2$ and $\mathcal{F}_{L/2}^{(2)} \simeq (J/\Gamma)^2/2$, which implies that entanglement entropy and fluctuations related quantities depend on the square of the measurement rate but not on the subsystem size, which corresponds to the proper behavior in the area-law phase. The number of bosons provides the most interesting results as they are easy to measure experimentally. Even when using a \textit{feedback} measurement that does not conserve the total number of bosons, the average quantity $N_{L}^{(2)}=L/2$ remains constant in the area-law phase. Higher orders are expected to yield the same constant value due to the symmetry of perturbation and ground state, indicating a fixed trajectory-averaged total number of bosons $N_\alpha=\sum_{\ell=1}^{L} \alpha_\ell $ for any $\Gamma >0$. However, this may not hold for a generic dimension $d$ and spatial profile. In the enlarged space, fourth-order perturbation theory reveals states that can be mapped to trajectories in the original circuit with different total numbers of bosons (also checked numerically), see App.~\ref{appsec:averaged_observable_non_conservation}. Thus, starting from a defined boson number state, a system governed by a non-Hermitian Hamiltonian exhibits states with different total boson numbers when perturbed by an anti-Hermitian Hamiltonian. This implies that the distribution of individual trajectories, as depicted in Fig.~\ref{fig:postselection_nopostelection}, carries information about the measurement rate $\Gamma$, even though the mean total number of bosons remains constant. For a small measurement rate, we can expect a state resembling an ergodic phase, where all basis states are expected to be visited equally, resulting in a Gaussian distribution of the total boson number, confirmed numerically in both the enlarged space and original circuit. In contrast, the area-law phase features trajectories following a delta distribution centered at $N_\alpha$. 

Another interesting quantity is the trajectory-averaged number of bosons at a single location in the chain. Deep in the area-law phase, these are given by $N_{\ell}^{(2)} \simeq (J/\Gamma)^2/2 $ and $N_{\ell'}^{(2)} \simeq 1-(J/\Gamma)^2/2 $, for even $\ell$ and odd $\ell'$ sites, respectively, see Eqs.~\eqref{N_l_even}-\eqref{N_l_odd}. While the specific value does not provide information about the phase due to monotonic changes for any $\Gamma > 0$, studying their distribution is meaningful in the sense explained for the total number of bosons. For two-dimensional subsystems i.e. hard-core bosons, there are only two possible values for all system sizes. For a small measurement rate, the number of bosons follows a uniform distribution, while for a large measurement rate, it forms a delta distribution centered at $\alpha_\ell=0 $ and $\alpha_{\ell'}=1 $ for even $\ell $ and odd $\ell' $ sites, respectively. Since for this observable, there is no size-dependent effect, we can expect the phase transition to coincide for a value $\Gamma_c$ for which the distribution of the total number of bosons fits equally well for both theoretical distributions. Note that while these quantities do not undergo the same phase transition as the corresponding entanglement phase transition for individual trajectories, they coincide because the statistics of the states in each phase are connected to a simple observable since the feedback probability is one. 

Interestingly, by determining the value of $\Gamma$ that aligns the observed distribution with the theoretical ones, we can derive a rough estimate of the critical measurement rate $\Gamma_c$ associated with the phase transition. For this purpose, we use a general distance measure, such as $d(\textnormal{obs},\textnormal{theo})=\sum_n^{N_{\rm bins}} |\textnormal{obs}(n)-\textnormal{theo}(n)|^s $, for any positive integer $s$. For an even site, the two theoretical distributions coincide for $p_0=3/4$ and $p_1=1/4$, where $p_0$ and $p_1$ refer to the proportion of results with $0$ and $1$ boson, respectively. Taking into account that $N_{\ell}^{(2)} \simeq (J/\Gamma)^2/2 $, we find $\Gamma_c^{d}/J=\sqrt{2}$ (or $p_c^{d} \approx 0.03$ for comparison with numerical results). Note that this estimate may change when considering higher orders in perturbation theory. 

\subsubsection{Standard measurements}
For the \textit{standard} measurement, we observe similar scaling for $S_{L/2}^{(2)} \approx (\overline{J}/\Gamma)^2$ and $\mathcal{F}_{L/2}^{(2)} \approx (J/\Gamma)^2/2$. The total number of bosons is constant $N_{L}^{(2)}=L/2 $, but in this case, it is a conserved quantity and remains constant for all trajectories, as we will show below. However, the number of bosons at a single site is also constant $N_{\ell}^{(2)} \simeq 1/2 $, although individual trajectories will have different values. While the \textit{standard} measurement yields the same statistical behavior for different phases, the measured observable value is constant regardless of the measurement rate. On the other hand, the quantities obtained through the replica method and statistical arguments in Sec.~\ref{sec:statistics} are averaged over trajectories. In the following section, numerical simulations are conducted to analyze the distributions of measured observables and study individual trajectories.

\section{Numerical simulations}\label{sec:numerics}
\begin{figure*}
\includegraphics[width=\linewidth]{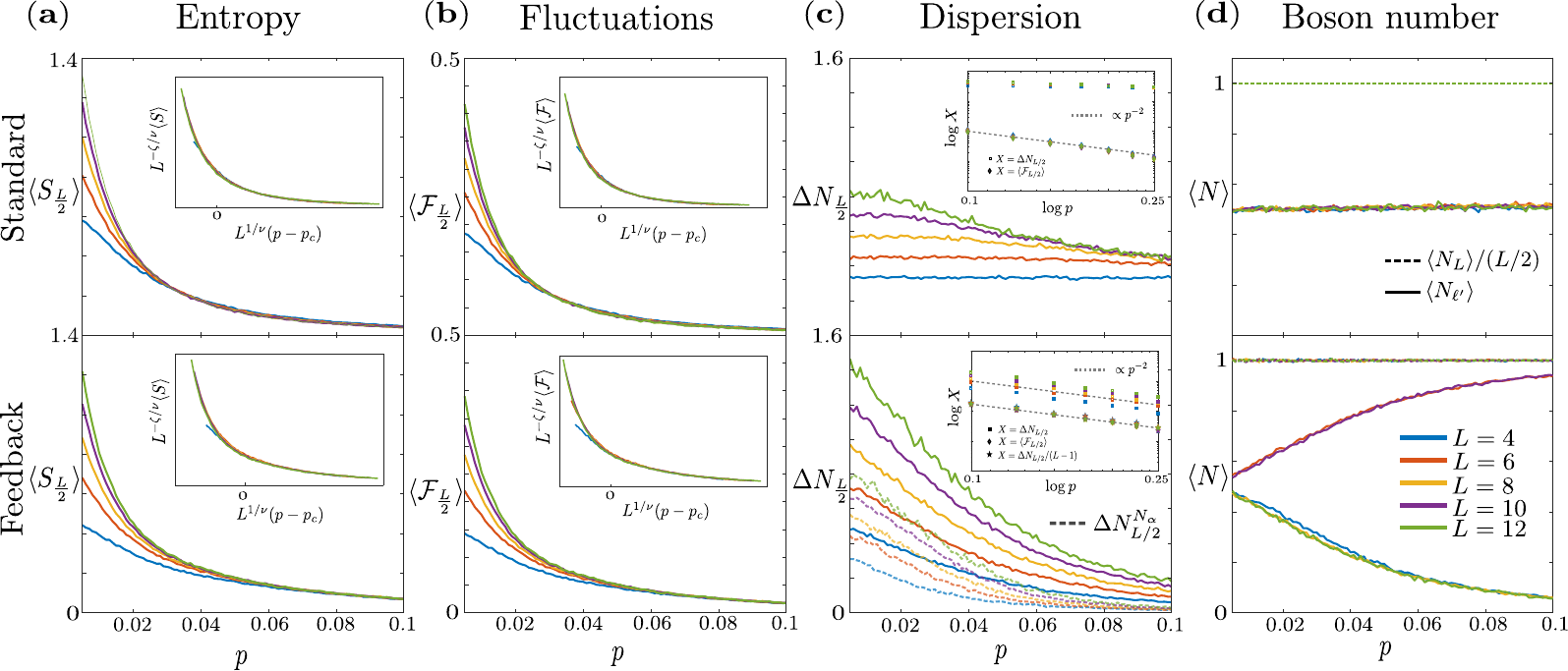}
\caption{\label{fig:numerical_results_1} Numerical simulations for a repeatedly-measured transmon array in the hard-core boson limit using the \textit{standard} and \textit{feedback} measurements for system sizes $L=4,6,8,10,$ and $12$. (a) von Neumann entanglement entropy and (b) fluctuations of the half-chain number operator as a function of the measurement probability $p$. The insets show the finite-size scaling analysis using the Ansatz $L^{-\zeta/\nu} \braket{S/\mathcal{F}}=f[L^{-1/\nu}(p-p_c) ]$, where $p_c$ is the critical parameter, $\zeta$ and $\nu$ are the scaling exponents and $f[x]$ is an unknown function~\citep{houdayer04}. (c) Dispersion of the half-chain boson number as a function of the measurement probability~$p$. For the \textit{feedback} measurements we also include the dispersion in the specific total photon number sectors $N_{\overline{\alpha}}$ defined at $t=T$ (light-colored dashed lines), see App.~\ref{appsec:circuit_average}. The insets show the comparison of the fluctuations of the number operator (diamonds) and the dispersion in the number of bosons (squares) with the $p^{-2}$ scaling (gray dashed lines) for larger values of the measurement probability. For the \textit{feedback} measurements we also show the dispersion divided by $L-1$ (stars), see App.~\ref{appsec:circuit_average}. (d) The full-chain $\braket{N_L}$ (dashed lines) and the center-site $\braket{N_{\ell'}}$ (solid lines) boson number as a function of the measurement probability~$p$. The results are averaged over $10^4$ iterations except for $L=12$, where the iteration count is $5\cdot10^3$. The other calculation parameters are $TJ=20$ and $dtJ=0.02$.}
\end{figure*}

Once we have analytically demonstrated the utility of using \textit{feedback} measurements to define observables that contain statistical information about the system without the need for explicit post-selection, we now demonstrate this using numerical simulations. We evolve the chain of $L$ transmons unitarily by using methods of exact diagonalization for a time step $dt$ after which each site is measured with probability $p$. This whole block is repeated for $T/dt$ times, where $T$ is chosen long enough to guarantee a steady state. The observables are computed at the end of time evolution $t=T$. The initial state of the system is the product state $\ket{10}^{\otimes L/2}$, which has a definite total number of bosons given by $N=L/2$. The results shown below correspond to a chain of two-dimensional subsystems (qubits). The numerical results for higher-dimensional sub-systems with  \textit{standard} measurements are shown in App.~\ref{appsec:numerical_simulation}. To reduce the burden of trajectory averaging in the numerical simulations we assume no disorder in the parameters.

\subsection{Phase transitions in transmon arrays by repeated standard and feedback measurements}
We are now interested in the von Neumann half-chain entanglement entropy $\braket{S_{L/2}}$, the fluctuations $\braket{\mathcal{F}_{L/2}}$ and the dispersion $\Delta N_{L/2}$ of the half-chain number operator $\hat N_{L/2}$, as well as the expectation values of the whole chain and single-site number operators, $\hat N_L$ and $\hat N_\ell$, respectively. Although all the quantities are calculated from the same set of simulations, there are crucial differences regarding post-selection. In the cases of the von Neumann entropy $\braket{S_{L/2}}$ and the fluctuations of the number operator $\braket{\mathcal{F}_{L/2}}$, we compute them for each final state of each iteration of the simulation, which are actual quantum trajectories that generally can correspond to a superposition. Then we average all the results over the iterations. This reflects the need to account for explicit post-selection, i.e.,~keep track of the result of each measurement to know the exact final state. However, in the case of the boson number~$\braket{N}$ and its dispersion~$\Delta N$, we do not calculate the expectation values for the final states, but we emulate an experimentally realizable non-post-selected measurement by projecting the final state using Born's rule and then directly averaging these outcomes from all iterations. The averaged values are obtained directly from the distribution of the outcomes by $\textnormal{mean}[N]=\braket{N}$ and $\textnormal{var}[N]=\Delta N $, in the sense explained in Fig.~\ref{fig:postselection_nopostelection}. This implies that, if two trajectories are at $t=T$ in the same superposition state, it is possible to obtain different outcomes for the $\hat N$ measurement.

\subsubsection{Half-chain entanglement entropy and number fluctuation}
Figure~\ref{fig:numerical_results_1} shows the numerical results for system sizes $L=4-12$ as a function of the measurement probability using the two types of measurements for a large number of iterations of the circuit. The von Neumann entropy of the half of the chain  $\braket{S_{L/2}}$ has values dependent on subsystem size for small values of $p$ and collapses to the same size-independent values for a given $p_c$ that tend to zero, thus suggesting the existence of a transition from a size-dependent to a size-independent phase, see Fig.~\ref{fig:numerical_results_1}(a). As expected, the collapsing behavior of the iteration-averaged fluctuation of the number operator in the half of the chain of transmons $\braket{\mathcal{F}_{L/2}}$ as a function of the measurement probability coincides with that of the von Neumann entropy [Fig.~\ref{fig:numerical_results_1}(b)]. 

The values obtained in the finite-size scaling analysis [insets of Fig.~\ref{fig:numerical_results_1}(a)-(b)] for the critical parameter $p_c $ and scaling exponents $\nu$ and $\zeta$ are the following: for the \textit{standard} measurement, $p_c^{S,\rm{st}}=0.022 \pm 0.002$, $\nu^{S,\rm{st}}=2.57 \pm 0.15$, $\zeta^{S,\rm{st}}=0.53 \pm 0.06$ and $p_c^{\mathcal{F},\rm{st}} =0.024 \pm 0.002$, $\nu^\mathcal{F,\rm{st}}=2.9 \pm 0.2$, $\zeta^\mathcal{F,\rm{st}}=0.50 \pm 0.08$; and for the \textit{feedback} measurement, $p_c^{S,\rm{fb}}=0.032 \pm 0.003$, $\nu^{S,\rm{fb}}=3.4 \pm 0.2$, $\zeta^{S,\rm{fb}}=0.9 \pm 0.1$ and $p_c^{\mathcal{F},\rm{fb}} =0.028 \pm 0.004$, $\nu^\mathcal{F,\rm{fb}}=3.8 \pm 0.4$, $\zeta^\mathcal{F,\rm{fb}}=1.01 \pm 0.11$; for the von Neumann entropy and the fluctuation of the number operator in both cases. The analyzed systems are small, so the results of the finite-size scaling analysis should be considered approximations. In that sense, we can assume that both \textit{standard} and \textit{feedback} measurements yield the same critical parameters, as one could expect, at least when there is an absorbing state~\citep{piroli2023}. 

Furthermore, due to the small system size, we can not rule out the possibility that the observed phase transition is a Berezinskii-Kosterlitz-Thouless (BKT) phase transition induced by the measurements rather than a canonical MIPT~\citep{alberton21}. In this case, instead of a volume-law phase, there is a sub-extensive phase where the entanglement entropy scales logarithmically with system size. This is plausible because the hard-core bosons model studied can be transformed into a free fermions model using the Jordan-Wigner transformation~\citep{Rigol2005}, which is known to undergo a BKT phase transition with \textit{standard}~\citep{alberton21} and \textit{feedback}~\citep{buchhold22} measurements. However, it is important to note that this issue is not trivial as the transformation introduces non-local correlations~\citep{Rigol2005} that may impact the properties of entanglement entropy\cite{alberton21}.

\subsubsection{Half-chain number dispersion, full-chain and single-site~boson~numbers}
For the \textit{standard} measurement scenario, the dispersion of the half-chain boson number $\Delta N_{L/2}$ does not exhibit a smooth dependence on the subsystem size as a function of the measurement probability $p$. This behavior differs from the fluctuations, compare Fig.~\ref{fig:numerical_results_1}(b-c). In the case of the \textit{feedback} measurement and restricted only to the iterations with a total number of bosons $N_{\overline{\alpha}}=L/2$ at $t=T$ [as described in Eq.~\eqref{averaged_DeltaN}], there is a collapse of the curves for the dispersion of the half-chain boson number. However, it occurs for a higher measurement probability than in the case of $\braket{S_{L/2}} $ and $\braket{\mathcal{F}_{L/2}} $, being the size-dependent phase overestimated and not giving useful information about the exact location of the critical point. 

To have a better understanding of the averaged observables, we can compute analytically the dispersion and fluctuation by directly averaging over the circuit realizations, through Eqs.~\eqref{averaged_observables_2}-\eqref{averaged_observables_DN}. The following results are obtained for a large measurement probability close to a perfectly measured system, as described in App.~\ref{appsec:circuit_average}. Trajectory-averaged fluctuation of the number operator in the half of the chain, considering a \textit{feedback} measurement with spatial pattern $\overline{\alpha}={\alpha_1,\alpha_2,...,\alpha_L }$ and up to second order in $x\equiv(1-p)$, is given by
\begin{equation}
    \braket{\mathcal{F}_{\frac{L}{2}}} \approx x^2 \left[(\alpha_{\frac{L}{2}}+1)\alpha_{\frac{L}{2}+1}+\alpha_{\frac{L}{2}}(\alpha_{\frac{L}{2}+1}+1)\right]  \left( \frac{J}{\Gamma}\right)^2.
    \label{averaged_observables_F_L2_comp}
\end{equation}
We can also obtain the dispersion of the number of bosons in the half of the chain, which is not a trajectory-averaged quantity but the variance of the number of bosons of all circuit iterations. At high measurement rates, it is enough to consider up to the first order in $x\equiv (1-p)$ to observe that there is a dependency on the size of the system,
\begin{equation}
    \Delta N_{\frac{L}{2}} \approx x g(\overline{\alpha},L)\left(\frac{J}{\Gamma}\right)^2,
    \label{DeltaN_nonconserving}
\end{equation}
where the exact expression of the function $g(\overline{\alpha},L)$ can be found in Eq.~\eqref{DN_nonconserving_app}. However, this dependency is due to the non-conservation of the total number of bosons of the \textit{feedback} measurement. Selecting only the iterations where the total number of bosons coincides with $N_{\overline{\alpha}}=\sum_{\ell=1}^{L}\alpha_\ell=N_T$, we recover the same behavior as for the fluctuation up to the second order,
\begin{equation}
    \Delta N_{\frac{L}{2}}^{N_{\overline{\alpha}}} \approx x^2 \left[(\alpha_{\frac{L}{2}}+1)\alpha_{\frac{L}{2}+1}+\alpha_{\frac{L}{2}}(\alpha_{\frac{L}{2}+1}+1)\right]  \left( \frac{J}{\Gamma}\right)^2.
    \label{averaged_DeltaN}
\end{equation}
Note that this last result does not imply post-selection as such. It only requires measuring the number of bosons at each site at the end of time evolution as a measure of the observable of interest, keeping the results with a total number $N_{\overline{\alpha}}$ and calculating the variance of the number distribution of bosons for the half of the chain in a similar way as in Ref.~\citep{ippoliti21c}. Note that $\overline{\braket{\mathcal{F}}}$ and $\Delta N$ are essentially different quantities and have the same behavior for the \textit{feedback} measurement, but not for the \textit{standard} measurement case deep in the area-law phase. However, it is expected that $ \Delta N$ overestimates the size-dependent phase. It is important to keep in mind that $\overline{\braket{\mathcal{F}}}$ behaves similarly to $\overline{\braket{S}}$. Therefore, the interest in studying quantities such as $\Delta N^{N_{\overline{\alpha}}}$, which is equal to $\overline{\braket{\mathcal{F}}}$ for a high measurement rate, lies in measuring observables that do not necessitate post-selection, which could indirectly measure the entanglement.

\begin{figure*}
\includegraphics[width=\linewidth]{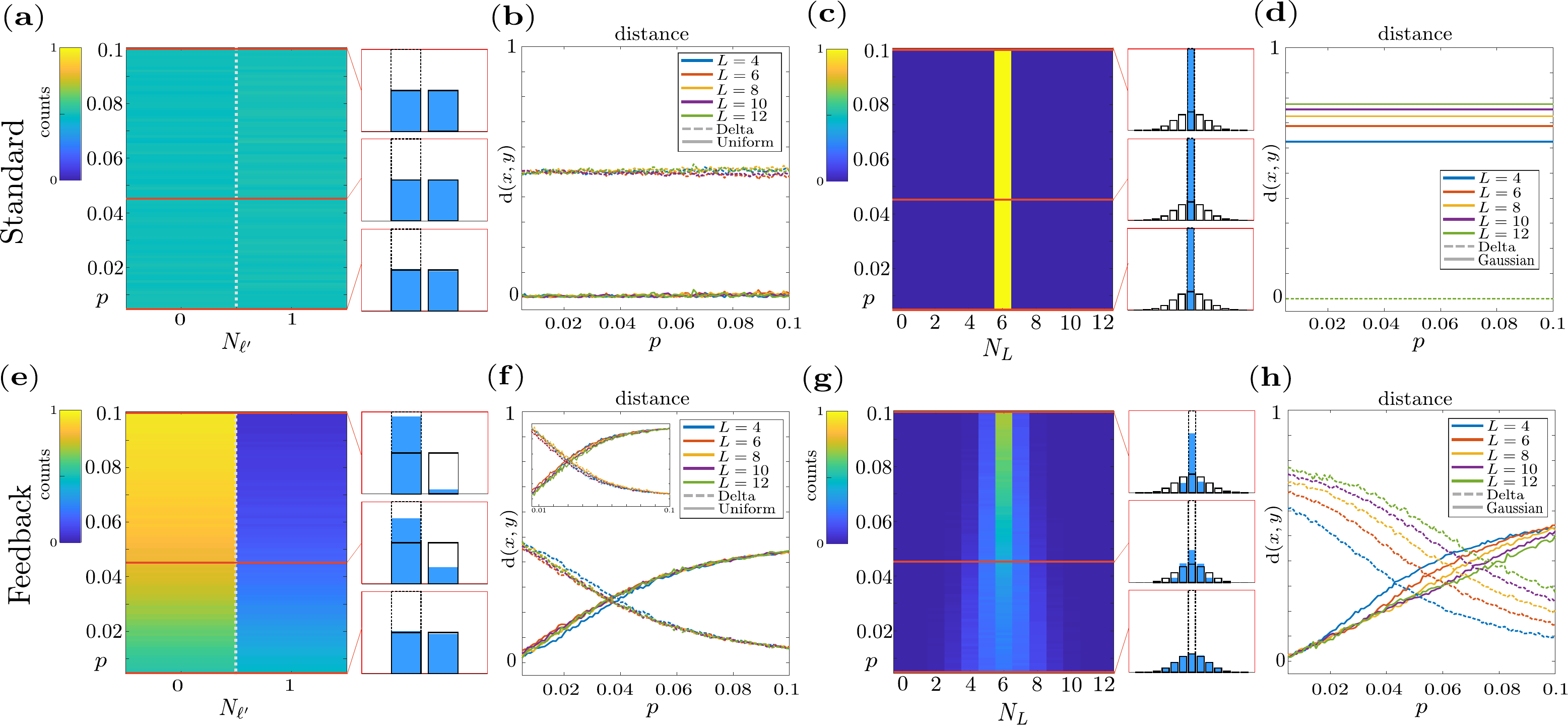}
\caption{\label{fig:numerical_results_2} Boson number distributions for \textit{standard} (a)-(d) and \textit{feedback} (e)-(h) measurements. (a),~(e) Single-site boson number distribution at the middle of the chain as a function of the measurement probability~$p$. The small panels compare the boson number distributions with the theoretical uniform and delta distributions (solid and dashed lines) at $p=\{0.005, 0.045, 0.1\}$. (b),~(f)~The distance between the single-site number distributions with respect to theoretical uniform and delta distributions. The insets show the same analysis for the data of the sector $N^{\overline{\alpha}}$. (c),~(g) The distribution of the full-chain boson number as a function of the measurement probability $p$. The small panels compare the total boson number distribution with the theoretical Gaussian and delta distributions (solid and dashed lines) at~$p=\{0.005, 0.045, 0.1\}$.  (d),~(h) The distance between the distributions with respect to theoretical delta and Gaussian as a function of measurement probability~$p$. See App.~\ref{appsec:numerical_simulation} for more details on the distributions. The results are computed for the same parameters as in Fig.~\ref{fig:numerical_results_1}.}
\end{figure*}

In the insets of Fig.~\ref{fig:numerical_results_1}(c), we show the numerical simulation-based comparison of $\braket{\mathcal{F}_{L/2}} $ and $\Delta N_{L/2}$ for larger measurement probabilities for the total dispersion of Eq.~\eqref{DeltaN_nonconserving}, thus representing the behavior in area-law phase. For the \textit{feedback} measurement, both quantities scale as $p^{-2}\propto (J/\Gamma)^2$ as predicted by the replica method for the fluctuation and by simple statistical arguments for the fluctuation and the dispersion. Interestingly, although both analytical approaches were performed at the limit of the fully measured system, the results agree with the numerical simulations for smaller measurement probabilities, but still in the area-law phase. Note that if we had considered all the iterations, that is, including cases with a different total number of bosons, the dispersion would scale in a similar way but with a factor $L-1$ depending on the size of the system as described in Eq.~\eqref{DeltaN_nonconserving}. 

The mean value of the full-chain boson number~$\braket{N_L}$, Fig.~\ref{fig:numerical_results_1}(d), results in constant values for both types of measurements. However, there are fluctuations around this value in the case of \textit{feedback} measurements, as predicted. Finally, we also show the number of bosons averaged over iterations at a single site $\braket{N_ {\ell'}}$ in the middle of the chain $\ell'=L/2$ [solid lines in Fig.~\ref{fig:numerical_results_1}(d)], whose value is constant~$\sim 1/2$ for the \textit{standard} measurement and increases from $0$ to $1/2$ for even sites (decreases from $1$ to $1/2$ for odd sites) in the case of the \textit{feedback} measurement, agreeing quite with good the result of the replica method.

\subsection{Boson number distributions}
Both measurement types conserve the full-chain boson number on average, as shown in Fig.~\ref{fig:numerical_results_1}(d). However, the mechanisms for the conservation are fundamentally different. Since the Bose-Hubbard Hamiltonian commutes with the full-chain photon number, the state before each measurement is in a definite full-chain photon number sector. The \emph{standard} measurement, that is, the measurement of local photon number, induces naturally no transitions in the full-chain photon number. Thus, we expect that the full-chain photon number has no dependence on the measurement probability for the \emph{standard} measurement. In contrast, the \emph{feedback} measurement, which includes the feedback operator projecting to $\ket{0}$ at odd sites and $\ket{1}$ at even sites, induces transitions between the photon number sectors. The transitions remove and add photons equally likely on average, thus the total photon number averaged over trajectories stays at the value of half-filling. However, for \emph{feedback} measurement, we expect that the statistical distribution of photon numbers is a function of the measurement probability. In other words, the fact that the phase transition occurs in the statistical properties of the circuit dynamics suggests studying the distribution of the measurement results of such observables, which ultimately depends on the distribution of the states over the Fock space~\citep{altland22}. 

In Fig.~\ref{fig:numerical_results_2} we show the numerical results regarding the boson number distributions and their fit to two theoretical distributions for the \textit{feedback} and \textit{standard} measurements. We do not consider any averaged quantity, but we take into account each measurement outcome for each iteration without post-selecting them, see App.~\ref{appsec:numerical_simulation} for more details. In the case of a single-site number operator $N_{\ell'}$, the \textit{standard} measurements fit the uniform distribution for all measurement probabilities [Fig.~\ref{fig:numerical_results_2}(a) and Fig.~\ref{fig:numerical_results_2}(b)], while distribution of the \textit{feedback} measurement fits a uniform distribution for small measurement probability $p$ and to a delta distribution for large $p$, with a crossing point between $p \sim 0.035 $ and $p \sim 0.04$ [Fig.~\ref{fig:numerical_results_2}(e) and Fig.~\ref{fig:numerical_results_2}(f)]. In the case of the total number of bosons, we show that for the \textit{feedback} measurement, there is an inversion in the fitting of the two distributions for different system sizes [Fig.~\ref{fig:numerical_results_2}(g) and Fig.~\ref{fig:numerical_results_2}(h)] and there is some size-dependent effect in the fittings. For the \textit{standard} measurement, all trajectories have the same total number of bosons, because the measurement preserves this symmetry thus fitting perfectly to the delta distribution~[Fig.~\ref{fig:numerical_results_2}(c) and Fig.~\ref{fig:numerical_results_2}(d)].

In the case of the feedback measurement in the area-law phase, the high measurement probability $p$ produces product states close to $\ket{\alpha_1,\alpha_2,...,\alpha_L}$. In the volume-law phase, due to ergodicity, the stationary states will correspond to states that can have a contribution by all base vectors with equal probability. Although an individual stationary state is a superposition of states belonging to the same sector, due to the presence of \textit{feedback} measurements that do not preserve the total number of bosons, different stationary states (i.e.~trajectories and iterations) could belong to sectors other than the initial state. In this sense, we measure the fit of the observed distributions to the two distributions of the extreme cases $p=\{0,1\}$ for $N_L$ and $N_{\ell'}$: in the area-law phase ($p=1$) there is a delta distribution corresponding to the value of the observable in the state $\ket{\alpha_1,\alpha_2 ,...,\alpha_L}$; and in the size-dependent phase ($0 < p \ll 1$) the distribution of the value of the observable corresponds to the one existing for a uniform distribution of the base vectors. 

Note that the theoretical distributions in the states yield different distributions for the number of bosons depending on the measurement type. In the size-dependent phase, the distribution for $N_L$ is a Gaussian centered at $N_T=L/2$ and uniform for $N_{\ell'}$, but in both cases, these distributions are computed directly by considering a uniform distribution in the vectors of the basis. In the area-law phase, the distributions are $\delta_{N_L,L/2} $ for $N_L$ and $\delta_{N_{\ell'},\alpha_{\ell'}} $ for $N_{\ell'}$ (located at $N=0$ or $N=1$, depending on the parity of the site $\ell'$ in the middle of the chain). This point is crucial because the constancy of the fitting $N_{\ell'}$ under the \textit{standard} measurement does not necessarily imply an absence of a phase transition. As in the case of the \textit{feedback} measurements, the steady states are distributed uniformly over the basis states for the ergodic phase corresponding to a uniform distribution in the number of bosons. But in the area-law phase, we must consider the states corresponding to the $2^L$ eigenstates of the \textit{standard} measurement so that the stationary states are uniformly distributed over the basis vectors, as in the case of the size-dependent phase, yielding the same results in both phases. 

In Fig.~\ref{fig:numerical_results_2}(e-h) we observe that the transition from the size-dependent phase to the size-independent phase is visible in the photon number distributions for the \emph{feedback} measurements demonstrating the change in photon number statistics. In this particular case, the crossing point of the distance measures for the photon number distribution is between $p \approx 0.035$ and $p \approx 0.04$, being in good agreement with the value of the critical parameter obtained by finite-size scaling for the iteration-averaged von Neumann entropy $p_c^{S,pre}=0.032 \pm 0.003$. We can also compare this critical parameter with the rough estimate by the replica method $p_c^{d} \approx 0.03$, which are in good agreement with the numerical results even having considered corrections only up to second order in $\overline{J}/\Gamma$ in the non-Hermitian perturbation theory. In the inset of Fig.~\ref{fig:numerical_results_2}(f), we show the same result for the data selected for the sector $N_{\overline{\alpha}}$, where there is a crossing point around $p \sim 0.03$.

Let us now consider a modified feedback measurement, where there is an additional probability $p_{\rm F}$ for applying feedback after each measurement. The \emph{standard} measurement corresponds $p_{\rm F}=0$ and the \emph{feedback} measurement corresponds to $p_{\rm F}=1$. Lowering $p_{\rm F}$ from $1$ shifts the crossing point of the distributions but does not change the critical point of the entanglement phase transition (data not shown), in the sense described in Refs.~\citep{odea22,ravindranath22}. Thus, this counterexample shows that the photon number distribution cannot be used as a general direct indicator for the entanglement phase transition.

\section{Discussion and Conclusions}
\label{sec:discussion}
In this work, we have presented a new perspective on the experimental realization of entanglement phase transitions in individual quantum trajectories induced by measurements, by means of which we can have access to information about the phase in which the system is located by monitoring simple-to-measure observables without post-selecting trajectories, using quantum measurements, denoted as \textit{feedback} measurement, in which the post-measurement state is determined in advance. For this, we have considered a superconducting circuit consisting of interacting transmons, Eq.~\eqref{Hamiltonian_BH}, on which projective measurements are applied probabilistically, Eq.~\eqref{Trotter_expansion_BH_measurements}. To access the statistical information of the dynamics where the phase transition is observed, we have used the replica method, thus obtaining analytically an effective non-Hermitian Hamiltonian, Eq.~\eqref{H_eff}, which can be interpreted as describing the dynamics of interacting bosons in an enlarged space, Eq.~\eqref{H_eff_enlarged}. Utilizing non-Hermitian perturbation theory, Eq.~\eqref{perturbation_correction}, we conducted calculations to elucidate the behavior of various quantities in the area-law phase. These quantities include those contingent on explicit post-selection, such as entanglement entropy and fluctuations of the number operator, as well as those not requiring explicit post-selection, such as the boson number. These calculations were conducted for both the \textit{standard} and \textit{feedback} measurement scenarios. Using simple statistical arguments, we have also introduced a post-selection independent quantity that describes the dispersion of the boson number, which behaves similarly to the fluctuations of the numerical operator in the area law phase. Through numerical simulations, we have demonstrated the existence of a phase transition in the entanglement properties of interacting transmons, observed in both \textit{standard} and \textit{feedback} measurements, and exhibiting similar critical properties. In the context of \textit{feedback} measurements,  we can indirectly estimate the phase and the value of the critical parameter from the distribution of the number of bosons measured in a single transmon without the need to post-select trajectories and without considering absorbing states in the feedback pattern.

The dynamics of our system, consisting of interacting transmons, can be described using the Bose-Hubbard model. To simplify our analysis and utilize available analytical tools like the replica method, we have designed a hybrid circuit involving unitary gates and measurements. In experimental systems, blocking the natural unitary dynamics of the system becomes necessary during measurements. This can be achieved by inhibiting the hopping interaction between the transmons, ensuring the isolation of the bosons during measurement procedures. This can be achieved either by detuning the transmon frequencies~\cite{Karamlou22} or having tunable couplers~\cite{arute2019}. Throughout the work, all the results are expressed as a function of the mean value of the hopping rate $J$. The choice of this parameter in the experimental set determines the other parameters and critical values. These hybrid circuits are models of open quantum circuits interacting with an environment, representing the volume-law phase where the circuit is useful for computation/communication purposes. This encourages further investigation into the limit of continuous measurements, enabling the study of the Bose-Hubbard model without creating an artificial circuit. Consequently, the knowledge gained here can be applied to understand the natural dynamics of transmons as open quantum systems, aiding in comprehending the quantum-to-classical transition.

One of the biggest issues facing the experimental implementation of an~MIPT is the need to perform an explicit post-selection of all the trajectories generated by the dynamics of the circuit in the different experiments in order to calculate the relevant quantities~\citep{Potter2022}. In this work, we propose using \textit{feedback} measurements to estimate the critical parameter value of the entanglement phase transition from simple observables. It is important to note that our findings are focused on the entanglement phase transition in individual quantum trajectories induced by \textit{feedback} measurements, which results in a corresponding change in the statistics of simple observables. This change can be utilized to identify the underlying phase transition and the critical parameter. However, we do not assert that simple observables can fully address the entanglement phase transition in individual trajectories. We also do not claim that observing the pattern in the statistics of simple observables is unequivocally connected to the entanglement phase transition in individual trajectories. Furthermore, we do not suggest the existence of a phase transition in the simple observables or the averaged density matrix that can be linked to the aforementioned entanglement phase transition. Subsequently, we will discuss possible phase transitions in the averaged density matrix. These \textit{feedback} measurements consist of standard projective measurements after which we access their outcomes classically and we apply a unitary gate to bring the system to a predetermined state. We have found through numerical simulations, and analytical analysis in the high-rate measurement regime, that fluctuations in the number operator have coinciding behavior with the entanglement entropy. This holds true for both \textit{standard} and \textit{feedback} measurements. Similar outcomes have been observed in \textit{standard} measurements~\citep{moghaddam2023}, where fluctuations have been proven to provide an exponential shortcut compared to measuring entanglement entropy, thereby reducing the cost of post-selection.

It is also interesting to compare our results using \textit{feedback} measurements for studying simple observables with other recent works involving some sort of feedback after the measurements~\citep{buchhold22, iadecola22,odea22, ravindranath22, piotr2023, piroli2023}. In these works, the averaged density matrix undergoes an absorbing state phase transition (APT), which generally belongs to a different universality class than the MIPT occurring in the entanglement of individual quantum trajectories. Unlike our model, these works assume the existence of an absorbing state that remains unchanged under the action of the unitary circuit. Once absorbed, the state cannot change, and the timescale to reach this state reveals the APT at the critical point~\citep{piotr2023,ravindranath22,piroli2023,odea22}. In our case, there is no absorbing state since the unitary dynamics modify the state selected by the measurements, and it is not clear to which universality class it belongs. Also, the order parameter has a characteristic behavior along the APT, being constant in the absorbing phase and depending on the measurement parameter on the non-absorbing phase~\citep{piotr2023,odea22,piroli2023}. In our case, analytical and numerical results show that quantities, e.g. the boson number at a single site, depend on the measurement probability for any non-zero probability. It is intriguing to explore whether the phase transitions in the averaged density matrix, with and without absorbing states, belong to the same universality class (and the absorbing state is not a necessary condition); or if they belong to different universality classes. Our approach eliminates the need for obtaining the averaged density matrix avoiding quantum state tomography. Additionally, by analyzing the results based on the sectors of the total number of bosons, we gain distinct and valuable information about the statistics using simple observables [Eq.\eqref{averaged_DeltaN}, Fig.\ref{fig:numerical_results_1}(c), and the inset in Fig.~\ref{fig:numerical_results_2}(c)]. This suggests the possibility that the \textit{feedback} measurements include critical results for different sectors in an overlapping way.

For future work, it will be interesting to conduct numerical simulations of larger systems using tensor network methods to quantitatively examine critical parameters and scaling exponents of the phase transition and to which universality classes belong in different conditions; we still have to determine whether the hard-core bosons undergo a BKT or a MIPT phase transition. Transmons provide an ideal setting for this investigation as they allow for the modification of various parameters defining different system types. For instance, studying the effect of increasing anharmonicity, that is, on-site interactions, or introducing disorder in the Bose-Hubbard Hamiltonian parameters, which scales as $dt^2$ as demonstrated, could offer a more diverse range of phases~\citep{Orell2019,mansikkamaki21, yamamoto2023}. It would also be interesting to study how the dimension of the local subsystems affects the critical properties of the~MIPT. Here we have seen preliminary results that arrays of two-dimensional subsystems with dimension, such as qubits or hard-core bosons, have a smaller critical parameter than the case where the dimension of the subsystem is larger in the case of \textit{standard} measurements (see Fig.~\ref{fig:numerical_results_1} and Fig.~\ref{fig:numerical_results_3}), as predicted for qudits~\citep{bao20}. Although in transmons, the role of the anharmonicity could be non-trivially crucial as Fig.~\ref{fig:numerical_results_3}(d) suggests. In our numerical simulations of higher dimensional transmons, we used a relatively large value of $U/J=5$, which is experimentally feasible, compared to the value of $U/J=0.14$ used in a similar model~\citep{tang20}. In that model, a phase transition from volume-law to area-law behavior was observed, with critical parameters close to those of random unitary circuit models~\citep{vasseur19}. Considering that our results in the limit of hard-core bosons ($U \to \infty$) possibly indicate a BKT phase transition, it would be interesting to investigate the influence of on-site interaction strength in arrays of system with higher local dimensions, that is, for qudits.

\section*{Acknowledgements}
We are grateful to Sami Laine, Olli Mansikkamäki, Teemu Ojanen, and Tuure Orell for useful discussions. We acknowledge financial support from the Kvantum Institute of the University of Oulu, the Academy of Finland under Grants Nos. 316619 and 320086, and the Scientific Advisory Board for Defence (MATINE), Ministry of Defence of Finland.

\newpage

\appendix
\section{Suzuki-Trotter expansion}
\label{appsec:trotter}
In this appendix, we summarize the standard Suzuki-Trotter expansion procedure, after which we include a measurement layer. We will show how to create a hybrid circuit consisting of unitary evolution and interleaved probabilistic measurements, starting from a unitary Hamiltonian that describes the natural dynamics of the system. We use a Suzuki-Trotter decomposition of the Hamiltonian, Eq.~\eqref{Hamiltonian_BH}, to design a unitary layer in terms of two-site gates~\citep{Pierkarska2018, Jaschke2018, Orell2019, Sieberer19, Barbiero2020, kargi21}, and then we add a layer of measurements (in a similar way as in Ref.~\citep{tang20}), but without the necessity of considering matrix product states~\cite{vidal03}. Let us start by considering that, after a time interval $T$, the state of the system is given by
\begin{equation}
\ket{\psi_T}=e ^{-i\hat{H} T} \ket{\psi_0},
\label{time_evolution}
\end{equation}
where $\ket{\psi_0}$ is any initial state and we have made $\hat{H}/\hbar \equiv \hat{H}$. The Hamiltonian~\eqref{Hamiltonian_BH} can be decomposed into two terms $ \hat{H}=\hat{H}_\textnormal{odd}+\hat{H}_\textnormal{even} $, where $\hat{H}_\textnormal{odd} \equiv \sum_{\text{odd } \ell}^L  \hat{H}_\ell$ and $\hat{H}_\textnormal{even} \equiv \sum_{\text{even } \ell}^L  \hat{H}_\ell$, and $\hat{H}_\ell$ is given by Ref.~\citep{vidal04}
\begin{equation}
 \hat{H}_\ell \equiv   \omega_{\ell}\hat{n}_{\ell}+ J_{\ell} (\hat{a}_{\ell}^{\dagger} \hat{a}_{\ell+1}^{ }+\text{h.c.} ) - \frac{U_{\ell}}{2} \hat{n}_{\ell} (\hat{n}_{\ell}-\hat{I} ),
 \label{Hamiltonian_Hl}
\end{equation}
where $[\hat{H}_{\ell_o} ,\hat{H}_{\ell'_o}]=[\hat{H}_{\ell_e} ,\hat{H}_{\ell'_e}]=0 $, for odd $(\ell_o,\ell'_o)$ and even $(\ell_e,\ell'_e)$ sites. Note that $ [\hat{H}_\textnormal{odd},\hat{H}_\textnormal{even} ] \neq 0$, but it does not affect the results~\citep{suzuki90}. Now, we can perform a Trotter expansion of order $q$ of the unitary time evolution operator $\hat{U}=e^{-i\hat{H}T}=e^{-i(\hat{H}_\textnormal{odd}+\hat{H}_\textnormal{even})T} $ for a small time step $dt >0 $ such that~\citep{suzuki90}
\begin{equation}
\left[ e^{-i(\hat{H}_\textnormal{odd}+\hat{H}_\textnormal{even})T} \right]^{T/dt} \approx \left[ f_q \left( \hat{U}_{H_o dt},\hat{U}_{H_e dt}  \right) \right] ^{T/dt},
\label{trotter_expansion}
\end{equation}
where $\hat{U}_{H_o dt} \equiv e^{-i\hat{H}_\textnormal{odd} dt} $, $\hat{U}_{H_e dt} \equiv e^{-i\hat{H}_\textnormal{even} dt} $ and $f_q (x,y)$ corresponds to the $q$ order expansion, where the first two orders are given by $f_1(x,y)=xy $ and $f_2(x,y)=x^{1/2}y x^{1/2}$. 

Therefore, we can express the approximation to the evolution operator $\hat{U}=e^{-i\hat{H}T} $ as a product of $\mathcal{O} \left( T/dt \right) $ operators $\hat{U}_{H_o dt} $ and $\hat{U}_{H_e dt} $, which are constituted by the product of two-body gates
\begin{equation}
\hat{U}_{H_o dt}= \prod_{\text{odd } \ell}^L e^{-i\hat{H}_\ell dt}, \quad  \hat{U}_{H_e dt}= \prod_{\text{even } \ell}^L e^{-i\hat{H}_\ell dt}.
\label{Hl_unitaries}
\end{equation}
The approximation to the time evolution in~\eqref{time_evolution} is obtained by applying iteratively $f_q ( \hat{U}_{H_o dt}, \hat{U}_{H_e dt} ) $ to $ \ket{\psi_0}$ a number of $\mathcal{O} (T/dt) $ times (involving $\mathcal{O} (T/dt) $ times the set of gates $ \hat{U}_{H_o dt} $ and $\hat{U}_{H_e dt}$). Considering $\ket{\tilde{\psi}_t} $ as the approximate evolved state at time $t$, the time evolution step is given by
\begin{equation}
\ket{\tilde{\psi}_{t+dt}}=f_q ( \hat{U}_{H_o dt}, \hat{U}_{H_e dt} )\ket{\tilde{\psi}_t},
\label{time_ev_dt}
\end{equation}
which introduces an error made by the order-$dt$ Trotter expansion~\eqref{trotter_expansion}, arising from neglecting corrections that scale as $\epsilon_{te} \sim (dt)^{2q} T^2 $ (for an error given by $ \epsilon_{te}(t) \equiv 1 - |   \braket{\psi_t | \tilde{\psi}_t}  |^2 $). Finally, we add a layer of measurement operators, which are going to be applied with a probability $p$, such that the effective time step of the circuit is expressed as
\begin{equation}
\ket{\tilde{\psi}_{t+dt}}=\prod_\ell^L M_\ell (p) f_q ( \hat{U}_{H_o dt}, \hat{U}_{H_e dt} )\ket{\tilde{\psi}_t},
\label{time_ev_measurement_dt}
\end{equation}
where $M_\ell (p)$ represents the measurements and $\ket{\tilde{\psi}_{t+dt}}$ needs to be renormalized after the measurements have been performed. It is important to note that the time scale $dt$ of the layer of measurements does not come from any approximation, but is set \textit{ad hoc}. We are keeping just the first order in $dt$ since first and second-order Trotter expansion approximations yield the same results in the posterior analysis since the effective Hamiltonian is obtained up to first order in $dt$, in such a way the Trotter expansion errors $\epsilon_{te} $ are minimal. For the same reason, in the subsequent analysis, we would obtain the same effective Hamiltonian if we had considered $q=2$.

\section{Replica method}
\label{appsec:replica_method}
Replica method is a mathematical tool that consists of considering a number of replicas $n$ of an object describing a system (such as a partition function or a density matrix), which allows us to more easily calculate certain averaged quantities. For example, let us consider that we are interested in computing the average of the free energy of a system $\overline{F[J]}=-k_BT\overline{\ln Z[J]} $, where the averaging is performed over a certain distribution in $J$. In practical examples, computing $\overline{\ln Z[J]}$ can be a cumbersome task, but we can use a trick by relying on the Taylor expansion, where we introduce artificially a replica index $n$, such that the relation $ \lim_{n \to 0} (Z^n-1)/n=\ln Z $ holds. This means that studying $(\overline{Z^n}-1)/n $ is analogous to studying $\overline{\ln Z}$ and in the \textit{replica limit} $n \to 0$ both quantities are identical. In the present case, this method is useful since it allows us to express the von Neumann entropy as
\begin{equation}
\overline{S_1}=-\overline{\tr \left[ \hat{\rho}^\prime \log \hat{\rho}^\prime \right]}=-\lim_{n \to 1} \frac{1}{n-1}\log \overline{\tr \hat{\rho}^{\prime n}},
\label{renyis_entropies}
\end{equation}
where the $n$-Rényi entropy is given by $ S_n=\left(1-n \right)^{-1} \tr \hat{\rho}^{\prime n} $ as usual. Note that we will work with a certain number of replicas and different $n$-Rényi entropies might be thought to have different critical properties, although numerical simulations indicate that this is not the case, and they share the same critical properties~\cite{bao20}. 

\subsection{Replicated space}
\label{appsec:replica_method_replicated_space}
Although in this article we follow the work done by Bao \textit{et al.}~\cite{bao21} quite faithfully, we present below all the details of the formalism for greater clarity since we have used a slightly different notation and some details of our models differ. In the last subsection, we present the arguments for studying the effective Hamiltonian as a modified Bose-Hubbard model in an enlarged space. Since we are interested in the ensemble of trajectories of the circuit dynamics, we start by labeling states for particular sequences of measurement outcomes $ m_{\bold{i}} $ and set of gate parameters $ \theta_{\bold{i}} $ (we will refer to this sequence also as trajectory):
\begin{equation}
\hat{\rho}_{m_{\bold{i}},\theta_{\bold{i}}}(t)=\hat{P}_{m_t} \hat{U}_{\theta_t}  \cdots  \hat{P}_{m_1} \hat{U}_{\theta_1} \hat{\rho}_0 \hat{U}_{\theta_1}^{\dagger} \hat{P}_{m_1}^\dagger  \cdots \hat{U}_{\theta_t}^{\dagger} \hat{P}_{m_t}^\dagger,
\label{densitevo}
\end{equation}
where $\hat{\rho}_0$ is the initial state, $\hat{U}_{\theta_{\bold{i}}} $ are the set of unitary evolution parameters, $ \hat{P}_{m_{\bold{i}}}$ are the projection operators associated with measurement outcomes $ m_\bold{i} $, and $\bold{i}$ refers to all the positions in the circuit space-time. The ensemble of states $\mathfrak{P}=\left\lbrace \hat{\rho}'_{m_{\bold{i}},\theta_{\bold{i}}}, p_{m_{\bold{i}},\theta_{\bold{i}}} \right\rbrace_{m_{\bold{i}},\theta_{\bold{i}}}$ is formed by the set of normalized quantum states $\hat{\rho}'_{m_{\bold{i}},\theta_{\bold{i}}} \equiv \hat{\rho}_{m_{\bold{i}},\theta_{\bold{i}}} /p_{m_{\bold{i}},\theta_{\bold{i}}} $ and its probability distributions $p_{m_{\bold{i}},\theta_{\bold{i}}}=p_{\theta_{\bold{i}}}p_{m_{\bold{i}}}(\theta) $. The probabilities include the probability distribution for the gate parameters $p_{\theta_\textbf{i}}$ and the measurement outcomes probability that depends on the gate distribution parameters on the state in a sense of Born's rule $ p_{m_\bold{i}} (\theta) $. Note that the states do not need to be fully measured as in the Eq.~\eqref{densitevo}, but $\mathfrak{P}$ includes also partially-measured states where we need to consider $\hat{P}_{m_i}=\hat{I} $ in those space-time events where no measurement was performed. Therefore, $ \mathfrak{P}$ includes all possible trajectories ranging from the linear unitary trajectories with no measurements to the trajectories where all the measurements have been performed, whose proportion in the ensemble will depend on the measurement probability parameter $p$. 

For studying the steady-state properties of this ensemble of states $\mathfrak{P}$, we consider the dynamics of $n$ copies of the density matrix, such that for a particular sequence of measurements and parameters, Eq.~\eqref{densitevo}, the system density matrix is $ \ket{\ket{\rho_{m_{\bold{i}},\theta_{\bold{i}}}^{(n)}}} \equiv \hat{\rho}_{m_{\bold{i}},\theta_{\bold{i}}}^{\otimes n} $. Similarly, we define operators acting in this replicated Hilbert space $ \mathcal{H}^{(n)}= \left( \mathcal{H} \otimes \mathcal{H}^* \right)^{\otimes n} $, such that the unitary and measurement operators should be treated in the following way
\begin{align}
\check{\mathcal{U}}_{\theta_i}^{(n)} \equiv ( \hat{U}_{\theta_i}  \otimes & \hat{U}_{\theta_i}^* )^{(n)} \to \nonumber \\
& \check{\mathcal{U}}_{\theta_i}^{(n)} \ket{\ket{\rho_{m_{\bold{i}},\theta_{\bold{i}}}^{(n)}}} \equiv ( \hat{U}_{\theta_i} \hat{\rho}_{m_{\bold{i}},\theta_{\bold{i}}} \hat{U}_{\theta_i}^{\dagger} )^{\otimes n},
\end{align}
\begin{align}
\check{\mathcal{M}}_{m_i}^{(n)} \equiv ( \hat{P}_{m_i} &\otimes  \hat{P}_{m_i}^\dagger )^{(n)} \to \nonumber \\
&\check{\mathcal{M}}_{m_i}^{(n)} \ket{\ket{\rho_{m_{\bold{i}},\theta_{\bold{i}}}^{(n)}}} \equiv ( \hat{P}_{m_i} \hat{\rho}_{m_{\bold{i}},\theta_{\bold{i}}} \hat{P}_{m_i}^\dagger )^{\otimes n},
\end{align}
while a regular operator $\hat{O}$ (i.e., to calculate observables), acts on the state in the following way
\begin{align}
\check{\mathcal{O}}^{(n)} \equiv ( \hat{O} \otimes & \hat{I} )^{(n)} \to \nonumber \\ 
& \check{\mathcal{O}}^{(n)} \ket{\ket{\rho_{m_{\bold{i}},\theta_{\bold{i}}}^{(n)}}} \equiv ( \hat{O} \hat{\rho}_{m_{\bold{i}},\theta_{\bold{i}}} )^{\otimes n}.
\label{replicated_operators_acts}
\end{align}

From this point on, we will work with the un-normalized averaged state of the ensemble $ \ket{\ket{ \rho^{(n)}}} = \sum_{m_{\bold{i}},\theta_{\bold{i}}} \ket{\ket{ \rho_{m_{\bold{i}},\theta_{\bold{i}}}^{(n)}}}=\sum_{m_{\bold{i}},\theta_{\bold{i}}} p_{m_{\bold{i}},\theta_{\bold{i}}}^n \ket{\ket{ {\rho'}_{m_{\bold{i}},\theta_{\bold{i}}}^{(n)}}} $, because it can be expressed as a linear function of the initial state at any time $t$, such that
\begin{align}
\ket{\ket{ \rho^{(n)}(t)}} &= \sum_{m_{\bold{i}},\theta_{\bold{i}}} \ket{\ket{ \rho_{m_{\bold{i}},\theta_{\bold{i}}}^{(n)}}} \nonumber \\
&=\sum_{m_{\bold{i}},\theta_{\bold{i}}} \check{\mathcal{M}}_{m_t} \check{\mathcal{U}}_{\theta_t} \cdots \check{\mathcal{M}}_{m_1} \check{\mathcal{U}}_{\theta_1} \ket{\ket{ \rho_0^{(n)}}} \nonumber \\
&\equiv \check{\mathcal{V}}(t) \ket{\ket{ \rho_0^{(n)}}},
\label{linear_evol_operator}
\end{align}
where the dynamics has been integrated into a linear operator $\check{\mathcal{V}}(t) $. In case of a continuous distribution for the gate parameters, we should have considered $\sum_{\theta_{\bold{i}}} \to \int {d\theta_\bold{i}} $. It is precisely this time evolution that we map exactly to an imaginary time evolution generated by an effective quantum Hamiltonian
\begin{equation}
\ket{\ket{ \rho^{(n)}(t)}}= e^{-t \check{\mathcal{H}}_{\textnormal{eff}}}\ket{\ket{ \rho_0^{(n)}}},
\label{state_evol}
\end{equation}
such that the properties of the averaged state of the ensemble $ \left.\ket{\rho^{(n)}(t)}\right\rangle$ in the long time limit are encoded in the ground state of $\check{\mathcal{H}}_{\textnormal{eff}} $, and we have considered $\hbar=1$. 

\subsection{Effective Hamiltonian in the replicated space}
\label{appsec:replica_method_effective_hamiltonian_recplicated}
To obtain the effective Hamiltonian in Eq.~\eqref{state_evol}, we need to compute the operator $\check{\mathcal{V}}(t) $ of Eq.~\eqref{linear_evol_operator}, for which we average at each space-time position $\ell^t \in \lbrace 1, ..., M \rbrace $ where $M=LT$. Each space-time position $\ell^t$ can be averaged independently, although some considerations need to be taken for averaging two-sites unitary gates. We can also average the measurement outcomes and unitary gate parameters distributions independently, obtaining $\check{\mathcal{M}}_\ell $ and $\check{\mathcal{U}}_\ell $, respectively. We assume that the gate parameters of the unitary evolution can take values from different Gaussian distributions with different mean values and variances: $\theta_\bold{i} \equiv \lbrace \omega_\bold{i},J_\bold{i},U_\bold{i}\rbrace  $: $\left\lbrace \omega, \sigma_\omega^2 \right\rbrace $ for the on-site energy $\omega_{\ell} $, $\left\lbrace J, \sigma_J^2 \right\rbrace $ for the hopping strength $J_{\ell} $ and $\left\lbrace U, \sigma_U^2 \right\rbrace $ for the interaction $U_{\ell} $. In this work, we consider will $n=2$ replicas; to simplify the notation, we will omit the superscript $(2)$ in the following calculations. Therefore, we can perform the averaging over the circuit realizations in the following way

\begin{widetext}
\begin{align}
\overline{\check{\mathcal{U}}_\ell} =&\overline{\hat{U}_\ell \otimes \hat{U}_\ell^* \otimes \hat{U}_\ell \otimes \hat{U}_\ell^*} \nonumber \\
=&\overline{e^{-i \omega_\ell \left( \check{\mathcal{n}}_{\ell} \right) dt} e^{-i J_\ell \left(\check{\mathcal{a}}_\ell^\dagger \check{\mathcal{a}}_{\ell+1}+\check{\mathcal{a}}_\ell \check{\mathcal{a}}_{\ell+1}^\dagger\right) dt}   e^{-i \frac{U_\ell}{2} \left(-\check{\mathcal{u}}_{\ell} \right) dt}}e^{\mathcal{O}(dt^2)} \nonumber \\
=&\left( \frac{1}{\sqrt{2 \pi \sigma_\omega^2}}\int_{- \infty}^{\infty} d\omega_\ell e^{- \frac{1}{2} \left(\frac{\omega_\ell-\overline{\omega}}{ \sigma_\omega}\right)^2}e^{-i \omega_\ell \left( \check{\mathcal{n}}_{\ell} \right) dt} \right)  \left( \frac{1}{\sqrt{2 \pi \sigma_J^2}}\int_{- \infty}^{\infty} dJ_\ell e^{- \frac{1}{2} \left(\frac{J_\ell-\overline{J}}{ \sigma_J}\right)^2}e^{-i J_\ell \left( \check{\mathcal{a}}_\ell^\dagger \check{\mathcal{a}}_{\ell+1}+\check{\mathcal{a}}_\ell \check{\mathcal{a}}_{\ell+1}^\dagger \right) dt} \right) \nonumber  \\
& \left( \frac{1}{\sqrt{2 \pi \sigma_U^2}}\int_{- \infty}^{\infty} dU_\ell e^{- \frac{1}{2} \left(\frac{U_\ell-\overline{U}}{ \sigma_U}\right)^2}e^{-i \frac{U_\ell}{2} \left( -\check{\mathcal{u}}_{\ell} \right) dt} \right) e^{\mathcal{O}(dt^2)} \nonumber\\
=&\left(e^{-i \omega \left( \check{\mathcal{n}}_{\ell}  \right) dt -\frac{1}{2} \sigma_\omega^2 \left(\check{\mathcal{n}}_{\ell}  \right)^2 dt^2}\right) \left( e^{-i J \left(\check{\mathcal{a}}_\ell^\dagger \check{\mathcal{a}}_{\ell+1}+\check{\mathcal{a}}_\ell \check{\mathcal{a}}_{\ell+1}^\dagger \right) dt -\frac{1}{2} \sigma_J^2 \left(\check{\mathcal{a}}_\ell^\dagger \check{\mathcal{a}}_{\ell+1}+\check{\mathcal{a}}_\ell \check{\mathcal{a}}_{\ell+1}^\dagger\right)^2 dt^2}\right) \nonumber \\
 &\left(e^{-i \frac{1}{2}U \left( -\check{\mathcal{u}}_{\ell}  \right) dt -\frac{1}{8} \sigma_U^2 \left(-\check{\mathcal{u}}_{\ell}  \right)^2 dt^2}\right)  e^{\mathcal{O}(dt^2)} \nonumber \\
=&e^{-i\check{\mathcal{A}}_\ell dt-\check{\mathcal{B}}_\ell dt^2}e^{\mathcal{O}(dt^2)},
\label{circuit_integral2}
\end{align}
where the operators of the replicated space are given by
\begin{align}
\check{\mathcal{A}}_\ell &= \omega\check{\mathcal{n}}_{\ell}+ J (\check{\mathcal{a}}_\ell^\dagger \check{\mathcal{a}}_{\ell+1}+\check{\mathcal{a}}_\ell \check{\mathcal{a}}_{\ell+1}^\dagger ) -\frac{1}{2}U \check{\mathcal{u}}_{\ell}, \\
\check{\mathcal{B}}_\ell &= \frac{1}{2} \sigma_\omega^2 \left(\check{\mathcal{n}}_{\ell}  \right)^2+\frac{1}{2} \sigma_J^2 (\check{\mathcal{a}}_\ell^\dagger \check{\mathcal{a}}_{\ell+1}+\check{\mathcal{a}}_\ell \check{\mathcal{a}}_{\ell+1}^\dagger)^2+ \frac{1}{8} \sigma_U^2 \left(\check{\mathcal{u}}_{\ell}  \right)^2, \\
\check{\mathcal{n}}_{\ell} &= \hat{n}_\ell \hat{I}\hat{I}\hat{I}-\hat{I}\hat{n}_\ell \hat{I}\hat{I}+\hat{I}\hat{I}\hat{n}_\ell \hat{I}-\hat{I}\hat{I}\hat{I}\hat{n}_\ell, \label{replicated_n} \\
\check{\mathcal{u}}_{\ell} &= [\hat{n}_\ell (\hat{n}_\ell-\hat{I} ) ]\hat{I}\hat{I}\hat{I}-\hat{I}[\hat{n}_\ell (\hat{n}_\ell-\hat{I} ) ]\hat{I}\hat{I} +\hat{I}\hat{I}[\hat{n}_\ell (\hat{n}_\ell-\hat{I} ) ]\hat{I}-\hat{I}\hat{I}\hat{I}[\hat{n}_\ell (\hat{n}_\ell-\hat{I} ) ], \label{replicated_u} \\
\check{\mathcal{a}}_\ell^\dagger \check{\mathcal{a}}_{\ell+1} &=( \hat{a}_{\ell}^\dagger \hat{I}\hat{I}\hat{I} ) (\hat{a}_{\ell+1}\hat{I}\hat{I}\hat{I} )-( \hat{I}\hat{a}_{\ell}^\dagger \hat{I}\hat{I} ) (\hat{I}\hat{a}_{\ell+1}\hat{I}\hat{I} )+( \hat{I}\hat{I}\hat{a}_{\ell}^\dagger \hat{I} ) (\hat{I}\hat{I}\hat{a}_{\ell+1}\hat{I} )-( \hat{I}\hat{I}\hat{I}\hat{a}_{\ell}^\dagger  ) (\hat{I}\hat{I}\hat{I}\hat{a}_{\ell+1} ), \label{replicated_ada} \\ 
\check{\mathcal{a}}_\ell \check{\mathcal{a}}_{\ell+1}^\dagger &=( \hat{a}_{\ell} \hat{I}\hat{I}\hat{I} ) (\hat{a}_{\ell+1}^\dagger \hat{I}\hat{I}\hat{I} )-( \hat{I}\hat{a}_{\ell} \hat{I}\hat{I} ) (\hat{I}\hat{a}_{\ell+1}^\dagger \hat{I}\hat{I} )+( \hat{I}\hat{I}\hat{a}_{\ell} \hat{I} ) (\hat{I}\hat{I}\hat{a}_{\ell+1}^\dagger \hat{I} )-( \hat{I}\hat{I}\hat{I}\hat{a}_{\ell}  ) (\hat{I}\hat{I}\hat{I}\hat{a}_{\ell+1}^\dagger ).
\label{replicated_aad}
\end{align}
\end{widetext}
Note that expressions for the operators have been simplified $\check{\mathcal{O}}=\hat{O}_1 \otimes \hat{O}_2 \otimes \hat{O}_3 \otimes \hat{O}_4 \equiv \hat{O}_1 \hat{O}_2 \hat{O}_3\hat{O}_4 $, where $\hat{O}_i$ and $\check{\mathcal{O}} $ are $d$-dimensional and $d^4$-dimensional operators, respectively, and should be understood in the sense of Eq.~\eqref{replicated_operators_acts}. In what follows, we will describe all the relevant steps followed in Eq.~\eqref{circuit_integral2}. In the first step, we have taken into account that $ e^{\hat{A}} \otimes e^{\hat{B}}=e^{\hat{A}\otimes \hat{I}+\hat{I}\otimes\hat{B}}$; note that the operators defined in~\eqref{Hamiltonian_Hl} include the four terms, each acting on different Hilbert spaces, i.e.~they commute. Note that some of the replicated operators do not commute between them: $[\check{\mathcal{n}}_{\ell},\check{\mathcal{a}}_\ell^\dagger \check{\mathcal{a}}_{\ell+1}+ \check{\mathcal{a}}_\ell \check{\mathcal{a}}_{\ell+1}^\dagger] \neq 0$ and $[\check{\mathcal{u}}_{\ell},\check{\mathcal{a}}_\ell^\dagger \check{\mathcal{a}}_{\ell+1}+ \check{\mathcal{a}}_\ell \check{\mathcal{a}}_{\ell+1}^\dagger] \neq 0$. Therefore, we have consider that $e^{\hat{A}dt+\hat{B}dt+\hat{C}dt}=e^{\hat{A}dt}e^{\hat{B}dt}e^{\hat{C}dt}e^{\mathcal{O}(dt^2)} $, where the term $ e^{\mathcal{O}(dt^2)} $ includes all the commutators of order $dt^2$ arising from the Baker-Campbell-Hausdorff (BCH) formula. In the second step, we perform a standard Gaussian integration considering the means and variances of the different parameters, and that for each integration the operators commute with themselves. In the last step,  we consider again the BCH formula and group all the $\mathcal{O}(dt^2) $ terms in $e^{\mathcal{O}(dt^2)}$. Note that, since the following commutators vanish $[(\check{\mathcal{n}}_{\ell}+\check{\mathcal{n}}_{\ell+1})^2,(\check{\mathcal{a}}_\ell^\dagger \check{\mathcal{a}}_{\ell+1}+ \check{\mathcal{a}}_\ell \check{\mathcal{a}}_{\ell+1}^\dagger)^2] $, $[(\check{\mathcal{n}}_{\ell}+\check{\mathcal{n}}_{\ell+1})^2,\check{\mathcal{a}}_\ell^\dagger \check{\mathcal{a}}_{\ell+1}+ \check{\mathcal{a}}_\ell \check{\mathcal{a}}_{\ell+1}^\dagger] $ and $[\check{\mathcal{n}}_{\ell}+\check{\mathcal{n}}_{\ell+1},(\check{\mathcal{a}}_\ell^\dagger \check{\mathcal{a}}_{\ell+1}+ \check{\mathcal{a}}_\ell \check{\mathcal{a}}_{\ell+1}^\dagger)^2] $, it is true that $e^{\mathcal{O}(dt^2)}=I  $ if there is no interactions (i.e.~$ U_\ell=0 $) and we make the change $ \check{\mathcal{n}}_\ell \to \frac{1}{2}(\check{\mathcal{n}}_\ell+\check{\mathcal{n}}_{\ell+1})$. This implies that $\overline{\check{\mathcal{U}}_\ell(U_\ell=0)}=e^{-i\check{\mathcal{A}}_\ell dt-\check{\mathcal{B}}_\ell dt^2} $, and we can have a better physical interpretation of how the variance in the distribution of parameters will affect the subsequent analysis; in the case of $U_\ell \neq 0 $ the result is~\eqref{circuit_integral2}, where the terms involving the variance of the parameters are of the same order as a complicate term involving different commutators, and the interpretation is less clear. It is important to note that to study the $dt^2$ terms properly, we need to perform the Trotter expansion up to second order (i.e.~considering $q=2$ and $f_2(x,y)=x^{1/2}yx^{1/2}$). Interestingly, non-disorder systems yield the same result up to the first order without requiring a Trotter expansion.

For the average over the measurement results, we define the following operators, which include the average over the possible outcomes,
\begin{equation}
\braket{\check{\mathcal{M}}_{\ell}}=\left(1-\Gamma_{\ell} d t \right) \check{\mathcal{I}}+\Gamma_{\ell} d t \sum_{n}^d \check{\mathcal{P}}_{\ell,n},
\end{equation}
where $ \Gamma _{\ell} $ may be viewed as the rate at which a measurement of a certain type is performed, $\check{\mathcal{I}}=\hat{I}\hat{I}\hat{I}\hat{I} $ is the identity operator in the replicated space, and the projection operator is
\begin{align}
   \check{\mathcal{P}}_{\ell,n}=&\hat{P}_{\ell,n}\hat{P}^\dagger_{\ell,n}\hat{P}_{\ell,n}\hat{P}^\dagger_{\ell,n},  
\end{align}
where $\hat{P}_{\ell,n}$ are the projectors described in the main text, different for each type of measurement. As discussed in App.~\ref{appsec:trotter}, we introduce \textit{ad hoc} the factor $dt$, but it does not derive from any Trotter expansion as the one of the Bose-Hubbard Hamiltonian. 

Finally, we obtain the transfer matrix between the state of the system at $t+dt$ and $t$ by averaging the time evolution over the gate parameters and the probabilities $p=\Gamma dt $ of applying measurement operators. For this, we only need to average over the different sites in space $\ell \in L$ for a single time $t$, i.e.~just the elements from a time step,
\begin{align}
\braket{\braket{ \rho_m(t+dt) || \check{\mathcal{T}} || \rho_m(t)}}= \nonumber \\
\braket{\braket{ \rho_m(t+dt) || \prod_{\ell'}^L \braket{\check{\mathcal{M}}_{\ell'}} \prod_{\ell}^L \overline{\check{\mathcal{U}}_{\ell}} ||  \rho_m(t)}} .
\end{align}
Considering terms up to the first order in $dt$, the transfer matrix becomes
\begin{align}
\check{\mathcal{T}}=&\prod_{\ell'}^L \braket{\check{\mathcal{M}}_{\ell'}} \prod_{\ell}^L \overline{\check{\mathcal{U}}_{\ell}} \nonumber \\
=& \prod_{\ell'}^L \left[\check{\mathcal{I}}-\Gamma_{\ell'}  d t\left(1- \sum_{n}^d \check{\mathcal{P}}_{\ell',n}\right)  \right] \nonumber \\
&\times \prod_\ell^L \left[\check{\mathcal{I}}-i\check{\mathcal{A}}_\ell dt\right]+\mathcal{O}\left( dt^2\right) \nonumber\\
=&\check{\mathcal{I}}-d t\sum_{\ell'}^L \Gamma_{\ell'} \left(\check{\mathcal{I}}- \sum_{n}^d \check{\mathcal{P}}_{\ell',n}\right) \nonumber \\
&-i dt \sum_\ell^L \check{\mathcal{A}}_\ell+\mathcal{O}\left( dt^2\right) \nonumber\\
=&\check{\mathcal{I}}-\check{\mathcal{H}}_{\mathcal{M}} d t-i\check{\mathcal{H}}_{\mathcal{U}} d t+\mathcal{O}(d t ^2) \nonumber\\
 \simeq & e^{- d t \left( \check{\mathcal{H}}_{\mathcal{M}} +i \check{\mathcal{H}}_{\mathcal{U}}\right)},
\label{transfer_exponential}
\end{align}
from which we obtain the effective Hamiltonian, which is given by
\begin{align}
\check{\mathcal{H}}_{\textnormal{eff}}=&\sum_{\ell}^L \Gamma_{\ell} \left(\check{\mathcal{I}}- \sum_{n}^d \check{\mathcal{P}}_{\ell,n} \right)+ \label{H_eff_app}\\
&+i  \left[\omega\check{\mathcal{n}}_{\ell}+ J \left(\check{\mathcal{a}}_\ell^\dagger \check{\mathcal{a}}_{\ell+1}+\check{\mathcal{a}}_\ell \check{\mathcal{a}}_{\ell+1}^\dagger \right) -\frac{1}{2}U \check{\mathcal{u}}_{\ell} \right]. \notag
\end{align}
Taking into account the expression of Eq.~\eqref{state_evol}, we deduce that the state of the system in the long time limit corresponds to the ground state of this effective Hamiltonian. Therefore, the main task will be to find the explicit expression for the ground state as a function of the different parameters and use it to obtain different quantities using the equations that we derive below. Note that this state does not represent any physical state but the ensemble of trajectories produced by the circuit dynamics. Note also that we have obtained a non-Hermitian Hamiltonian $\check{\mathcal{H}}_{\textnormal{eff}}=\check{\mathcal{H}}_\mathcal{M}+i\check{\mathcal{H}}_{\mathcal{U}}$, where $\check{\mathcal{H}}_{\mathcal{U}}$ is Hermitian, and the hermicity of $\check{\mathcal{H}}_{\mathcal{M}}$ will depend on the type of measurement (i.e.~\textit{standard} or \textit{feedback}).

\subsection{Replicated observables}
\label{appsec:replica_method_replicated_observables}
Using this formalism, we can compute different types of quantities. Generally, we are going to be interested in obtaining the average $k$-moment of an observable $\hat{O}$ over the measurement results and gate parameters:
\begin{align}
\overline{\left\langle O_k \right\rangle} &=\overline{\sum_n p_n \left[\tr \left( \hat{O}\hat{\rho}_{n}^{\prime} \right)\right]^k} \nonumber \\
&=\sum_{m_\bold{i},\theta_\bold{i}} p_{m_\bold{i},\theta_\bold{i}} \left[\frac{\tr \left( \hat{O}\hat{\rho}_{m_\bold{i},\theta_\bold{i}} \right)}{\tr \left( \hat{\rho}_{m_\bold{i},\theta_\bold{i}} \right)} \right]^k \nonumber \\
&=\lim_{n \to 1}
\frac{\braket{\braket{ \mathcal{I}^{(n)} ||  \check{\mathcal{O}}_k^{(n)} || \rho^{(n)}}}}{\braket{\braket{ \mathcal{I}^{(n)} || \rho^{(n)}} }} \nonumber \\
&=\lim_{n \to 1}O_k^{(n)},
\label{replicated_observables}
\end{align}
where $ \overline{\cdot} $ refers to the average over the set of gate parameters (that should be understood as an integral in case of a continuous variable distribution) and $\left\langle \cdot \right\rangle $ to the average over measurements outcomes. From now on, we will refer to this average $ \overline{\braket{\cdot}}$ as the trajectory average. For the expression of the operator $\hat{O}$, we need to consider its $k$-th moment acting on $n$ replicas and therefore it is defined as $\check{\mathcal{O}}_k^{(n)}=\left[ \bigotimes_{i=1}^k (\hat{O}\otimes \hat{I}) \right] \otimes (\hat{I} \otimes \hat{I})^{(n-k)}$. Note that we have introduced the norm of the replicated quantum state (i.e.~the trace of the density matrix) by an inner product between $ \ket{\ket{\rho_{m_\bold{i},\theta_\bold{i}}^{(n)}}}$ and a reference state $ \ket{\ket{\mathcal{I}}}$ such that $\braket{\braket{ \mathcal{I}^{(n)} || \rho_{m_\bold{i},\theta_\bold{i}}^{(n)}}}= \tr \left( \hat{\rho}_{m_\bold{i},\theta_\bold{i}}^{\otimes n} \right)=\tr \left(  \hat{\rho}_{m_\bold{i},\theta_\bold{i}} \right)^n $, where $ \braket{\braket{ \mu|| \sigma}} \equiv \tr \left( \hat{\mu}^\dagger \hat{\sigma} \right) $ for arbitrary states $ \hat{\mu} $ and  $ \hat{\sigma} $ in the replicated Hilbert space, and the reference state is given by
\begin{equation}
    \bra{\bra{\mathcal{I}^{(n)}}}=\sum_{\lbrace \alpha_\ell \rbrace} \bra{\bra{\alpha_1 \alpha_1, \alpha_2 \alpha_2, ... , \alpha_n \alpha_n}},
    \label{replicated_reference}
\end{equation}
where the two copies of each $\alpha_\ell$ refers to the \textit{ket} and \textit{bra}.

Using this framework, we can also express subsystems purities by introducing the trajectory-averaged $k$-th purity of a subsystem $A$ defined as $\overline{\braket{\mu_{k,A}}}=\overline{\sum_{m_\bold{i}} p_{m_\bold{i}} \tr(\hat{\rho}_{A,{m_\bold{i}},\theta_\bold{i}}^k)/(\tr \hat{\rho}_{m_\bold{i},\theta_\bold{i}})^k}=\sum_{m_\bold{i},\theta_\bold{i}} p_{m_\bold{i},\theta_\bold{i}} \tr(\hat{\rho}_{A,{m_\bold{i}},\theta_\bold{i}}^k)/(\tr \hat{\rho}_{m_\bold{i},\theta_\bold{i}})^k$ in the form of~\eqref{replicated_observables} by using the subsystem cyclic permutation operator $\check{\mathcal{C}}_{k,l,A}^{(n)}= \left[\sum_{[\alpha_ i]} \bigotimes_{i=1}^k (\ket{\alpha_{i+1}}\bra{\alpha_i} \otimes \hat{I})\right]\otimes (\hat{I} \otimes \hat{I})^{(n-k)}$ (where $ \alpha_{k+1} \equiv \alpha_1$ and $\ket{\alpha_i} $ runs over all basis states of subsystem $A$). Note that the subscript $l$ (left) refers to the fact that it acts to the left of the density matrices by permuting the \textit{kets} of the subsystem $A$ between replicas and acting as the identity outside of it; we can define the analogous $\check{\mathcal{C}}_{k,r,A}^{(n)} $ acting on \textit{bras}. In this way, the trajectory-averaged purity is given by $ \overline{\braket{\mu_{k,A}}}=\lim_{n \to 1} \mu_{k,A}^{(n)}=\lim_{n \to 1}\braket{\braket{ \mathcal{I}^{(n)} ||  \check{\mathcal{C}}_{k,l/r,A}^{(n)} || \hat{\rho}^{(n)}}}/\braket{\braket{ \mathcal{I}^{(n)} || \hat{\rho}^{(n)}} } $. 

Returning to the case we are considering of $n=2$ replicas, we have that the first $k=1$ and second $k=2$ moments of the observables are given, respectively, by
\begin{align}
   O_1^{(2)}=\frac{\braket{\braket{ \mathcal{I}^{(2)} ||  \check{\mathcal{O}}_1^{(2)} || \hat{\rho}^{(2)}}}}{\braket{\braket{ \mathcal{I}^{(2)} || \hat{\rho}^{(2)}} }}=\frac{\sum_{m_\bold{i},\theta_\bold{i}} p_{m_\bold{i},\theta_\bold{i}}^2 \braket{\hat{O}}_{m_\bold{i},\theta_\bold{i}}}{\sum_{m_\bold{i},\theta_\bold{i}} p_{m_\bold{i},\theta_\bold{i}}^2}, \\
   O_2^{(2)}=\frac{\braket{\braket{ \mathcal{I}^{(2)} ||  \check{\mathcal{O}}_2^{(2)} || \hat{\rho}^{(2)}}}}{\braket{\braket{ \mathcal{I}^{(2)} || \hat{\rho}^{(2)}} }}=\frac{\sum_{m_\bold{i},\theta_\bold{i}} p_{m_\bold{i},\theta_\bold{i}}^2 \braket{\hat{O}}_{m_\bold{i},\theta_\bold{i}}^2}{\sum_{m_\bold{i},\theta_\bold{i}} p_{m_\bold{i},\theta_\bold{i}}^2},
\end{align}
and the quantity related to the trajectory-averaged second moment of the purity
\begin{equation}
    \mu_{2,A}^{(2)}=\frac{\sum_{m_\bold{i},\theta_\bold{i}} p_{m_\bold{i},\theta_\bold{i}}^2 \tr (\hat{\rho}_{A,m_\bold{i},\theta_\bold{i}}^{\prime 2})}{\sum_{m_\bold{i},\theta_\bold{i}} p_{m_\bold{i},\theta_\bold{i}}^2}.
\end{equation}
Therefore, working with two replicas implies that the probability of the trajectories is weighted by a probability distribution $p_{m_\bold{i},\theta_\bold{i}}^{(2)}=p_{m_\bold{i},\theta_\bold{i}}^2/(\sum_{m_\bold{i},\theta_\bold{i}} p_{m_\bold{i},\theta_\bold{i}}^2) $. Therefore, the quantities that we are going to study to address the phase transition are the objects $O_k^{(n)}$, which corresponds to the exact trajectory-averaged quantum mechanical observables only in the replica limit $n \to 1$. It has been shown, at least for the von Neumann entropy, that although they are not identical, both share critical properties in the~MIPT~\cite{bao20}. Thus, the \textit{observables} that we are going to be considering in the article related to the replica method, in the sense explained above in~\eqref{replicated_observables}, are
\begin{align}
    S_{A}^{(2)}&=-\log \frac{\braket{\braket{ \mathcal{I}^{(2)} ||  \check{\mathcal{C}}_{2,l/r, A}^{(2)} || \rho^{(2)}}}}{\braket{\braket{ \mathcal{I}^{(2)}  || \rho^{(2)}}, }}=-\log \mu_{2,A}^{(2)} \label{observable_entropy} \\
    (N_A)_1^{(2)}&= \frac{\braket{\braket{ \mathcal{I}^{(2)}  ||  \check{\mathcal{N}}_1^{(2)} || \rho^{(2)}}}}{\braket{\braket{ \mathcal{I}^{(2)}  || \rho^{(2)}} }} = \frac{\tr(\hat{\rho}_1\hat{N}_A )\tr(\hat{\rho}_2)}{\tr(\hat{\rho}_1 )\tr(\hat{\rho}_2)},
    \label{observable_number} \\
    \mathcal{F}_{A}^{(2)}&= \frac{\braket{\braket{ \mathcal{I}^{(2)} ||  (\check{\mathcal{N}}^2)_1^{(2)}-\check{\mathcal{N}}_2^{(2)} || \rho^{(2)}}}}{\braket{\braket{ \mathcal{I}^{(2)} || \rho^{(2)}} }} \nonumber \\
    &=\frac{\tr(\hat{\rho}_1\hat{N}_A^2 )\tr(\hat{\rho}_2)}{\tr(\hat{\rho}_1 )\tr(\hat{\rho}_2)}-\frac{\tr(\hat{\rho}_1\hat{N}_A )\tr(\hat{\rho}_2\hat{N}_A)}{\tr(\hat{\rho}_1 )\tr(\hat{\rho}_2)}, \label{observable_fluctuation}
\end{align}
where all the quantities correspond to trajectory-averaged quantities in the proper replica limit $n \to 1$: Eq.~\eqref{observable_entropy} is the von Neumann entropy $\overline{\braket{S_A}}$, Eq.~\eqref{observable_number} to the number of bosons $\overline{\braket{(N_A)_1}}$, Eq.~\eqref{observable_fluctuation} to the fluctuation of the number operator $\overline{\braket{\mathcal{F}_A}}$; in all cases refer to a subsystem $A$. The quantity $ S_A^{(2)}$ is the $2$-conditional Renyi entropy, and it is related to the purity by $ e^{-S_A^{(2)}}=\mu_{2,A}^{(2)}$, taking into account Eq.~\eqref{renyis_entropies}, and considering the classical measurement device as a part of the extended system~\cite{bao20,bao21}.

\subsection{Effective Hamiltonian in the enlarged space}
\label{appsec:replica_method_effective_hamiltonian_enlarged}
As we discussed in the main text, one of the main problems that arise when designing an experimental set-up to simulate an~MIPT is related to post-selection, which refers to the need to monitor each trajectory (that is, the result of each measurement performed during the dynamics and the values of each parameter in each unitary gate), to obtain the expectation values associated with them. Furthermore, this implies that for less frequent trajectories, the expectation values are extremely difficult to measure. As noted in~\cite{bao21}, this issue is related to the fact that the simplest observable for diagnosing an~MIPT needs to be associated, at least, with the second moment $k=2$ of the density matrix in Eq.~\eqref{replicated_observables}. But, as explained above, we can also consider observables, Eq.~\eqref{observable_number}, involving the first moment $k=1$ of the density matrix and, therefore, not requiring post-selection, to gain some indirect insight into the phase transition (Fig.~\ref{fig:postselection_nopostelection}). 

As we discussed in the main text, the effective Hamiltonian of Eq.~\eqref{H_eff_app} can be interpreted as an effective Bose Hubbard dynamics in an enlarged space of $4L$ sites, with interacting terms arising from measurements (Fig.~\ref{fig:enlarged_space_scheme}) as $\hat{H}_{\textnormal{eff}}=\hat{H}^M+i\hat{H}^{\rm BH} $~\eqref{H_eff}, where
\begin{align}
    \hat{H}^M&=\Gamma \sum_{l}^{L} \left(\hat{I}-\sum_{n=0}^{d-1} \hat{P}_{l,n} \hat{P}_{l+L,n} \hat{P}_{l+2L,n} \hat{P}_{l+3L,n} \right),\\
    \hat{H}^{\rm BH}&=\sum_{l}^{4L} W_l\left[ \omega \hat{n}_{l}+J\left( \hat{a}_l^\dagger \hat{a}_{l+1} + \text{h.c.} \right) -\frac{U}{2}\hat{n}_l (\hat{n}_l-\hat{I}) \right],
\end{align}
and
\begin{equation}
W_l=\begin{cases}
+1, l \in \left[1,L\right] \cup \left[2L+1,3L\right]  \\
-1, l \in \left[L+1,2L\right] \cup \left[3L+1,4L\right]
\end{cases} .
\end{equation}
For obtaining the effective Hamiltonian in the enlarged space we have vectorized the replicated density matrices in the sense described in Algorithm II in~\citep{kamakari2022}. The density matrices $\ket{\ket{\rho^{(2)}}} \equiv \hat{\rho}_1 \otimes \hat{\rho}_2= \ket{\alpha_1}\bra{\alpha_2}\otimes \ket{\beta_1}\bra{\beta_2} \to \ket{\phi}=\ket{\alpha_1}\otimes\ket{\alpha_2}\otimes\ket{\beta_1}\otimes\ket{\beta_2}$, for which we consider that the replicated operators in~\eqref{H_eff_app} that act on the density matrices as $ \hat{O}_1 \otimes \hat{O}_2^\dagger \otimes \hat{O}_3 \otimes \hat{O}_4^\dagger (\hat{\rho}_1 \otimes \hat{\rho}_2)= \hat{O}_1\hat{\rho}_1\hat{O}_2^\dagger \otimes  \hat{O}_3\hat{\rho}_2\hat{O}_4^\dagger $, need to be be understood as $\hat{O}_1 \otimes \hat{O}_2 \otimes \hat{O}_3 \otimes \hat{O}_4 (\ket{\alpha_1} \otimes \ket{\alpha_2} \otimes \ket{\beta_1} \otimes \ket{\beta_2})= \hat{O}_1\ket{\alpha_1} \otimes \hat{O}_2 \ket{\alpha_2} \otimes  \hat{O}_3\ket{\beta_1} \otimes \hat{O}_4 \ket{\alpha_2} $. As discussed in Eq.~\eqref{replicated_observables}, for computing the different observables will be crucial to use the reference state $ \bra{\bra{\mathcal{I}}}$ defined in Eq.~\eqref{replicated_reference}. Upon vectorization, this reference state is given by
\begin{align}
    \bra{\bra{\mathcal{I}}} & \to    \label{Ref_Pn1}\\
    &\bra{I}=\sum_{\substack{\alpha_1 \dots \alpha_L \\ \beta_1 \dots \beta_L}} \bra{\alpha_1 \cdots \alpha_L, \alpha_1 \dots \alpha_L, \beta_1 \dots \beta_L, \beta_1 \dots \beta_L},\notag 
\end{align}
where $\alpha_i,\beta_j=0,1,..,d-1$. This implies that when calculating the observables we only need to take into account states that do not vanish in the inner product with $\bra{I}$, therefore, we must look for states with the same quantum numbers in the first and second blocks, and the same for the third and fourth blocks (i.e.~$\ket{02,02,11,11}$ or $\ket{20,20,02,02}$ for a $L=2$ and $d \geq 3$ case). It is important to note that we are not interested in calculating quantum mechanical observables in this enlarged space (i.e.~$\braket{\phi|\hat{O}|\phi} $), but rather a 2-replica quantity that represents the observable averaging at the proper replica limit in the original circuit
\begin{equation}
    \frac{\bra{\bra{\mathcal{I}^{(2)}}} \check{\mathcal{O}}^{(2)} \ket{\ket{\rho^{(2)}}}}{\braket{\braket{\mathcal{I}^{(2)} || \rho^{(2)}}}} \to \frac{\braket{I|\hat{O}|\phi}}{\braket{I|\phi}},
    \label{observables_enlarged}
\end{equation}
where $ \check{\mathcal{O}}^{(2)}=\sum_{\ell_1,\ell_2,\ell_3,\ell_4}^L \hat{O}_{\ell_1,1} \otimes \hat{O}_{\ell_2,2}^\dagger \otimes \hat{O}_{\ell_3,3} \otimes \hat{O}_{\ell_4,4}^\dagger$ and $\hat{O}=\sum_{l}^{4L} \hat{O}_l $. However, to find the ground state of $\hat{H}_\textnormal{eff}$ we will calculate the expectation value of the energy in the usual way $ \mathfrak{Re}\braket{\tilde{\phi}| \hat{H}_\textnormal{eff}|\phi} $, where we will need to consider a bi-orthogonal basis as explained in the next subsection.

\section{Non-Hermitian perturbation theory}
\label{appsec:non-Hermitian}
In this appendix, we will describe in detail how to use perturbation theory in the case of a non-Hermitian Hamiltonian following the work done by Sternheim and Walker~\citep{sterheim1972}. Additionally, we include our new result of the calculation of wave renormalization for completeness, extending our analysis to the second order. When obtaining the eigenenergies and eigenvectors of a non-Hermitian operator, we face several issues that prevent us from applying standard techniques in quantum mechanics, such as the non-orthogonality of the eigenvectors. In this case, we can use the bi-orthogonal quantum mechanical formalism~\citep{sterheim1972, Brody_2013}, in which we need to obtain the eigenstates and eigenenergies for the operator and its Hermitian conjugate. Let us focus on the simple case of $\hat{H}_{\textnormal{eff}}=\hat{H}^M+i\hat{H}^{\rm BH}$, where $\hat{H}^M$ and $\hat{H}^{\rm BH}$ are Hermitian and following equations are satisfied
\begin{align}
    \hat{H}_{\textnormal{eff}}\ket{\phi_m}&=E_m\ket{\phi_m},  &(\hat{H}_{\textnormal{eff}})^\dagger \ket{{\varphi}_m}=&\varepsilon_m\ket{{\varphi}_m} \label{Heffbi}, \\
    \bra{\phi_m}(\hat{H}_{\textnormal{eff}})^\dagger&=E_m^*\bra{\phi_m}, &\bra{{\varphi}_m} \hat{H}_{\textnormal{eff}} =&{\varepsilon}_m^*\bra{{\varphi}_m} \label{Heffdaggerbi}.
\end{align}
Note that in the case of $ [\hat{H}^M,\hat{H}^{\rm BH} ] \neq 0$, the orthogonality of the eigenstates $ \braket{\phi_m|\phi_n}=0$ for $E_m \neq E_n$, no longer holds. But if the condition $ {\varepsilon}_n=E_n^*$ is hold, $ \big\{ {\varphi}_m,\phi_m \big\} $ forms a bi-orthogonal set such that $ \braket{{\varphi}_m|\phi_n}=\delta_{m,n}$ and $\hat{I}=\sum_m \ket{\phi_m}\bra{{\varphi}_m}$. As shown below, we also consider non-Hermitian measurements, for which equations~\eqref{Heffbi} and~\eqref{Heffdaggerbi} do not necessarily hold. However, the non-Hermitian measurements considered are real non-symmetric, i.e.~$\hat{H}^M \neq (\hat{H}^M )^\dagger$ and $\hat{H}^M=(\hat{H}^M)^*$, and we can prove that any real non-symmetric operator $\hat{O}$ can be expressed as $\hat{O}=\hat{O}_1 + i \hat{O}_2$, where $ \hat{O}_1$ and $\hat{O}_2$ are Hermitian. Therefore, we can use the same bi-orthogonal formalism to obtain the basis for the real non-symmetric measurement operator $\hat{H}^M$.

We will be particularly interested in obtaining the ground state of $H_{\textnormal{eff}}$ as a function of the measurement rate $\Gamma$, which allows us to calculate how different quantities behave in different cases. We can consider two regimes: $\Gamma/J \ll 1 $ where $\hat{H}^{\rm M}$ acts as a perturbation; and $\Gamma/J \gg 1$ where $\hat{H}^{\rm BH}$ acts as an imaginary perturbation (where we have made $\hat{H}_\textnormal{eff}/J\equiv \hat{H}_\textnormal{eff} $). As we explain below, we will focus on the area-law phase, since it is relatively easy to obtain the non-degenerate ground state of $\hat{H}^M$ and understand how the imaginary perturbation acts on it. For $\Gamma \ll 1$ the situation is more complicated, since the unperturbed term $ i \hat{H}^{\rm BH} $ is imaginary and, therefore, the ground state is not well defined).

Let us consider the area-law phase regime, where the perturbation parameter is defined as $\lambda=J/\Gamma \ll 1 $. Taking into account~\eqref{Heffbi} and~\eqref{Heffdaggerbi}, it is evident that we need to obtain the perturbation analysis for both the Hamiltonian $\hat{H}_{\textnormal{eff}}$ and its Hermitian conjugate $(\hat{H}_{\textnormal{eff}})^\dagger$. Therefore, we consider two Schrödinger equations $ \hat{H}_{\textnormal{eff}}\ket{\phi}=(\hat{H}^M+i\lambda \hat{H}^{\rm BH} )\ket{\phi}=E\ket{\phi} $ and $ \hat{H}_{\textnormal{eff}}^\dagger \ket{{\varphi}}=\left[(\hat{H}^M)^\dagger-i\lambda \hat{H}^{\rm BH} \right]\ket{{\varphi}}={\varepsilon}\ket{{\varphi}} $. Then, expanding the energies and states in power series of $\lambda$,
\begin{align}
    \left(\hat{H}^M+i\lambda \hat{H}^{\rm BH} \right) \sum_r \lambda^r\ket{\phi^{(r)}} & =\sum_{r'} \lambda^{r'} E^{(r')}\sum_r \lambda^r\ket{\phi^{(r)}},     \label{H_expanded_lambda} \\
    \left[(\hat{H}^M)^\dagger-i\lambda \hat{H}^{\rm BH} \right] \sum_r \lambda^r\ket{{\varphi}^{(r)}} &=\sum_{r'} \lambda^{r'} {\varepsilon}^{(r')}\sum_r \lambda^r\ket{{\varphi}^{(r)}}.
    \label{Hdagger_expanded_lambda}
\end{align}
For normalization purposes of general states $ \ket{\phi}=\sum_n c_n \ket{\phi_n^{(0)}}$ and $ \ket{\varphi}=\sum_n d_n \ket{\varphi_n^{(0)}}$  , we will define the associated states $ \ket{\tilde{\phi}}=\sum_n c_n \ket{\varphi_n^{(0)}} $ and $\ket{\tilde{\varphi}}=\sum_n d_n \ket{\phi_n^{(0)}} $, respectively \citep{Brody_2013}. The normalization is then given by $\braket{\tilde{\phi}|\phi}=\sum_n c_n^* c_n$ and $\braket{\tilde{\varphi}|\varphi}=\sum_n d_n^* d_n$, and the expectation values of an observable $\hat{O}$ by $\braket{\tilde{\phi}|\hat{O}|\phi}$ and $\braket{\tilde{\varphi}| \hat{O}|\varphi}$. Then the renormalization of the state in the second order of the perturbative analysis $Z$ is given by $Z\braket{\tilde{\phi}|\phi}=1$ and $Z\braket{\tilde{\varphi}|\varphi}=1$. We will be interested in studying how the ground state $ \ket{\phi_0}$ is perturbed, therefore, up to the second order in $\lambda$, we have the equations for the Hamiltonian
\begin{align}
    [E^{(0)}-\hat{H}^M]\ket{\phi_0^{(0)}}=&0 , \\
    [E^{(0)}-\hat{H}^M]\ket{\phi_0^{(1)}}+[E^{(1)}-i\hat{H}^{\rm BH}]\ket{\phi_0^{(0)}}=&0, \\
    [E^{(0)}-\hat{H}^M]\ket{\phi_0^{(2)}}+[E^{(1)}-i\hat{H}^{\rm BH}]\ket{\phi_0^{(1)}} \quad \quad &\nonumber \\
    +E^{(2)}\ket{\phi_0^{(0)}}=&0,
\end{align}
and for its Hermitian conjugate
\begin{align}
    [\varepsilon^{(0)}-(\hat{H}^M)^\dagger]\ket{{\varphi}_0^{(0)}}=&0, \\
    [\varepsilon^{(0)}-(\hat{H}^M)^\dagger]\ket{{\varphi}_0^{(1)}}+[\varepsilon^{(1)}+i\hat{H}^{\rm BH}]\ket{{\varphi}_0^{(0)}}=&0, \\
    [\varepsilon^{(0)}-(\hat{H}^M)^\dagger]\ket{{\varphi}_0^{(2)}}+[\varepsilon^{(1)}+i\hat{H}^{\rm BH}]\ket{{\varphi}_0^{(1)}} \quad & \nonumber \\
    +\varepsilon^{(2)}\ket{{\varphi}_0^{(0)}}=&0.
\end{align}
Therefore, we need to obtain the bi-orthogonal basis of the Hamiltonian $\hat{H}^M$. Note that in the case of a Hermitian $\hat{H}^M$, the bi-orthogonal basis is given by the usual basis of the number of bosons operator $\hat{N}=\sum_l^{4L} \hat{n}_l $, while in the case of a non-Hermitian $\hat{H}^M$, the bi-orthogonal basis needs to be found.

The normalized corrections for the non-degenerate state $\ket{\phi_\alpha} $ up to second order in $\lambda$ are given by
\begin{align}
    \ket{\phi_\alpha}=&\left[1-\frac{\lambda^2}{2}\sum_{n \neq \alpha} \left(\frac{ iV_{{n}\alpha}}{E_\alpha^{(0)}-E_n^{(0)} } \right)^* \frac{iV_{{n}\alpha}}{(E_\alpha^{(0)}-E_n^{(0)} )} \right]\ket{\phi_\alpha^{(0)}} \nonumber \\
    &\quad +i\lambda \sum_{n \neq \alpha} \frac{V_{{n}\alpha}}{(E_\alpha^{(0)}-E_n^{(0)} )}\ket{\phi_n^{(0)}} \nonumber \\
    &\quad -\lambda^2 \sum_{n,m \neq \alpha} \frac{V_{{n}m}V_{{m}\alpha}}{(E_\alpha^{(0)}-E_n^{(0)} )(E_\alpha^{(0)}-E_m^{(0)} )}\ket{\phi_n^{(0)}} \nonumber \\
    &\quad +\lambda^2 \sum_{n \neq \alpha} \frac{V_{{\alpha}\alpha}V_{{n}\alpha}}{(E_\alpha^{(0)}-E_n^{(0)} )^2}\ket{\phi_n^{(0)}},\label{perturbed_state_renorm} \\
   E_\alpha=&E_\alpha^{(0)}+i\lambda V_{{\alpha}\alpha}-\lambda^2 \sum_{n \neq \alpha} \frac{V_{{\alpha}n}V_{{n}\alpha}}{( E_\alpha^{(0)}-E_n^{(0)} )},
\end{align}
and for the non-degenerate state $\ket{{\varphi}_\alpha} $ are given by
\begin{align}
    \ket{\varphi_\alpha}=&\left[1-\frac{\lambda^2}{2}\sum_{n \neq \alpha} \left(\frac{ -iV_{{n}\alpha}}{\varepsilon_\alpha^{(0)}-\varepsilon_n^{(0)} } \right)^* \frac{-iV_{{n}\alpha}}{(\varepsilon_\alpha^{(0)}-\varepsilon_n^{(0)} )} \right]\ket{\varphi_\alpha^{(0)}} \nonumber \\
    &\quad -i\lambda \sum_{n \neq \alpha} \frac{V_{{n}\alpha}}{(\varepsilon_\alpha^{(0)}-\varepsilon_n^{(0)} )}\ket{\varphi_n^{(0)}} \nonumber \\
    &\quad -\lambda^2 \sum_{n,m \neq \alpha} \frac{V_{{n}m}V_{{m}\alpha}}{(\varepsilon_\alpha^{(0)}-\varepsilon_n^{(0)} )(\varepsilon_\alpha^{(0)}-\varepsilon_m^{(0)} )}\ket{\varphi_n^{(0)}} \nonumber \\
    &\quad +\lambda^2 \sum_{n \neq \alpha} \frac{V_{{\alpha}\alpha}V_{{n}\alpha}}{(\varepsilon_\alpha^{(0)}-\varepsilon_n^{(0)} )^2}\ket{\varphi_n^{(0)}}\\
   \varepsilon_\alpha=&\varepsilon_\alpha^{(0)}-i\lambda V_{{\alpha}\alpha}-\lambda^2 \sum_{n \neq \alpha} \frac{V_{{\alpha}n}V_{{n}\alpha}}{( \varepsilon_\alpha^{(0)}-\varepsilon_n^{(0)} )},
\end{align}
where the matrix elements are $V_{{a}b}=\braket{{\varphi}_a^{(0)} | \hat{H}^{\rm BH}| \phi_b^{(0)} } $, and  $ \{ {\varphi}_m^{(0)},\phi_m^{(0)} \} $ is the bi-orthogonal basis of the unperturbed Hamiltonian $\hat{H}^M$ that defines the identity $ \hat{I}=\sum_m \ket{\phi_m^{(0)}}\bra{{\varphi}_m^{(0)}}$. Note that even-order perturbation terms are real, while odd-order perturbation terms are imaginary. The imaginary disturbance has a different effect on the energy correction compared to the real case: the second-order correction to the state exhibits a sign opposite to that in the standard Hermitian case.

\section{Trajectory-averaged observables for $n=2$ replicas}
\label{appsec:averaged_observable}
In this appendix, we describe how we calculate the trajectory-averaged quantities of the original circuit for $n=2$ replicas using the second-order perturbed ground state in the enlarged space. Therefore, since we are going to use the normalization given in Eq.~\eqref{replicated_observables}, we will not use here the wave-renormalized state in Eq.~\eqref{perturbed_state_renorm}, for the sake of simplicity.

\subsection{Non-physical measurement: case $\ket{1111111...} $ and $d\geq 3$}
In the first case, and solely for clarification purposes, we consider a non-realistic model where, instead of proper measurement, we just introduce a projective operator to a particular subspace $ \hat{P}_{l,n}^{\textnormal{np}}=\ket{\alpha_l}\bra{\alpha_l}$, where $\alpha_l$ can be any number of bosons dependent on the site. Note that it is not a proper measurement since $ \sum_{n=0}^{d-1}\hat{P}_{l,n}^{\textnormal{np}\dagger} \hat{P}_{l,n}^{\textnormal{np}}  \neq \hat{I} $, although $\hat{H}^M$ is Hermitian and conserves the number of bosons (i.e.~we can use the boson number basis of $\hat{N}=\sum_l^{4L} \hat{n}_l $ as the unperturbed basis) and has a non-degenerate ground state given by $ \ket{\phi_0^{\rm{np}(0)}}=\left( \ket{\alpha_1 \alpha_2 \dots \alpha_L} \right)^{\otimes 4}$ (Note that in this case $\ket{\phi_i^{\rm{np}(0)}} \equiv \ket{\varphi_i^{\rm{np}(0)}} $). Nevertheless, this procedure could be related to \textit{standard} measurement where unwanted measurement outcomes are discarded~\citep{tang20}.

For explanation purposes, we will present the explicit result for the $L=4$ and $d=3$ case, with projections to the $n=1$ boson subspace at each position, because it is illustrative and gives us clear insight for understanding higher orders in perturbation theory, which can be useful to tackle the size-dependent phase. Therefore, the ground state is given by $\ket{\phi_0^{(0)}}=\ket{1}^{\otimes 4L}$, and "second-order" implies that there are two jumps resulting in $1)$ two sites with $n=0$ and two sites with $n=2$ bosons, and $2)$ two jumps within same positions resulting in the ground state. Although there can be a large number of combinations we only need to consider those that do not vanish in the inner product with $\bra{I}$, because we are interested in computing a very specific set of quantities~\eqref{observables_enlarged}.  As we explain above, we consider just the hopping terms $\hat{V}=\sum_l^{16}W_l(\hat{a}_l^\dagger \hat{a}_{l+1} + \text{h.c.})$ and study $\hat{V}\ket{\phi_0^{\rm{np}(0)}} $ and $\hat{V}^2\ket{\phi_0^{\rm{np}(0)}} $ where $\ket{\phi_0^{\rm{np}(0)}}=\ket{1111,1111,1111,1111}$ such that

\begin{widetext}
\begin{align}
\hat{V} \ket{1111,1111,1111,1111}=&\sqrt{2} \Big( \ket{0211,1111,1111,1111}+\ket{1021,1111,1111,1111}+\ket{1102,1111,1111,1111} \Big. \nonumber \\
&+\ket{2011,1111,1111,1111}+\ket{1201,1111,1111,1111}+\ket{1120,1111,1111,1111} \nonumber \\
&-\ket{1111,0211,1111,1111}-\ket{1111,1021,1111,1111}-\ket{1111,1102,1111,1111} \nonumber \\
&-\ket{1111,2011,1111,1111}-\ket{1111,1201,1111,1111}-\ket{1111,1120,1111,1111} \nonumber \\
&+\ket{1111,1111,0211,1111}+\ket{1111,1111,1021,1111}+\ket{1111,1111,1102,1111} \nonumber \\
&+\ket{1111,1111,2011,1111}+\ket{1111,1111,1201,1111}+\ket{1111,1111,1120,1111} \nonumber \\
&-\ket{1111,1111,1111,0211}-\ket{1111,1111,1111,1021}-\ket{1111,1111,1111,1102} \nonumber \\
&\Big. -\ket{1111,1111,1111,2011}-\ket{1111,1111,1111,1201}-\ket{1111,1111,1111,1120} \Big)
\end{align}

\begin{align}
\hat{V}^2 \ket{1111,1111,1111,1111}=&-(\sqrt{2})^2 2 \Big(\ket{0211,0211,1111,1111}+\ket{1021,1021,1111,1111}+\ket{1102,1102,1111,1111} \Big. \nonumber \\
&+\ket{2011,2011,1111,1111}+\ket{1201,1201,1111,1111}+\ket{1120,1120,1111,1111} \nonumber \\
&+\ket{1111,1111,0211,0211}+\ket{1111,1111,1021,1021}+\ket{1111,1111,1102,1102} \nonumber \\
&\Big. +\ket{1111,1111,2011,2011}+\ket{1111,1111,1201,1201}+\ket{1111,1111,1120,1120} \Big) \nonumber \\
&-(\sqrt{2})^2 2 \Big(\ket{0211,1111,1111,0211}+\ket{1021,1111,1111,1021}+\ket{1102,1111,1111,1102} \Big. \nonumber \\
&+\ket{2011,1111,1111,2011}+\ket{1201,1111,1111,1201}+\ket{1120,1111,1111,1120} \nonumber \\
&+\ket{1111,0211,0211,1111}+\ket{1111,1021,1021,1111}+\ket{1111,1102,1102,1111} \nonumber \\
&\Big. +\ket{1111,2011,2011,1111}+\ket{1111,1201,1201,1111}+\ket{1111,1120,1120,1111} \Big) \nonumber \\
&+(\sqrt{2})^2 24 \ket{1111,1111,1111,1111}+\ket{\overline{I}} \nonumber \\
=&-4\ket{\Phi}-4\ket{\Psi}+48\ket{\phi_0^{\rm{np}(0)}}+\ket{\overline{I}},
\end{align}

\end{widetext}
where $\ket{\overline{I}}$ includes all the terms that vanish in every inner product with $\bra{I}$~\eqref{Ref_Pn1} that we will compute~\eqref{observables_enlarged}, i.e.~$\braket{I|\overline{I}}=\braket{I|\hat{C}_{A,l/r}|\overline{I}}=0 $, where $A\in[1,2]$. $\ket{\Phi}$ includes all the terms that do not vanish in the inner product, i.e.~$\braket{I|\Phi} =12$, and will be used for computing observables, and $ \braket{I|\phi_0^{\rm{np}(0)}} =1$.  $\ket{\Psi}$ includes terms which do not vanish upon application of permutation operator, i.e.~$ \braket{I|\Psi}=0$ and $ \braket{I|\hat{C}_{A,l/r}|\Psi} =4$. Note that there are also terms in $\ket{\Phi}$ that do not vanish when permuted, i.e.~$ \braket{I|\hat{C}_{A,l/r}|\Phi} =4$. We can compute explicitly how the permutation acts on $\ket{\Phi} $
\begin{align}
    \hat{C}_{A,l} \ket{0211,0211,1111,1111}=\ket{1111,0211,0211,1111}, \nonumber \\
    \hat{C}_{A,l} \ket{1021,1021,1111,1111}=\ket{1121,1021,1011,1111}, \nonumber \\
    \hat{C}_{A,l} \ket{1102,1102,1111,1111}=\ket{1102,1102,1111,1111}, \nonumber \\    
    \hat{C}_{A,l} \ket{1111,1111,0211,0211}=\ket{0211,1111,1111,0211}, \nonumber \\
    \hat{C}_{A,l} \ket{1111,1111,1021,1021}=\ket{1011,1111,1121,1021}, \nonumber \\
    \hat{C}_{A,l} \ket{1111,1111,1102,1102}=\ket{1111,1111,1102,1102} ,
    \label{entro_1}
\end{align}
and on $ \ket{\Psi}$
\begin{align}
    \hat{C}_{A,l} \ket{0211,1111,1111,0211}=\ket{1111,1111,0211,0211}, \nonumber \\
    \hat{C}_{A,l} \ket{1021,1111,1111,1021}=\ket{1121,1111,1011,1021}, \nonumber \\
    \hat{C}_{A,l} \ket{1102,1111,1111,1102}=\ket{1102,1111,1111,1102}, \nonumber \\
    \hat{C}_{A,l} \ket{1111,0211,0211,1111}=\ket{0211,0211,1111,1111} , \nonumber \\
    \hat{C}_{A,l} \ket{1111,1021,1021,1111}=\ket{1011,1021,1121,1111}, \nonumber \\
    \hat{C}_{A,l} \ket{1111,1102,1102,1111}=\ket{1111,1102,1102,1111}, 
    \label{entro_2}
\end{align}
where there are only $4$ terms (third and sixth equations in \eqref{entro_1}, and first and third equations in \eqref{entro_2}) that do not vanish in the inner product, and we have computed only jumps \textit{to the right}, therefore we need to multiply it by $2$. Note that we have considered the \textit{left} permutation $\hat{C}_{A,l}$ that originally permute the kets between the two replicas and analogous sites within the first and third blocks of the enlarged space; we would have obtained the same result if we had used the \textit{right} permutation $\hat{C}_{A,r}$ that originally permute the bras between the two replicas and analogous sites within the second and fourth blocks of the enlarged space. The unperturbed energies of the two types of states are computed as $E_0^{(0)}=\braket{\phi_0^{\rm{np}(0)}|\hat{H}^M|\phi_0^{\rm{np}(0)}}=0$ for the ground state and $E_\alpha^{(0)}=\braket{\phi_\alpha^{\rm{np}(0)}|\hat{H}^M|\phi_\alpha^{\rm{np}(0)}}=2\Gamma$ for the rest of the states. Finally, we obtain the normalized second-order correction to the ground state 
\begin{equation}
    \ket{\phi_0^{(2)}} \equiv (1-6\lambda^2)\ket{\phi_0^{\rm{np}(0)}}+\lambda^2\left(\ket{\Phi}+\ket{\Psi} \right),
\end{equation}
where we have included in the state only those terms that do not vanish in the inner product defined in the replicated space, as explained above. Although we have assumed that $d=3$, it is easy to notice that the dimension of the subsystem up to the second order is not relevant, since it is necessary to have similar states in blocks $1$ and $2$ on the one hand and blocks $3$ and $4$ on the other, and any state with a site with more than $2$ bosons vanishes in the inner product at second-order perturbation theory since it involves only two jumps. For the same reason, we know that odd-order terms also vanish. 

Following the same procedure as for the $L=4$ case, we can obtain the expression of the second order correction to the ground state for an arbitrary $L$ number of transmons, for which we compute the averaged observables for $n=2$ replicas, given by
\begin{align}
    S_{L/2}^{(2)}&=-\log \left[ 1-\frac{4\lambda^2}{1+2(L-1)\lambda^2}\right] \approx 4\lambda^2 , \\
    F_{L/2}^{(2)}& = \frac{2\lambda^2}{1+2(L-1)\lambda^2} \approx 2\lambda^2, \\ 
    N_{L/2}^{(2)}&=\frac{L}{2}, \\
    N_{L}^{(2)}&=L.
\end{align}
Note that $S_{L/2}^{(2)}$ and $F_{L/2}^{(2)}$ do not depend on the size and scale similar regarding the perturbation parameter.

\subsubsection{Understanding the size-dependent phase}
Although we did not study the size-dependent phase by the replica method we can make reasonable estimations in this non-physical case. First, we can expect the size-dependent phase to appear at $\left(J/\Gamma\right)^{r \geq 4}$, for which we must take into account that they are new states arising at order $ r$ in systems of size $L \geq r$ that are not present in systems of size $L=r-2$, whose inner product with $\bra{I}$ is not zero, but does not have permutation symmetry. Thus, as the order increases in perturbation theory, new states emerge for larger systems, leading to higher entropy in ground states. Consequently, at the $r$th order in perturbation theory, systems composed of $L \geq r$ transmons exhibit identical entropy, being larger than systems consisting of $L=r-2$ transmons.

We can see this more clearly by studying the case of measurements that are projected to $n=1$ at each position, whose entropy is given by 
\begin{equation}
S_{L/2}^{(2)}(r)=- \log \left\lbrace \frac{1+2[(L-1)-2]\lambda^2+\sum_{n =4}^r f_n^{ps}(L)\lambda^n}{1+2(L-1)\lambda^2+\sum_{n=4}^r f_n(L)\lambda^n} \right\rbrace,
\label{entropy_volume}
\end{equation}
with $f_n^{ps}(L) < f_n(L)$. We have computed this quantity explicitly up to the second order where in the denominator is the function $f_2 (L)=2(L-1)$ which accounts for the states that do not vanish in the inner product with $\braket{I | \phi}$ and in the numerator the function $f_2^{ps}(L)=2(L-1)-4$ which accounts for those states that do not vanish when permuted $\bra{I} \hat{C}_{L/2}^{(2)} \ket{\phi}$. Note that these results are the same for every system's size. The reason is that up to the second-order there can be two different types of states no matter the size of the system: the ground state and states produced by hoppings in different blocks. But at the fourth order, new states appear, some of them shared by every system size but others that are not present in $L=2$, and those are the states with two hoppings in different positions in different blocks. We can see this explicitly (without taking into account all the permutations within blocks and hoppings)
\begin{align}
\mathcal{O}(\lambda^2) \quad L=2 \quad &\ket{11,11,11,11}, \notag \\
          &\ket{02,02,11,11}, \\
L=4 \quad &\ket{1111,1111,1111,1111}, \notag \\
          &\ket{0211,0211,1111,1111},  \\
\mathcal{O}(\lambda^4) \quad L=2 \quad &\ket{11,11,11,11}, \notag \\
          &\ket{02,02,11,11},  \notag \\
          &\ket{02,02,02,02}, \\
L=4 \quad &\ket{1111,1111,1111,1111}, \notag\\
          &\ket{0211,0211,1111,1111}, \notag\\
          &\ket{0211,0211,0211,0211}, \notag\\
          &\ket{0202,0202,1111,1111}, \label{volume_new_states}
\end{align}
where we can see that new states~\eqref{volume_new_states} appear for $L \geq 4$. Note that this new states has an unperturbed energy of $E_\beta^{(0)}=4$, different to the $E_\alpha^{(0)}=2\Gamma$ of the previous states. This implies that at the fourth order, the entropy of the system of $L \geq 4$ will include new terms in the numerator and denominator of~\eqref{entropy_volume} such that
\begin{equation}
    \frac{f_4^{ps}(L=2)}{f_4(L=2)} > \frac{f_4^{ps}(L \geq 4)}{f_4(L\geq4)}.
\end{equation}
Therefore, it is expected that the entropy for $L=2$ will be smaller than for other system sizes. We can infer inductively that a similar phenomenon will occur at the sixth order between $L=4$ and $L \geq 6$, and so forth. It is important to note that this analysis pertains to a non-physical measurement, as the calculations for \textit{feedback} measurements become considerably more complex due to the need for a bi-orthogonal basis.

\subsection{\textit{Feedback} measurement}
In the second case, we consider the \textit{feedback} measurement described in the main text. Because of the non-hermicity of $\hat{H}^M$ that does not conserve the total number of bosons, we need to obtain explicitly its full bi-orthogonal basis as described in Eqs.~\eqref{biortogonal1_states} and~\eqref{biortogonal2_states}. As a proof of concept, we obtain explicitly the bi-orthogonal basis for the simplest case consisting of two two-dimensional transmons, where we set the measurements as $\hat{P}_{1,m}^\textnormal{fb}=\ket{1}\bra{m}$ and $\hat{P}_{2,m}^\textnormal{fb}=\ket{0}\bra{m}$, i.e.~we make projections to $n=1$ and $n=0$ at sites $\ell=1$ and $\ell=2$, respectively, in the original circuit (i.e.~$l=1,3,5,7$ and $l=2,4,6,8$ in the enlarged space). Since we cannot use the basis of $\hat{N}=\sum_l^{8} \hat{n}_l $, we obtain explicitly the eigenstates and eigenenergies the Hamiltonian $\hat{H}^M$ 
\begin{align}
    \ket{\Phi_1^{\rm{fb}(0)}}&=\ket{\overline{1}}\ket{\overline{0}},   &E_1^{(0)}=&0, \\
    \ket{\Phi_2^{\rm{fb}(0)}}&=\left(\ket{\overline{0}}-\ket{\overline{1}}\right) \ket{\overline{0}}, &E_2^{(0)}=&\Gamma, \\
    \ket{\Phi_3^{\rm{fb}(0)}}&=\ket{\overline{1}} \left(\ket{\overline{1}}-\ket{\overline{0}}\right), &E_3^{(0)}=&\Gamma,\\
    \ket{\Phi_{4-17}^{\rm{fb}(0)}}&=\ket{\overline{1}} \ket{i_2j_2k_2l_2}, &E_{4-17}^{(0)}=&\Gamma,\\
    \ket{\Phi_{18-31}^{\rm{fb}(0)}}&=\ket{i_1j_1k_1l_1} \ket{\overline{0}},  &E_{18-31}^{(0)}=&\Gamma, \\
    \ket{\Phi_{32}^{\rm{fb}(0)}}&=\left(\ket{\overline{0}}-\ket{\overline{1}}\right) \left(\ket{\overline{1}}-\ket{\overline{0}}\right), &E_{32}^{(0)}=&2\Gamma, \\
    \ket{\Phi_{33-46}^{\rm{fb}(0)}}&=\left(\ket{\overline{0}}-\ket{\overline{1}}\right) \ket{i_2j_2k_2l_2}, &E_{33-46}^{(0)}=&2\Gamma \\
    \ket{\Phi_{47-60}^{\rm{fb}(0)}}&=\ket{i_1j_1k_1l_1} \left(\ket{\overline{1}}-\ket{\overline{0}}\right), &E_{47-60}^{(0)}=&2\Gamma,  \\
    \ket{\Phi_{61-256}^{\rm{fb}(0)}}&=\ket{i_1j_1k_1l_1} \ket{i_2j_2k_2l_2}, &E_{61-256}^{(0)}=&2\Gamma,
\end{align}
and its Hermitian conjugate $(\hat{H}^M)^\dagger$
\begin{align}
    \ket{{\Theta}_1^{\rm{fb}(0)}}&=\left(\ket{\overline{0}}+\ket{\overline{1}}\right) \left(\ket{\overline{0}}+\ket{\overline{1}}\right), &\varepsilon_1^{(0)}=&0, \\
    \ket{\Theta_2^{\rm{fb}(0)}}&=\ket{\overline{0}} \left(\ket{\overline{0}}+\ket{\overline{1}}\right), &\varepsilon_2^{(0)}=&\Gamma,  \\
    \ket{\Theta_3^{\rm{fb}(0)}}&=\left(\ket{\overline{0}}+\ket{\overline{1}}\right) \ket{\overline{1}}, &\varepsilon_3^{(0)}=&\Gamma,  \\
    \ket{\Theta_{4-17}^{\rm{fb}(0)}}&=\left(\ket{\overline{0}}+\ket{\overline{1}}\right) \ket{m_2n_2p_2q_2}, &\varepsilon_{4-17}^{(0)}=&\Gamma, \\
    \ket{\Theta_{18-31}^{\rm{fb}(0)}}&=\ket{m_1n_1p_1q_1} \left(\ket{\overline{0}}+\ket{\overline{1}}\right),  &\varepsilon_{18-31}^{(0)}=&\Gamma,   \\
    \ket{\Theta_{32}^{\rm{fb}(0)}}&=\ket{\overline{0}}\ket{\overline{1}}, &\varepsilon_{32}^{(0)}=&2\Gamma,   \\
    \ket{\Theta_{33-46}^{\rm{fb}(0)}}&=\ket{\overline{0}}\ket{m_2n_2p_2q_2}, &\varepsilon_{33-46}^{(0)}=&2\Gamma , \\
    \ket{\Theta_{47-60}^{\rm{fb}(0)}}&=\ket{m_1n_1p_1q_1} \ket{\overline{1}}, &\varepsilon_{47-60}^{(0)}=&2\Gamma,   \\
    \ket{\Theta_{61-256}^{\rm{fb}(0)}}&=\ket{m_1n_1p_1q_1} \ket{m_2n_2p_2q_2}, &\varepsilon_{61-256}^{(0)}=&2\Gamma,
\end{align}
where the subscript is an index for the basis' number, the eigenenergies are the same for both cases (i.e.~$E_m=\varepsilon_m^*=\varepsilon_m$) and $i_\ell, j_\ell, k_\ell, l_\ell,m_\ell, n_\ell, p_\ell, q_\ell=0,1$. Once we have the full bi-orthogonal unperturbed basis, we use perturbation theory up to second order (as we did for the simple example of the non-physical measurement) to compute the different quantities for half of the original system; using the same recipe, we can generalize the bi-orthogonal basis for a general system size $L$, such that
\begin{align}
    S_{L/2}^{(2)}&=-\log \left[ 1-\frac{\lambda^2}{1+\frac{1}{2}(L-1)\lambda^2}\right] \approx \lambda^2, \\
    F_{L/2}^{(2)}&=\frac{1}{2}\frac{\lambda^2}{1+\frac{1}{2}(L-1)\lambda^2} \approx \frac{\lambda^2}{2}, \\
    N  _{L}^{(2)}&=\frac{L}{2},
    \label{case10_gen_Ntotal} \\
    N  _{l \textnormal{ even}}^{(2)}&=\frac{\frac{1}{2}\lambda^2}{1+\frac{1}{2}(L-1)\lambda^2} \approx \frac{1}{2}\lambda^2,
    \label{N_l_even} \\
    N  _{l \textnormal{ odd}}^{(2)}&=1-\frac{\frac{1}{2}\lambda^2}{1+\frac{1}{2}(L-1)\lambda^2} \approx 1-\frac{1}{2}\lambda^2,
    \label{N_l_odd}
\end{align}
where $\lambda=J/\Gamma$. Although not shown in the previous subsection, we obtain the same results as in the non-physical projective measurement. As we will discuss later, this is not expected to hold to the fourth order or higher, where the non-conservation in the total boson number will presumably affect the statistics. Interestingly, up to the second order in $\lambda$, the average total number of bosons~\eqref{case10_gen_Ntotal} is constant even using a non-conserving measurement.

\subsubsection{Understanding the non-conservation of the total number of bosons}
\label{appsec:averaged_observable_non_conservation}
The relevant aspect of this type of measurement is that while it does not conserve the total number of bosons, for a high measurement rate (i.e.~deep in the area-law phase), the steady-state behavior of the average of the ensemble of trajectories has a constant total number of bosons. Therefore, if we measure the total number of bosons and average the quantity for a large number of experiments we will obtain a constant number for different values of the measurement probability parameter $p$ as long as we are deep in the area-law phase. It is important to note that this is not a symmetry, which implies that if we initialize the system with a state with definite quantum numbers with respect to the number of bosons, the dynamics always produce states with the same total number of bosons. This effect in the averaged value is dependent on the set of single subspace projectors used in the \textit{feedback} measurement (i.e.~the total number of bosons is $N=\sum_{l=1}^{4L} \alpha_l $ for the set of projectors $ \hat{P}_{l,m}^\textnormal{fb}=\ket{\alpha_l}\bra{m} $), regardless of the initial state, which can even be in a superposition. We can understand this phenomenon for $L=2$ two-dimensional systems by considering that, up to the second order, starting the system from the ground state $\ket{\Phi_1^{\rm{fb}(0)}}=\ket{10101010} $ and applying the hopping term twice, states of the form $ \ket{\Phi_{61-256}^{\rm{fb}(0)}}=\ket{0_i1_j} \ket{0_k1_l} $ may appear, specifically the states $\ket{01011010}$ and $\ket { 10100101}$, which are the only ones that contribute to the number of bosons (besides the ground state). Note, however, that this is not a general case: the averaged total number of bosons can change depending on the subsystem dimension and the spatial pattern of the \textit{feedback} measurement; the constant effect we are studying is due to the symmetries we have considered.

Although we do not calculate corrections up to higher orders, we can predict that for orders $\geq 4$, states with a different total number of bosons can appear in the dynamics, since at order $\lambda^r $ we need to consider terms like
\begin{equation}
    \left[ \sum_{n \neq 1} \frac{\ket{\Phi_n^{\rm{fb}(0)}}\bra{{\Theta}_n^{\rm{fb}(0)}}\hat{H}^{\rm BH}}{E_1^{(0)}-E_n^{(0)}} \right]^r \ket{\Phi_1^{\rm{fb}(0)}}.
\end{equation}
In this sense, we can consider, for example, that at the fourth order, the system can perform four shopping, and the state $\ket{\overline{0}}\ket{\overline{1}} \equiv \ket{01010101}$, which is not an eigenstate of $\hat{H}^M$, appears in the dynamics, which have non-vanishing inner products with the eigenstates of $(\hat{H}^M)^\dagger $: $\ket{{\Theta}_{2}^{\rm{fb}(0)}} $, $\ket{{\Theta}_{3}^{\rm{fb}(0)}} $ and $\ket{{\Theta}_{32}^{\rm{fb}(0)}} $ (note that we cannot take into account the ground state $\ket{{\Theta}_1^{\rm{fb}(0)}}$). Considering the bi-orthogonal states, at fourth order, the perturbed state will include $ \ket{\Phi_2^{\rm{fb}(0)}}=\ket{00000000}-\ket{10101010}$, $ \ket{\Phi_3^{\rm{fb}(0)}}=-\ket{10101010}+\ket{11111111}$ and $ \ket{\Phi_{32}^{\rm{fb}(0)}}=-\ket{00000000}+\ket{01010101}-\ket{10101010}-\ket{11111111}$. The new states $\ket{00000000}$ and $\ket{11111111}$ contain a different total number of bosons. Note that although for a general system, the trajectory-averaged total number of bosons can now vary, we can consider a symmetric case where this value remains constant, as we verified numerically for the case of two-dimensional systems with a measurement pattern defined by filling factor $N/L=1/2$. In the numerical simulations, we observe that for a high measurement probability, only the state $\ket{10}$ exists, where the total number of bosons is constant ($N=1$). However, with a small probability, states $\ket{00}$, $\ket{01}$, and $\ket{11}$ appear, resulting in a non-constant total number of bosons.

\section{Numerical simulation}
\label{appsec:numerical_simulation}

\begin{figure*}
\includegraphics[width=\linewidth]{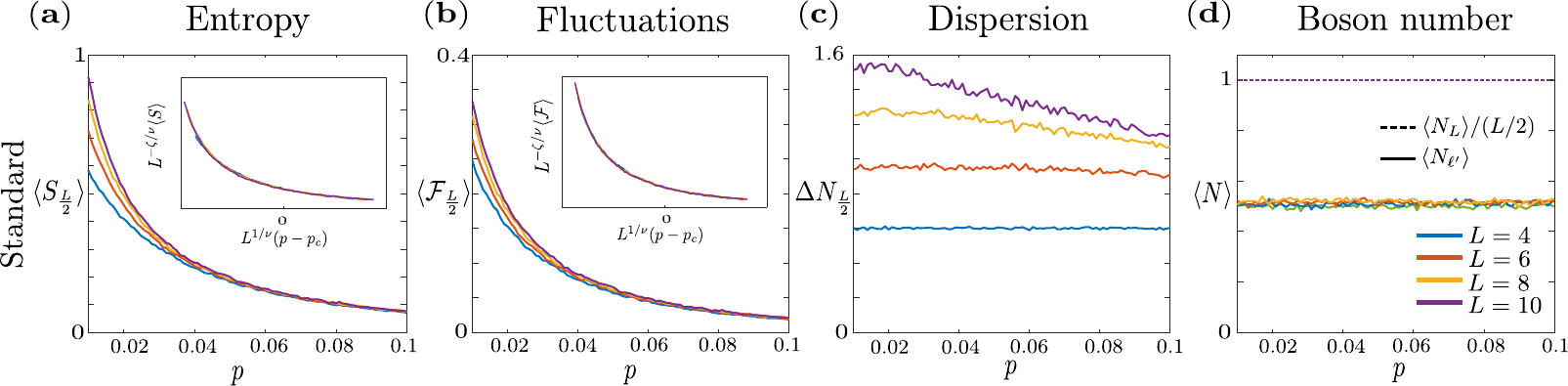}
\caption{\label{fig:numerical_results_3} Numerical simulations for arrays of subsystems with local dimensions $L/2+1$ using \textit{standard} measurements for different system sizes $L=4,6,8,10$ using $U/J=5$. (a) von Neumann entropy and (b) fluctuations of the number operator in the half of the chain of transmons averaged over iterations as a function of the measurement probability $p$. The insets show the finite-size scaling analysis using the Ansatz $L^{-\zeta/\nu} \braket{S/\mathcal{F}}=f[L^{-1/\nu}(p-p_c) ] $ where $p_c$ is the critical parameter, $\zeta$ and $\nu$ are the scaling exponents and $f[x]$ is an unknown function~\citep{houdayer04}. (c) Dispersion in the number of bosons of the half of the chain of transmons. (d) Number of bosons in the complete chain of transmons $ \braket{N_L}$ (colored dashed lines) and at one single site in the middle of the chain $\braket{N_{\ell'}}$ (colored solid lines). For computing $\braket{S}$ and $\braket{\mathcal{F}}$ we considered post-selection of trajectories, while for the computing $\braket{N}$ and $\Delta N$ we did not take into account any post-selection. The results are computed for $10^4$ circuit iterations.}
\end{figure*}

In this appendix, we describe the details of the numerical simulation emphasizing the subtleties of the \textit{feedback} measurements protocol and how we compute relevant statistics. The results were obtained using the Julia~\citep{bezanson17} programming language. The simulated hybrid system, which represents the circuit, is built by a series of time steps consisting of a unitary evolution followed by probabilistic projective measurements at each site for an initial state $\ket{\psi_0}=\ket{10}^{\otimes L/2}$. For each time step, the system of $L$ transmons is evolved unitarily under the Hamiltonian of the Eq.~\eqref{Hamiltonian_BH} for a period of time $dt=0.02 J^{-1}$, using exact diagonalization for small systems and the numerically exact Krylov subspace method~\citep{saad92} for larger systems. After the unitary evolution, we evaluate in each site the probability of performing a measurement or not depending on the measurement probability $p=\Gamma dt$. Note that the physical relevant parameter in the numerical simulations is $p$, but this relation is useful for comparing with the analytical results. We determine the measurements' results by evaluating the probabilities based on Born's rule. Note that for the \textit{feedback} measurements, the resulting state is projected to the predetermined state of each site, which corresponds to the initial state $ \ket{\overline{\alpha}}=\ket{\psi_0}$. Therefore, these two events define a time step after which the state needs to be renormalized due to the non-linear effect of the measurements. For each set of parameters, we repeat the simulation several times ($10^4$ for $L \leq 10$ and $5\cdot 10^3$ for $L=12$), which we have defined as iterations. The simulations are executed for a total time of $T=20 J^{-1}$  for the array of two-dimensional subsystems and $T=30 J^{-1}$ for the higher-dimensional subsystems array. In each case, these times correspond to the steady state, defined as the time when the iteration-averaged observables stabilize across all system sizes and ranges of measured probabilities.

Regarding the circuit-averaged observables in the steady state $T$, we differentiate two types: postselected and non-postselected. It is also important to point out that in the analytical results, the average is performed over all the trajectories weighted by their probabilities, but in the numerical simulations, the averaging is performed over iterations of the circuit, each of them corresponding to a trajectory, which may appear repeated or not appear at all, for frequent and infrequent trajectories, respectively (the incidence, in this case, represents the weighting). For the postselected quantities, we evaluate the complete state of the system $\ket{\psi^i(T)}$, that is, the pure state that is given generally by a superposition on the basis of the number of bosons for a particular iteration of the circuit $i$. We do this for each iteration and then calculate the von Neumann entropy and the observables' fluctuation defined by the Hermitian operator's expectation values. For non-postselected quantities, we obtain the value of the corresponding observable using Born's rule on the final state of each iteration, that is, performing a probabilistic projection on one of the basis vectors in the number of bosons that constitutes the final state $\ket{\tilde{\psi^i}}$. Therefore, the circuit-averaged quantities are given by 
\begin{align}
    \braket{S_{L/2}} & =-\frac{1}{M}\sum_{i=1}^{M}  \Tr \left(\hat{\rho}^i(T)_{[1,\frac{L}{2}]} \ln \hat{\rho}^i(T)_{[1,\frac{L}{2}]}\right),\\
    \braket{\mathcal{F}_{L/2}}& =\frac{1}{M} \sum_{i=1}^{M} \left [ \bra{\psi^i(T)}\hat{N}_{[1,\frac{L}{2}]}^2\ket{\psi^i(T)} \right.  \nonumber \\
    & \qquad\qquad \left. -\braket{\psi^i(T)|\hat{N}_{[1,\frac{L}{2}]}|\psi^i(T)}^2\right],\\
    \braket{N}& =\frac{1}{M}\sum_{i=1}^{M} \bra{\tilde{\psi^i}}  \hat{N}_{A}  \ket{\tilde{\psi^i}},\\
    \label{N_num_mean}
    \Delta N_{L/2}& =\frac{1}{M}\sum_{i=1}^{M}\left[\bra{\tilde{\psi^i}} \hat{N}_{[1,\frac{L}{2}]}  \ket{\tilde{\psi^i}} \right.\nonumber \\
    & \qquad \quad \left. - \Big (M^{-1}\sum_{j=1}^{M}\bra{\tilde{\psi^j}}  \hat{N}_{[1,\frac{L}{2}]}  \ket{\tilde{\psi^j}} \Big)\right]^2  ,
\end{align}
where $M$ is the total number of iterations, $\hat{\rho}_{[1,\frac{L}{2}]}=\Tr_{[\frac{L}{2}+1,L]}\hat{\rho}$, and $\ket{\tilde{\psi^i}}$ is obtained for each iteration and type of observable from the pure state $\ket{\psi^i(T)} $, after projecting the corresponding final state at $T$ following the Born's rule in the subspace of said observable, in such a way that we thus emulate a real quantum measurement for the particular observable. For calculating the distance between the distribution of the number of bosons for all the iterations considered in Eq.~\eqref{N_num_mean} and theoretical distributions, we used the simple metric $d(\text{obs},\text{theo})=(1/2) \sum_{n=0}^{N_T} |\text{obs}(n)-\text{theo}(n)|$. Similar results were obtained for Kullback-Leibler divergence and Bhattacharyya distance; data not shown.

For performing the collapse of the curves for the von Neumann entropy and fluctuation in the number operator we realized a finite-size scaling analysis. After assigning a set of values for the parameters $\{p_c, \nu,\zeta \}$ for the scaling law $L^{-\zeta/\nu} \braket{S/\mathcal{F}}=f[L^{-1/\nu}(p-p_c) ]$, we quantify the quality of the collapse following straightforwardly the method derived by Houdayer and Hartmann~\citep{houdayer04} based on a work of Kawashima and Ito~\citep{kawashima93}, where we minimize the function 
\begin{align}
    S=\frac{1}{\mathcal{N}} \sum_{i,j} \frac{(y_{ij}-Y_{ij})^2}{dy^2_{ij}+dY^2_{ij}},
    \label{S_fssa}
\end{align}
where $(x_{ij},y_{ij},dy_{ij})$ are the data points after scaling, $\mathcal{N}$ is the number of points, and $Y_{ij}$ and $dY_{ij}$ are the estimated position and standard error of the master curve at $x_{ij}$. Using standard in-built functions in MATLAB, we minimize iteratively the value of Eq.~\eqref{S_fssa}, and the errors for the set of parameters $\{p_c, \nu,\zeta \} $ were obtained individually for each parameter $b$ by $\max(\delta b_l,\delta b_u) $ where $S(b)+1=S(b-\delta b_l)=S(b+\delta b_u)$. 

All the parameters of the numerical simulations are expressed in terms of the hopping rate $J$, which corresponds to the mean value of a given distribution. Note that although we explicitly show the mean value of the on-site energy $\omega$, it could be ignored by switching to a rotating coordinate system with $\hat{U}=e^{-i\omega t \sum_\ell^L{\hat{n}_\ell}} $ since all transmons have the same mean energy and it does not affect many-body dynamics. Note also that during hybrid circuit dynamics, the total number of bosons may change, but it is a conserved quantity during the unitary evolutions. The interaction strength $U$ arising from anharmonicity does not affect two-dimensional systems and for that reason was omitted from the main text numerical simulations, although it is explicitly included in higher-dimensional systems from the appendix.

\subsection{Arrays of subsystems with higher dimensions}
In the main text, we show the numerical simulations of hybrid circuits implementing different types of measurements of transmons modeled as two-dimensional systems, that is, qubits. As described in Sec.~\ref{sec:model}, the dynamics of interacting transmons can be modeled by the attractive Bose-Hubbard model where, in principle, each site can host an infinite number of excitations, constituting an infinite-dimensional system, although as the number of excitations increases, higher-order corrections must be considered. However, experimentally the number of excitations per site is limited to $\sim 10$~\cite{koch2007}. In this subsection, we show the results of numerical simulations of transmons of local dimension $L/2+1$ with \textit{standard} measurements, where we restrict the dynamics to a particular sector of the total number of bosons, thus considerably reducing the size of the Hilbert space. However, we do not implement \textit{feedback} measurements, since we would need to consider all sectors, and the total dimension $d^L$ becomes quite large and unmanageable with our numerical methods for the minimum sizes required to calculate relevant quantities, i.e.~$ L=4,6,8,10$. 

The numerical results are shown in Fig.~\ref{fig:numerical_results_3} for the experimental feasible parameter $U/J=5$. We obtain essentially the same result as in the two-dimensional case of Fig.~\ref{fig:numerical_results_1}, although the critical parameters may depend on the value of $U/J$. For the chosen value $U/J=5 $, we obtained the critical parameters $p_c^{S,st}=0.057 \pm 0.007$, $\nu^{S,st}=7.7 \pm 0.9$, $\zeta^{S,st}=0.6 \pm 0.2$ and $p_c^{\mathcal{F},st} =0.057 \pm 0.013$, $\nu^\mathcal{F,st}=14 \pm 3$, $\zeta^\mathcal{F,st}=1.3 \pm 0.5$, for entanglement entropy and the fluctuation of the number operator, respectively. The dispersion in the number of bosons and the mean value in the number of bosons show similar behavior to the two-dimensional case. However, it is interesting to note that $\braket{N_{\ell'}} \sim 1/2$, even considering larger dimensions; this phenomenon could be interpreted as the value of $U/J $ is large enough for evolution to remain in the same boson number manifold than the initial state i.e.~the one in which the bosons are not stacked in the same location~\citep{mansikkamaki22}.

\section{Average over circuit realizations}
\label{appsec:circuit_average}
In this appendix, we explain in detail the new results obtained in Sec. \ref{sec:statistics} on calculating trajectory-averaged quantities in the area-law phase by using simple statistical and combinatorial arguments. As previously discussed, the observables described by Eq.~\eqref{replicated_observables} are connected to trajectory-averaged quantities in the proper replica limit. However, to gain a clearer intuition about the averaging process, it is helpful to comprehend each probability distribution and its corresponding physical significance. First, note that the average value of an observable $\overline{\left\langle O_k \right\rangle}$ has four different sources of randomness. The first is related to the probability of performing a measurement $p$, and it is the critical parameter of the~MIPT, such that it describes the probabilities for different scenarios; there is a probability $p^{b}(1-p)^{M-b}$ of performing $b$ measurements, where $M=LT$ is the maximum number of possible measurements in the whole circuit space-time, and $0 \leq b \leq M$. The other three sources of randomness arise from probability distributions: quantum mechanical uncertainty for a superposition state in the last measurement for obtaining the result of observables, the measurement outcomes of the measurements performed during the dynamics, and the values of the parameters of every unitary gate. The first arises from the inevitable probabilistic nature of performing a particular measurement of an observable, which will give us the expectation value for the final state of the circuit. The second source arises from the probabilistic result of each of the measurements made \textit{during the time evolution}, representing the probability of each trajectory, given by Born's rule; therefore, there will be trajectories that are more frequent than others. The last source of randomness refers to the fact that each gate parameter can have different values from a probability distribution. The variation of these values will lead to the production of different unitary evolution operators. Consequently, for each set of parameter values, distinct states will emerge, thereby influencing the probabilities of measurement outcomes according to Born's rule.

For experimental purposes, we could think of schemes of increasing complexity. Let us assume that the circuit has fixed parameter values and that there are no measurements during the evolution. If we measure an observable $\hat{O}$ at long times, and we could repeat the experiment several times, we would obtain a mean value corresponding to the expectation value $ \braket{\hat{O}}=\tr(\hat{O}\hat{\rho}^{\prime}) $ and a fluctuation corresponding to the uncertainty $ \mathcal{F}=\braket{\hat{O}^2}-\braket{\hat{O}}^2 $ for only one trajectory. If we include measurements during the time evolution and let the unitary gates have different values, each time we run the circuit we will obtain a trajectory defined by the outcomes of every time evolution measurement $m_\bold{i} $ and a set of parameters for unitary gates $\theta_\bold{i}  $ with an associated expectation value $\braket{\hat{O}}_{m_\bold{i},\theta_\bold{i}}$ and variance $\mathcal{F}_{m_\bold{i},\theta_\bold{i}}$. Due to the randomness in the measurements and circuit parameters during time evolution, obtaining experimentally $ \braket{\hat{O}}_{m_\bold{i},\theta_\bold{i}}$ and $\mathcal{F}_{m_\bold{i},\theta_\bold{i}}$ will require to perform different experiments maintaining the same measurement outcomes $m_\bold{i} $ and circuit parameters $\theta_\bold{i}$. In this sense, as we introduced in the main text, we can express the trajectory-averaged observable as
\begin{align}
    \overline{\langle O_k \rangle}= \sum_{b=0}^M p^{b} (1-p)^{M-b}\langle O_k^b \rangle_{\textnormal{m},\theta} ,
    \label{averaged_observables_app}
\end{align}
where
\begin{align}
\langle O_k^b \rangle_{\substack{\textnormal{m},\theta}} =  \int d\theta_{\bold{i}}   \sum_{\bold{b}}^{\binom{M}{b}} \sum_{m_{\bold{b}}}^{d^b} p_{m_\bold{b},\theta_{\bold{i}}} [\tr(\hat{O}\hat{\rho}^{\prime}_{m_{\bold{b}},\theta_\bold{i}} )]^k,
\label{averaged_observables_distribution_app}
\end{align}
where $p_{m_{\bold{b}},\theta_{\bold{i}}} \equiv p_{\theta_\bold{i}} p_{m_\bold{b}}(\theta) $ indicating the Born's rule dependency on the gate parameters, $\binom{M}{b}=\frac{M!}{b!(M-b)!}$, $\bold{b}$ are the arrays comprising all the possible $\binom{M}{b}$ combinations of arranging $b$ measurements in the total $M=LT$ positions of space-time, and $\hat{O}$ can be any operator. All the probability distributions are normalized in different ways. For the probability $p$ of performing a measurement at the sites of the circuit, we have
\begin{align}
    1=&\sum_{b=0}^M p^{b} (1-p)^{M-b} \sum_{\bold{b}}^{\binom{M}{b}}  \equiv \sum_{b=0}^M p^{b} (1-p)^{M-b} {\binom{M}{b}} \nonumber \\
    =& p^{\rm M}+M p^{M-1}(1-p)+\frac{M(M-1)}{2}p^{M-2}(1-p)^2 \nonumber \\
    &+\cdots+p(1-p)^{M-1}+(1-p)^M,
    \label{p_normalization_app}
\end{align}
where all the possibilities of performing and not performing measurements (and in which specific location in space-time) are represented. Second, we have that for a particular set of performed measurements $\bold{b}$ and a set of gate parameters $\theta_{\bold{i}} $ the normalization of the measurement outcomes is
\begin{align}
    1= \sum_{m}^{d^b} p_{m_\bold{b}}(\theta).
\end{align}
Third, the gate parameters are normalized such that
\begin{equation}
   1=\int d\theta_{\bold{i}} p_{\theta_\bold{i}}
\end{equation}

As described in the main text in Sec. \ref{sec:statistics}B, we are interested in calculating the following quantities related to the number of bosons under certain regimes
\begin{align}
    \langle N_k \rangle =& \sum_{b=0}^M p^{b} (1-p)^{M-b} \left\{  \sum_{\bold{b}}^{\binom{M}{b}} \sum_{m}^{d^b} p_{m_\bold{b}} [\tr(\hat{N}\hat{\rho}^{\prime}_{m_{\bold{b}}} )]^k \right\},
    \label{averaged_observables_2_app} \\
    \langle \mathcal{F} \rangle=& \sum_{b=0}^M  p^{b}  (1-p)^{M-b} \nonumber \\
    & \times \left\{  \sum_{\bold{b}}^{\binom{M}{b}} \sum_{m}^{d^b} p_{m_\bold{b}} \left[\tr(\hat{N}^2\hat{\rho}^{\prime}_{m_{\bold{b}}} )-[\tr(\hat{N}\hat{\rho}^{\prime}_{m_{\bold{b}}} )]^2 \right] \right\},
    \label{averaged_observables_F_app} \\
    \Delta N \equiv & \langle (N^2)_1 \rangle -\langle N_1 \rangle^2 \nonumber \\
    =& \sum_{b=0}^M p^{b} (1-p)^{M-b} \left\{  \sum_{\bold{b}}^{\binom{M}{b}} \sum_{m}^{d^b} p_{m_\bold{b}} \tr(\hat{N}^2\hat{\rho}^{\prime}_{m_{\bold{b}}} ) \right\} \nonumber \\
    &- \left( \sum_{b=0}^M p^{b} (1-p)^{M-b} \left\{  \sum_{\bold{b}}^{\binom{M}{b}} \sum_{m}^{d^b} p_{m_\bold{b}} \tr(\hat{N}\hat{\rho}^{\prime}_{m_{\bold{b}}} ) \right\}\right)^2.
    \label{averaged_observables_DN_app}
\end{align}
Although the formula in Eq.~\eqref{averaged_observables_app} is extremely complicated for calculating trajectory-averaged quantities, we can obtain relevant analytical results in the high measurement regime $p \sim 1$, i.e.~deep in the area-law phase (and similarly in the measurement regime under $p \sim $0, i.e.~size-dependent phase). In a first approximation, we only consider the cases in Eq.~\eqref{p_normalization_app} for terms up to $\mathcal{O}(x^2)$, where $x=1-p$ is the probability of not measuring a single event in space-time. Therefore, we consider the unique possibility of performing $M$ measurements, the $M$ possibilities of performing $M-1$ measurements, and the $\frac{M(M-1)}{2}$ possibilities of performing $ M-2$ measurements. This implies that we must keep terms scaling as $p^M$, $p^{M-1}(1-p)$, and $p^{M-2}(1-p)^2$.

\begin{figure}
\includegraphics[width=\linewidth]{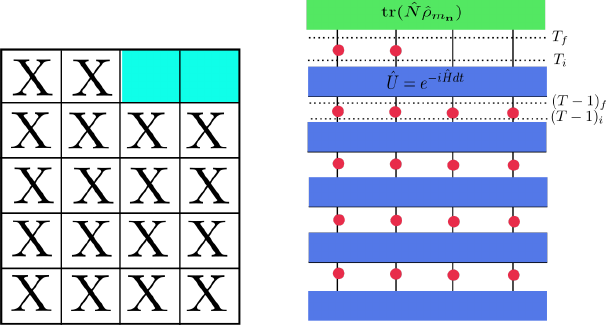}
\caption{\label{fig:circuit_average} Diagram followed to study the realizations of the circuit. Left: Example for a circuit of $L=4$ transmons with $T=5$ time step in which $M-2=LT-2=18$ measurements (represented by an $\text{X}$) were made. Right: Correspondence of the circuit with the states of the system in different time steps (Note that at every $t_f$, the state is renormalized.}
\end{figure}

The second approximation is related to the state; since we are going to study a highly measured system, we describe the unitary evolution for a small $dt$ (i.e., between measurements) of a product state given by the results of the measurements in the previous time step, paying attention to how the states are in $ T_i$ and $T_f$ (Fig.~\ref{fig:circuit_average}). Assuming that all measurements were made just before $(T-1)_f$ with the result $\overline{\beta}=\{ \beta_1,...,\beta_L \}$, we know that the state at $ (T-1)_f$ is given by the product state
\begin{align}
    \ket{\psi_{\overline{\beta}}(T-1)_f}=\ket{\beta_1,...,\beta_L},
    \label{circuit_averaged_state_T_1f_app}
\end{align}
so, expanding the time evolution operator for the Bose-Hubbard Hamiltonian of Eq.~\eqref{Hamiltonian_BH}
\begin{align}
    e^{-\frac{i\hat{H}dt}{\hbar}}&=\sum_{n=0}^{\infty} (-i)^n \left(\frac{\hat{H}}{\hbar} \right)^n dt^n \nonumber \\
    &=\hat{I}-i\frac{\hat{H}}{\hbar}dt-\frac{1}{2}\left(\frac{\hat{H}}{\hbar}\right)^2 dt^2+\cdots,
\end{align}
the state at $T_i$ is given by
\begin{align}
    \ket{\psi_{\overline{\beta}}}=\ket{\overline{\beta}}+\sum_{\overline{\gamma}}^{d^L} \sum_{p=1}^{\infty}(-i)^p c_{\overline{\gamma}}^{[p]} dt^p \ket{\overline{\gamma}},
    \label{state_Ti_app}
\end{align}
where $c_{\overline{\gamma}}^{[p]} $ are real and include all the possible prefactors for every Fock state of the basis. We are interested in calculating observables $\hat{O}$ for which the Fock are eigenstates, up to a certain order in $dt$. Therefore, we first obtain
\begin{align}
    &\braket{\psi_{\overline{\beta}}|\hat{O}|\psi_{\overline{\beta}}}=f(\overline{\beta})+\sum_{\overline{\gamma}}^{d^L} \sum_{p,q=1}^{\infty} i^p (-i)^q c_{\overline{\gamma}}^{[p]} c_{\overline{\gamma}}^{[q]} f(\overline{\gamma}) dt^p dt^q, \notag \\
    &\braket{\psi_{\overline{\beta}}|\psi_{\overline{\beta}}}=1+\sum_{\overline{\gamma}}^{d^L} \sum_{p,q=1}^{\infty}  i^p (-i)^q c_{\overline{\gamma}}^{[p]} c_{\overline{\gamma}}^{[q]} dt^p dt^q,
\end{align}
where $f(\overline{\gamma})$ is a function that depends on the Fock states $\ket{\overline{\gamma}}=\ket{\gamma_1,\gamma_2,...,\gamma_L} $. Therefore, the expectation values of the observables up to fourth order in $dt$ is given by
\begin{align}
    \frac{\braket{\psi_{\overline{\beta}}|\hat{O}|\psi_{\overline{\beta}}}}{\braket{\psi_{\overline{\beta}}|\psi_{\overline{\beta}}}}=&f(\overline{\beta})+dt^2 \left\{ \sum_{\overline{\gamma}}^{d^L} \left|c_{\overline{\gamma}}^{[1]} \right|^2 \left[ f(\overline{\gamma})-f(\overline{\beta})\right]\right\} \nonumber \\
    &+dt^4 \left\{\left(1- \sum_{\overline{\gamma}}^{d^L} \left| c_{\overline{\gamma}}^{[2]} \right|^2\right) \right. \nonumber \\
    & \left. \times \left( \sum_{\overline{\gamma}}^{d^L} \left| c_{\overline{\gamma}}^{[2]} \right|^2 \left[ f(\overline{\gamma})-f(\overline{\beta})\right] \right) \right\}  \nonumber \\
    &- dt^4 \left\{ 2\sum_{\overline{\gamma}}^{d^L} \left| c_{\overline{\gamma}}^{[1]} \right|^2 \left| c_{\overline{\gamma}}^{[3]} \right|^2 f( \overline{\gamma}) \right\}.
    \label{observables_circuit_dt4_app}
\end{align}
So far, we have shown that for a general system that conserves the total number of excitations, the trajectory-averaged expectation values of the observables do not depend on the terms of the Hamiltonian for which the Fock states are eigenstates (i.e., the on-site energy and auto-interaction) up to second order in $dt$ because the summation term cancels. For the following subsections, we will keep the terms up to the second order in $dt$.

\subsection{\textit{Standard} measurements}
Although we are not going to be able to know the Born's probabilities $p_{m_\bold{n}} $ in Eq.~\eqref{averaged_observables_F_app}, we can obtain the scaling of $\braket{\mathcal{F}_{\frac{L}{2}}}$ taking into account that it vanishes for any state for which the first half of the chain can be expressed as a product state. From all the possible cases representing the different arrangement of $M$, $M-1$, and $M-2$ measurements, only the case where there were no measurements at the sites $L/2$ and $L/2+1$ at the last time step $T_f$ does not vanish, contributing to the trajectory-average such that
\begin{multline}
    \braket{\mathcal{F}_{L/2}}  \approx  x^2 \sum_{\beta_1 +\cdots+\beta_L=N_T}^{\binom{L+N_T-1}{N}} p_{m_{\overline{\beta}}} \left[(\beta_{L/2}+1)\beta_{L/2+1} \right. \\
    \left.+\beta_{L/2}(\beta_{L/2+1}+1)\right]  \left( \frac{J}{\Gamma}\right)^2, 
    \label{averaged_observables_F5_app}
\end{multline}
where we do not know the probabilities for the different trajectories $p_{m_{\overline{\beta}}} $ and we have considered that $p=\Gamma dt$. Notice that the summation is normalized so that $\sum_{\beta_1 +\cdots+\beta_L=N_T}^{\binom{L+N_T-1}{N_T}} p_{m_{\overline{\beta}}} = 1 $, where we have explicitly included that the total number of bosons is conserved, thus limiting the possible results of the measurements (considering that we have started from an initial state with a defined number of bosons $N_T$), therefore there is no contribution of $L$ to the total value but a dependency on the local dimension. We have proved that for a large probability of performing measurements (i.e.~considering terms up to $(1-p)^2$), $\braket{\mathcal{F}_{L/2}}$ (and, therefore, also $\braket{S_{L/2}}$) scales as $\left( \frac{J}{\Gamma}\right)^2$ and is independent of the size of the subsystem $L/2$ and dependent on the local dimension~$d$. 

\subsection{\textit{Feedback} measurements}
The problem of not knowing the probabilities associated with each trajectory arising in the case of \textit{standard} measurements disappears in the case of \textit{feedback} measurements, where the measurement outcomes are always the same following a predetermined pattern $\overline{\alpha}=\{\alpha_1,\alpha_2,...,\alpha_L \}$. However, since for a \textit{feedback} measurement, a \textit{standard} measurement must first be carried out before the conditional projection to the known states, the study is complicated by still having to take into account all possible cases, where the total number of bosons can change for different states.

To calculate $\braket{\mathcal{F}_{L/2}}$ we again take advantage of the knowledge that all cases in which arranged measurements produce product states for the first half vanishes, except for the same single case that for \textit{standard} measurements. Explicitly, the Born probabilities and possible states (with the number of bosons in the first half of the chain) for this particular case with probability $p^{M-2}(1-p)^2 $ are
\begin{widetext}
\begin{align}
    p_{sup}=&1-p_{N_{\frac{L}{2}}-1}-p_{N_{\frac{L}{2}}+1}-p_{N_{\frac{L}{2}}}:   \\
    & \quad \to \ket{\psi_{sup}}=\sqrt{1-|\epsilon_{\frac{L}{2},\frac{L}{2}+1}^{+-}|^2dt^2-|\epsilon_{\frac{L}{2},\frac{L}{2}+1}^{-+}|^2dt^2-p_{N_{\frac{L}{2}}-1}-p_{N_{\frac{L}{2}}+1}-p_{N_{\frac{L}{2}}}}\ket{\alpha_1,...,\alpha_L} \nonumber \\
    &  \quad \quad \quad \quad \quad \quad -i|\epsilon_{\frac{L}{2},\frac{L}{2}+1}^{+-}|dt \ket{\alpha_1,...,\alpha_{\frac{L}{2}}+1,...,\alpha_L}-i|\epsilon_{\frac{L}{2},\frac{L}{2}+1}^{-+}|dt\ket{\alpha_1,...,\alpha_{\frac{L}{2}}-1,...,\alpha_L} \label{p_sup_app} \\
    p_{N_{\frac{L}{2}}}=&\sum_{\ell=1}^{\frac{L}{2}-2} |\epsilon_{\ell,\ell+1}^{+-}|^2 dt^2 +\sum_{\ell=\frac{L}{2}+1}^{L-1}|\epsilon_{\ell,\ell+1}^{+-}|^2 dt^2+\sum_{\ell=1}^{\frac{L}{2}-2} |\epsilon_{\ell,\ell+1}^{-+}|^2 dt^2 +\sum_{\ell=\frac{L}{2}+1}^{L-1}|\epsilon_{\ell,\ell+1}^{-+}|^2 dt^2: \ket{\alpha_1,...,\alpha_L}, \quad N_{L/2}
    \label{p_N_app} \\ 
    p_{N_{\frac{L}{2}}-1}=&|\epsilon_{\frac{L}{2}-1,\frac{L}{2}}^{+-}|^2 dt^2: \ket{\alpha_1,...,\alpha_{\frac{L}{2}}-1,...,\alpha_L}, \quad N_{L/2}-1
    \label{p_Nmenos_app} \\
    p_{N_{\frac{L}{2}}+1}=&|\epsilon_{\frac{L}{2}-1,\frac{L}{2}}^{-+}|^2 dt^2: \ket{\alpha_1,...,\alpha_{\frac{L}{2}}+1,...,\alpha_L}, \quad N_{L/2}+1,
    \label{p_Nmas_app}
\end{align}
\end{widetext}
where $\epsilon_{\ell,\ell+1}^{+-} $ and $\epsilon_{\ell,\ell+1}^{-+} $ stand for $J\sqrt{(\alpha_\ell+1)\alpha_{\ell+1}} $ and $J\sqrt{\alpha_\ell(\alpha_{\ell+1}+1)} $, respectively; and $ \braket{\psi_{sup}|\psi_{sup}}=p_{sup}$. The states with a different number of bosons (Eqs.~\eqref{p_Nmenos_app} and~\eqref{p_Nmas_app}) are product states, therefore they will contribute to the fluctuation by lowering the Born's probability of the superposition state in Eq.~\eqref{p_sup_app}, such that the trajectory-averaged fluctuation is given by
\begin{align}
    \braket{\mathcal{F}} \approx & x^2 p_{sup} \left[(\alpha_{\frac{L}{2}}+1)\alpha_{\frac{L}{2}+1}+\alpha_{\frac{L}{2}}(\alpha_{\frac{L}{2}+1}+1)\right]  \left( \frac{J}{\Gamma}\right)^2 \nonumber \\
    \approx & x^2 \left[(\alpha_{\frac{L}{2}}+1)\alpha_{\frac{L}{2}+1}+\alpha_{\frac{L}{2}}(\alpha_{\frac{L}{2}+1}+1)\right]  \left( \frac{J}{\Gamma}\right)^2,
    \label{averaged_observables_F6_app}
\end{align}
where in the last step we have considered that $p_{sup}=1+\mathcal{O}[(J/\Gamma)^4]$.

Calculating $\Delta N_{L/2}$ up to the second order in $(1-p)$ is a challenging task, but obtaining it up to the first order is straightforward. Here, we can observe significant differences compared to the fluctuation. For the zeroth order, we have a product state $\ket{\alpha_1,...,\alpha_L}$ with probability $1$, but we need to split the $M$ possibilities of performing $M-1$ measurements in four different groups:
\begin{widetext}
\begin{align}
    &\#L(T-1),  &p_{N_{L/2}}\to&1: \ket{\alpha_1,...,\alpha_L}, \quad &&N_{L/2},& \\
    &\#L/2,   &p_{N_{L/2}}=&|\epsilon_{\ell,\ell+1}^{+-/-+}|^2 dt^2 \to 1: \ket{\alpha_1,...,\alpha_L}, \quad &&N_{L/2},& \\
    &\#1, &p_{N_{L/2}}=&1-p_{N_{L/2}+1}-p_{N_{L/2}-1}: \ket{\alpha_1,...,\alpha_L}, \quad &&N_{L/2},& \\
    & &p_{N_{L/2}+1}=&|\epsilon_{1,2}^{+-}|^2 dt^2: \ket{\alpha_1+1,...,\alpha_L}, \quad &&N_{L/2}+1,& \\
    & &p_{N_{L/2}-1}=&|\epsilon_{1,2}^{-+}|^2 dt^2: \ket{\alpha_1-1,...,\alpha_L}, \quad &&N_{L/2}-1,& \\
    &\#L/2-1,  &p_{N_{L/2}}=&1-p_{N_{L/2}+1}-p_{N_{L/2}-1}: \ket{\alpha_1,...,\alpha_L}, \quad &&N_{L/2},& \\
    & \quad \left\lbrace \ell \in [2,\frac{L}2] \right\rbrace &p_{N_{L/2}+1}=&(|\epsilon_{\ell-1,\ell}^{-+}|^2 +|\epsilon_{\ell,\ell+1}^{+-}|^2)dt^2: \ket{\alpha_1,...,\alpha_{\ell-1},\alpha_{\ell}+1,\alpha_{\ell+1},...,\alpha_L}, \quad &&N_{L/2}+1,&  \\
    & &p_{N_{L/2}-1}=&(|\epsilon_{\ell-1,\ell}^{+-}|^2 +|\epsilon_{\ell,\ell+1}^{-+}|^2)dt^2: \ket{\alpha_1,...,\alpha_{\ell-1},\alpha_{\ell}-1,\alpha_{\ell+1},...,\alpha_L}, \quad &&N_{L/2}-1,&
\end{align}
where $\#$ refers to the number of terms. Using Eq.~\eqref{averaged_observables_DN_app}, we obtain
\begin{align}
    \braket{N_{\frac{L}{2}}^2}=&  p^M N_{\frac{L}{2}}^2+p^{M-1}(1-p) L(T-1) N_{\frac{L}{2}}^2+p^{M-1}(1-p) \frac{L}{2}N_{\frac{L}{2}}^2 \nonumber \\
    &+p^{M-1}(1-p) \left[ (1-(|\epsilon_{1,2}^{+-}|^2 +|\epsilon_{1,2}^{+-}|^2)dt^2)N_{\frac{L}{2}}^2+|\epsilon_{1,2}^{+-}|^2 dt^2(N_{\frac{L}{2}}+1)^2+|\epsilon_{1,2}^{-+}|^2 dt^2(N_{\frac{L}{2}}-1)^2 \right] \nonumber \\
    & +p^{M-1}(1-p) \sum_{\ell=2}^{L/2} \left[ (1-(|\epsilon_{\ell-1,\ell}^{-+}|^2 +|\epsilon_{\ell,\ell+1}^{+-}|^2+|\epsilon_{\ell-1,\ell}^{+-}|^2 +|\epsilon_{\ell,\ell+1}^{-+}|^2)dt^2)N_{\frac{L}{2}}^2 \right. \nonumber \\
    &\left.+(|\epsilon_{\ell-1,\ell}^{-+}|^2 +|\epsilon_{\ell,\ell+1}^{+-}|^2)dt^2(N_{\frac{L}{2}}+1)^2+(|\epsilon_{\ell-1,\ell}^{+-}|^2 +|\epsilon_{\ell,\ell+1}^{-+}|^2)dt^2(N_{\frac{L}{2}}+1)^2 \right],
    \label{circuit_average_N2_app}
\end{align}
\begin{align}
    \braket{N_{\frac{L}{2}}}=&  p^M N_{\frac{L}{2}}+p^{M-1}(1-p) L(T-1) N_{\frac{L}{2}}+p^{M-1}(1-p) \frac{L}{2}N_{\frac{L}{2}} \nonumber \\
    &+p^{M-1}(1-p) \left[ (1-(|\epsilon_{1,2}^{+-}|^2 +|\epsilon_{1,2}^{+-}|^2)dt^2)N_{\frac{L}{2}}+|\epsilon_{1,2}^{+-}|^2 dt^2(N_{\frac{L}{2}}+1)+|\epsilon_{1,2}^{-+}|^2 dt^2(N_{\frac{L}{2}}-1) \right] \nonumber \\
    & +p^{M-1}(1-p) \sum_{\ell=2}^{L/2} \left[ (1-(|\epsilon_{\ell-1,\ell}^{-+}|^2 +|\epsilon_{\ell,\ell+1}^{+-}|^2+|\epsilon_{\ell-1,\ell}^{+-}|^2 +|\epsilon_{\ell,\ell+1}^{-+}|^2)dt^2)N_{\frac{L}{2}} \right. \nonumber \\
    &\left.+(|\epsilon_{\ell-1,\ell}^{-+}|^2 +|\epsilon_{\ell,\ell+1}^{+-}|^2)dt^2(N_{\frac{L}{2}}+1)+(|\epsilon_{\ell-1,\ell}^{+-}|^2 +|\epsilon_{\ell,\ell+1}^{-+}|^2)dt^2(N_{\frac{L}{2}}+1) \right],
    \label{circuit_average_N_app}
\end{align}
such that the dispersion is given by
\begin{align}
    \Delta N_{L/2}=\braket{N_{\frac{L}{2}}^2}-\braket{N_{\frac{L}{2}}}^2 &\approx x\left[|\epsilon_{1,2}^{+-}|^2+|\epsilon_{1,2}^{-+}|^2 +\sum_{\ell=2}^{L/2}(|\epsilon_{\ell-1,\ell}^{-+}|^2 +|\epsilon_{\ell,\ell+1}^{+-}|^2+|\epsilon_{\ell-1,\ell}^{+-}|^2 +|\epsilon_{\ell,\ell+1}^{-+}|^2)\right] \left(\frac{J}{\Gamma}\right)^2.
    \label{DN_nonconserving_app}
\end{align}
\end{widetext}
However, as it is easily seen in Eq.~\eqref{DN_nonconserving_app}, this effect is due to the non-conservation of the total number of bosons. If we separate the final results by the total number of bosons (in this case splitting the results into three groups of $N_{\overline{\alpha }}$, and $N_{\overline{\alpha}}+1$ and $ N_{\overline{\alpha}}-1$) we recover the same result as the fluctuation for the total number of bosons $N_T= N_{\overline{\alpha}}=\sum_{\ell=1}^ L \alpha_\ell$. We can calculate this quantity up to the second order in $(1-p)$, taking into account the possible distributions of measurements in the same way as we did with the \textit{standard} measurement. Thus, all cases produce product states except one whose states follow the probabilities described in Eqs.~\eqref{p_sup_app} and~\eqref{p_N_app}, although discarding the probabilities from Eqs.~\eqref{p_Nmas_app} and~\eqref{p_Nmenos_app}. Calculating the trajectory-averaged observables for the particular sector $N_{\overline{\alpha}}$ we have
\begin{widetext}
\begin{align}
    \braket{N^2_{L/2}}^{N_{\overline{\alpha}}}=&p^M N_{\frac{L}{2}}^2+M p^{M-1}(1-p)N_{\frac{L}{2}}^2+\left[ \frac{M(M-1)}{2}-1\right]p^{M-2} (1-p)^2 N_{\frac{L}{2}}^2 \nonumber \\
    &+p^{M-2} (1-p)^2\left[ (1-p_{N_{\frac{L}{2}}})\left(N_{L/2}^2+(|\epsilon_{\frac{L}{2},\frac{L}{2}+1}^{+-}|^2+|\epsilon_{\frac{L}{2},\frac{L}{2}+1}^{-+}|^2)dt^2+2N_{\frac{L}{2}}(|\epsilon_{\frac{L}{2},\frac{L}{2}+1}^{+-}|^2-|\epsilon_{\frac{L}{2},\frac{L}{2}+1}^{-+}|^2)dt^2 \right)+p_{N_{\frac{L}{2}}}N_{\frac{L}{2}}^2\right],
    \label{circuit_average_N2_sector_app} \\
    \braket{N_{L/2}}^{N_{\overline{\alpha}}}=&p^M N_{\frac{L}{2}}+M p^{M-1}(1-p)N_{\frac{L}{2}}+\left[ \frac{M(M-1)}{2}-1\right]p^{M-2} (1-p)^2 N_{\frac{L}{2}} \nonumber \\
    &+p^{M-2} (1-p)^2\left[ (1-p_{N_{\frac{L}{2}}})\left(N_{L/2}+(|\epsilon_{\frac{L}{2},\frac{L}{2}+1}^{+-}|^2-|\epsilon_{\frac{L}{2},\frac{L}{2}+1}^{-+}|^2)dt^2\right)+p_{N_{\frac{L}{2}}}N_{\frac{L}{2}}\right],
    \label{circuit_average_N_sector_app}
\end{align}
\end{widetext}
such that the dispersion of the number of bosons in the $N_{\overline{\alpha}} $ sector in half of the chain is given by
\begin{align}
    \Delta N_{L/2}^{N_{\overline{\alpha}}}= & \braket{N^2_{L/2}}^{N_{\overline{\alpha}}}-\left( \braket{N_{L/2}}^{N_{\overline{\alpha}}}\right)^2     \label{DN_conserving_app} \\
    \approx & x^2\left[(1-p_{N_{\frac{L}{2}}})(|\epsilon_{\frac{L}{2},\frac{L}{2}+1}^{+-}|^2+|\epsilon_{\frac{L}{2},\frac{L}{2}+1}^{-+}|^2) \right]\left( \frac{J}{\Gamma}\right)^2 \nonumber \\
    \approx & x^2 \left[(\alpha_{L/2}+1)\alpha_{L/2+1}+\alpha_{L/2}(\alpha_{L/2+1}+1)\right]  \left( \frac{J}{\Gamma}\right)^2. \notag
\end{align}
Note that in these approximations, the dispersion of Eq.~\eqref{DN_conserving_app} and the fluctuation of Eq.\eqref{averaged_observables_F6_app} are equivalent. However, it is worth noting that for higher orders, the dispersion may overestimate the size-dependent phase since $1-p_{N_{L/2}}>p_{sup}$. For an experimental implementation, we should measure the number of bosons at the $L$ sites, separate the results into different groups depending on the total number of bosons, and then calculate the dispersion for half of the chain. Note that for comparing these results with the replica method, we should take the limit $dt \to 0$, such that $\lim_{dt \to 0}x^2=\lim_{dt \to 0}(1-\Gamma dt)^2=1$. In this sense, the results obtained from the replica method are more general; however, in this case, the quantity has a clear physical interpretation.

\bibliography{references}

\begin{thebibliography}{96}%
\makeatletter
\providecommand \@ifxundefined [1]{%
 \@ifx{#1\undefined}
}%
\providecommand \@ifnum [1]{%
 \ifnum #1\expandafter \@firstoftwo
 \else \expandafter \@secondoftwo
 \fi
}%
\providecommand \@ifx [1]{%
 \ifx #1\expandafter \@firstoftwo
 \else \expandafter \@secondoftwo
 \fi
}%
\providecommand \natexlab [1]{#1}%
\providecommand \enquote  [1]{``#1''}%
\providecommand \bibnamefont  [1]{#1}%
\providecommand \bibfnamefont [1]{#1}%
\providecommand \citenamefont [1]{#1}%
\providecommand \href@noop [0]{\@secondoftwo}%
\providecommand \href [0]{\begingroup \@sanitize@url \@href}%
\providecommand \@href[1]{\@@startlink{#1}\@@href}%
\providecommand \@@href[1]{\endgroup#1\@@endlink}%
\providecommand \@sanitize@url [0]{\catcode `\\12\catcode `\$12\catcode `\&12\catcode `\#12\catcode `\^12\catcode `\_12\catcode `\%12\relax}%
\providecommand \@@startlink[1]{}%
\providecommand \@@endlink[0]{}%
\providecommand \url  [0]{\begingroup\@sanitize@url \@url }%
\providecommand \@url [1]{\endgroup\@href {#1}{\urlprefix }}%
\providecommand \urlprefix  [0]{URL }%
\providecommand \Eprint [0]{\href }%
\providecommand \doibase [0]{https://doi.org/}%
\providecommand \selectlanguage [0]{\@gobble}%
\providecommand \bibinfo  [0]{\@secondoftwo}%
\providecommand \bibfield  [0]{\@secondoftwo}%
\providecommand \translation [1]{[#1]}%
\providecommand \BibitemOpen [0]{}%
\providecommand \bibitemStop [0]{}%
\providecommand \bibitemNoStop [0]{.\EOS\space}%
\providecommand \EOS [0]{\spacefactor3000\relax}%
\providecommand \BibitemShut  [1]{\csname bibitem#1\endcsname}%
\let\auto@bib@innerbib\@empty
\bibitem [{\citenamefont {Potter}\ and\ \citenamefont {Vasseur}(2022)}]{Potter2022}%
  \BibitemOpen
  \bibfield  {author} {\bibinfo {author} {\bibfnamefont {A.~C.}\ \bibnamefont {Potter}}\ and\ \bibinfo {author} {\bibfnamefont {R.}~\bibnamefont {Vasseur}},\ }\bibinfo {title} {Entanglement dynamics in hybrid quantum circuits},\ in\ \href {https://doi.org/10.1007/978-3-031-03998-0_9} {\emph {\bibinfo {booktitle} {Entanglement in Spin Chains: From Theory to Quantum Technology Applications}}},\ \bibinfo {editor} {edited by\ \bibinfo {editor} {\bibfnamefont {A.}~\bibnamefont {Bayat}}, \bibinfo {editor} {\bibfnamefont {S.}~\bibnamefont {Bose}},\ and\ \bibinfo {editor} {\bibfnamefont {H.}~\bibnamefont {Johannesson}}}\ (\bibinfo  {publisher} {Springer International Publishing},\ \bibinfo {address} {Cham},\ \bibinfo {year} {2022})\ pp.\ \bibinfo {pages} {211--249}\BibitemShut {NoStop}%
\bibitem [{\citenamefont {Aharonov}(2000)}]{aharonov00}%
  \BibitemOpen
  \bibfield  {author} {\bibinfo {author} {\bibfnamefont {D.}~\bibnamefont {Aharonov}},\ }\bibfield  {title} {\bibinfo {title} {Quantum to classical phase transition in noisy quantum computers},\ }\href {https://doi.org/10.1103/PhysRevA.62.062311} {\bibfield  {journal} {\bibinfo  {journal} {Phys. Rev. A}\ }\textbf {\bibinfo {volume} {62}},\ \bibinfo {pages} {062311} (\bibinfo {year} {2000})}\BibitemShut {NoStop}%
\bibitem [{\citenamefont {Li}\ \emph {et~al.}(2018)\citenamefont {Li}, \citenamefont {Chen},\ and\ \citenamefont {Fisher}}]{li18}%
  \BibitemOpen
  \bibfield  {author} {\bibinfo {author} {\bibfnamefont {Y.}~\bibnamefont {Li}}, \bibinfo {author} {\bibfnamefont {X.}~\bibnamefont {Chen}},\ and\ \bibinfo {author} {\bibfnamefont {M.~P.~A.}\ \bibnamefont {Fisher}},\ }\bibfield  {title} {\bibinfo {title} {Quantum {Zeno} effect and the many-body entanglement transition},\ }\href {https://doi.org/10.1103/PhysRevB.98.205136} {\bibfield  {journal} {\bibinfo  {journal} {Phys. Rev. B}\ }\textbf {\bibinfo {volume} {98}},\ \bibinfo {pages} {205136} (\bibinfo {year} {2018})}\BibitemShut {NoStop}%
\bibitem [{\citenamefont {Li}\ \emph {et~al.}(2019)\citenamefont {Li}, \citenamefont {Chen},\ and\ \citenamefont {Fisher}}]{li19}%
  \BibitemOpen
  \bibfield  {author} {\bibinfo {author} {\bibfnamefont {Y.}~\bibnamefont {Li}}, \bibinfo {author} {\bibfnamefont {X.}~\bibnamefont {Chen}},\ and\ \bibinfo {author} {\bibfnamefont {M.~P.~A.}\ \bibnamefont {Fisher}},\ }\bibfield  {title} {\bibinfo {title} {Measurement-driven entanglement transition in hybrid quantum circuits},\ }\href {https://doi.org/10.1103/PhysRevB.100.134306} {\bibfield  {journal} {\bibinfo  {journal} {Phys. Rev. B}\ }\textbf {\bibinfo {volume} {100}},\ \bibinfo {pages} {134306} (\bibinfo {year} {2019})}\BibitemShut {NoStop}%
\bibitem [{\citenamefont {Chan}\ \emph {et~al.}(2019)\citenamefont {Chan}, \citenamefont {Nandkishore}, \citenamefont {Pretko},\ and\ \citenamefont {Smith}}]{chan19}%
  \BibitemOpen
  \bibfield  {author} {\bibinfo {author} {\bibfnamefont {A.}~\bibnamefont {Chan}}, \bibinfo {author} {\bibfnamefont {R.~M.}\ \bibnamefont {Nandkishore}}, \bibinfo {author} {\bibfnamefont {M.}~\bibnamefont {Pretko}},\ and\ \bibinfo {author} {\bibfnamefont {G.}~\bibnamefont {Smith}},\ }\bibfield  {title} {\bibinfo {title} {Unitary-projective entanglement dynamics},\ }\href {https://doi.org/10.1103/PhysRevB.99.224307} {\bibfield  {journal} {\bibinfo  {journal} {Phys. Rev. B}\ }\textbf {\bibinfo {volume} {99}},\ \bibinfo {pages} {224307} (\bibinfo {year} {2019})}\BibitemShut {NoStop}%
\bibitem [{\citenamefont {Skinner}\ \emph {et~al.}(2019)\citenamefont {Skinner}, \citenamefont {Ruhman},\ and\ \citenamefont {Nahum}}]{skinner19}%
  \BibitemOpen
  \bibfield  {author} {\bibinfo {author} {\bibfnamefont {B.}~\bibnamefont {Skinner}}, \bibinfo {author} {\bibfnamefont {J.}~\bibnamefont {Ruhman}},\ and\ \bibinfo {author} {\bibfnamefont {A.}~\bibnamefont {Nahum}},\ }\bibfield  {title} {\bibinfo {title} {Measurement-induced phase transitions in the dynamics of entanglement},\ }\href {https://doi.org/10.1103/PhysRevX.9.031009} {\bibfield  {journal} {\bibinfo  {journal} {Phys. Rev. X}\ }\textbf {\bibinfo {volume} {9}},\ \bibinfo {pages} {031009} (\bibinfo {year} {2019})}\BibitemShut {NoStop}%
\bibitem [{\citenamefont {Szyniszewski}\ \emph {et~al.}(2019)\citenamefont {Szyniszewski}, \citenamefont {Romito},\ and\ \citenamefont {Schomerus}}]{szyniszewski19}%
  \BibitemOpen
  \bibfield  {author} {\bibinfo {author} {\bibfnamefont {M.}~\bibnamefont {Szyniszewski}}, \bibinfo {author} {\bibfnamefont {A.}~\bibnamefont {Romito}},\ and\ \bibinfo {author} {\bibfnamefont {H.}~\bibnamefont {Schomerus}},\ }\bibfield  {title} {\bibinfo {title} {Entanglement transition from variable-strength weak measurements},\ }\href {https://doi.org/10.1103/PhysRevB.100.064204} {\bibfield  {journal} {\bibinfo  {journal} {Phys. Rev. B}\ }\textbf {\bibinfo {volume} {100}},\ \bibinfo {pages} {064204} (\bibinfo {year} {2019})}\BibitemShut {NoStop}%
\bibitem [{\citenamefont {Szyniszewski}\ \emph {et~al.}(2020)\citenamefont {Szyniszewski}, \citenamefont {Romito},\ and\ \citenamefont {Schomerus}}]{szyniszewski20}%
  \BibitemOpen
  \bibfield  {author} {\bibinfo {author} {\bibfnamefont {M.}~\bibnamefont {Szyniszewski}}, \bibinfo {author} {\bibfnamefont {A.}~\bibnamefont {Romito}},\ and\ \bibinfo {author} {\bibfnamefont {H.}~\bibnamefont {Schomerus}},\ }\bibfield  {title} {\bibinfo {title} {Universality of entanglement transitions from stroboscopic to continuous measurements},\ }\href {https://doi.org/10.1103/PhysRevLett.125.210602} {\bibfield  {journal} {\bibinfo  {journal} {Phys. Rev. Lett.}\ }\textbf {\bibinfo {volume} {125}},\ \bibinfo {pages} {210602} (\bibinfo {year} {2020})}\BibitemShut {NoStop}%
\bibitem [{\citenamefont {Lunt}\ and\ \citenamefont {Pal}(2020)}]{lunt20}%
  \BibitemOpen
  \bibfield  {author} {\bibinfo {author} {\bibfnamefont {O.}~\bibnamefont {Lunt}}\ and\ \bibinfo {author} {\bibfnamefont {A.}~\bibnamefont {Pal}},\ }\bibfield  {title} {\bibinfo {title} {Measurement-induced entanglement transitions in many-body localized systems},\ }\href {https://doi.org/10.1103/PhysRevResearch.2.043072} {\bibfield  {journal} {\bibinfo  {journal} {Phys. Rev. Research}\ }\textbf {\bibinfo {volume} {2}},\ \bibinfo {pages} {043072} (\bibinfo {year} {2020})}\BibitemShut {NoStop}%
\bibitem [{\citenamefont {Tang}\ and\ \citenamefont {Zhu}(2020)}]{tang20}%
  \BibitemOpen
  \bibfield  {author} {\bibinfo {author} {\bibfnamefont {Q.}~\bibnamefont {Tang}}\ and\ \bibinfo {author} {\bibfnamefont {W.}~\bibnamefont {Zhu}},\ }\bibfield  {title} {\bibinfo {title} {Measurement-induced phase transition: A case study in the nonintegrable model by density-matrix renormalization group calculations},\ }\href {https://doi.org/10.1103/PhysRevResearch.2.013022} {\bibfield  {journal} {\bibinfo  {journal} {Phys. Rev. Research}\ }\textbf {\bibinfo {volume} {2}},\ \bibinfo {pages} {013022} (\bibinfo {year} {2020})}\BibitemShut {NoStop}%
\bibitem [{\citenamefont {Bao}\ \emph {et~al.}(2020)\citenamefont {Bao}, \citenamefont {Choi},\ and\ \citenamefont {Altman}}]{bao20}%
  \BibitemOpen
  \bibfield  {author} {\bibinfo {author} {\bibfnamefont {Y.}~\bibnamefont {Bao}}, \bibinfo {author} {\bibfnamefont {S.}~\bibnamefont {Choi}},\ and\ \bibinfo {author} {\bibfnamefont {E.}~\bibnamefont {Altman}},\ }\bibfield  {title} {\bibinfo {title} {Theory of the phase transition in random unitary circuits with measurements},\ }\href {https://doi.org/10.1103/PhysRevB.101.104301} {\bibfield  {journal} {\bibinfo  {journal} {Phys. Rev. B}\ }\textbf {\bibinfo {volume} {101}},\ \bibinfo {pages} {104301} (\bibinfo {year} {2020})}\BibitemShut {NoStop}%
\bibitem [{\citenamefont {Jian}\ \emph {et~al.}(2020)\citenamefont {Jian}, \citenamefont {You}, \citenamefont {Vasseur},\ and\ \citenamefont {Ludwig}}]{jian20}%
  \BibitemOpen
  \bibfield  {author} {\bibinfo {author} {\bibfnamefont {C.-M.}\ \bibnamefont {Jian}}, \bibinfo {author} {\bibfnamefont {Y.-Z.}\ \bibnamefont {You}}, \bibinfo {author} {\bibfnamefont {R.}~\bibnamefont {Vasseur}},\ and\ \bibinfo {author} {\bibfnamefont {A.~W.~W.}\ \bibnamefont {Ludwig}},\ }\bibfield  {title} {\bibinfo {title} {Measurement-induced criticality in random quantum circuits},\ }\href {https://doi.org/10.1103/PhysRevB.101.104302} {\bibfield  {journal} {\bibinfo  {journal} {Phys. Rev. B}\ }\textbf {\bibinfo {volume} {101}},\ \bibinfo {pages} {104302} (\bibinfo {year} {2020})}\BibitemShut {NoStop}%
\bibitem [{\citenamefont {Choi}\ \emph {et~al.}(2020)\citenamefont {Choi}, \citenamefont {Bao}, \citenamefont {Qi},\ and\ \citenamefont {Altman}}]{choi20}%
  \BibitemOpen
  \bibfield  {author} {\bibinfo {author} {\bibfnamefont {S.}~\bibnamefont {Choi}}, \bibinfo {author} {\bibfnamefont {Y.}~\bibnamefont {Bao}}, \bibinfo {author} {\bibfnamefont {X.-L.}\ \bibnamefont {Qi}},\ and\ \bibinfo {author} {\bibfnamefont {E.}~\bibnamefont {Altman}},\ }\bibfield  {title} {\bibinfo {title} {Quantum error correction in scrambling dynamics and measurement-induced phase transition},\ }\href {https://doi.org/10.1103/PhysRevLett.125.030505} {\bibfield  {journal} {\bibinfo  {journal} {Phys. Rev. Lett.}\ }\textbf {\bibinfo {volume} {125}},\ \bibinfo {pages} {030505} (\bibinfo {year} {2020})}\BibitemShut {NoStop}%
\bibitem [{\citenamefont {Gullans}\ and\ \citenamefont {Huse}(2020)}]{gullans20}%
  \BibitemOpen
  \bibfield  {author} {\bibinfo {author} {\bibfnamefont {M.~J.}\ \bibnamefont {Gullans}}\ and\ \bibinfo {author} {\bibfnamefont {D.~A.}\ \bibnamefont {Huse}},\ }\bibfield  {title} {\bibinfo {title} {Dynamical purification phase transition induced by quantum measurements},\ }\href {https://doi.org/10.1103/PhysRevX.10.041020} {\bibfield  {journal} {\bibinfo  {journal} {Phys. Rev. X}\ }\textbf {\bibinfo {volume} {10}},\ \bibinfo {pages} {041020} (\bibinfo {year} {2020})}\BibitemShut {NoStop}%
\bibitem [{\citenamefont {Fuji}\ and\ \citenamefont {Ashida}(2020)}]{fuji20}%
  \BibitemOpen
  \bibfield  {author} {\bibinfo {author} {\bibfnamefont {Y.}~\bibnamefont {Fuji}}\ and\ \bibinfo {author} {\bibfnamefont {Y.}~\bibnamefont {Ashida}},\ }\bibfield  {title} {\bibinfo {title} {Measurement-induced quantum criticality under continuous monitoring},\ }\href {https://doi.org/10.1103/PhysRevB.102.054302} {\bibfield  {journal} {\bibinfo  {journal} {Phys. Rev. B}\ }\textbf {\bibinfo {volume} {102}},\ \bibinfo {pages} {054302} (\bibinfo {year} {2020})}\BibitemShut {NoStop}%
\bibitem [{\citenamefont {Yang}\ \emph {et~al.}(2020)\citenamefont {Yang}, \citenamefont {Grankin}, \citenamefont {Sieberer}, \citenamefont {Vasilyev},\ and\ \citenamefont {Zoller}}]{yang20}%
  \BibitemOpen
  \bibfield  {author} {\bibinfo {author} {\bibfnamefont {D.}~\bibnamefont {Yang}}, \bibinfo {author} {\bibfnamefont {A.}~\bibnamefont {Grankin}}, \bibinfo {author} {\bibfnamefont {L.~M.}\ \bibnamefont {Sieberer}}, \bibinfo {author} {\bibfnamefont {D.~V.}\ \bibnamefont {Vasilyev}},\ and\ \bibinfo {author} {\bibfnamefont {P.}~\bibnamefont {Zoller}},\ }\bibfield  {title} {\bibinfo {title} {Quantum non-demolition measurement of a many-body {Hamiltonian}},\ }\href {https://doi.org/10.1038/s41467-020-14489-5} {\bibfield  {journal} {\bibinfo  {journal} {Nat. Commun.}\ }\textbf {\bibinfo {volume} {11}},\ \bibinfo {pages} {775} (\bibinfo {year} {2020})}\BibitemShut {NoStop}%
\bibitem [{\citenamefont {Rossini}\ and\ \citenamefont {Vicari}(2020)}]{rossini20}%
  \BibitemOpen
  \bibfield  {author} {\bibinfo {author} {\bibfnamefont {D.}~\bibnamefont {Rossini}}\ and\ \bibinfo {author} {\bibfnamefont {E.}~\bibnamefont {Vicari}},\ }\bibfield  {title} {\bibinfo {title} {Measurement-induced dynamics of many-body systems at quantum criticality},\ }\href {https://doi.org/10.1103/PhysRevB.102.035119} {\bibfield  {journal} {\bibinfo  {journal} {Phys. Rev. B}\ }\textbf {\bibinfo {volume} {102}},\ \bibinfo {pages} {035119} (\bibinfo {year} {2020})}\BibitemShut {NoStop}%
\bibitem [{\citenamefont {Ivanov}\ \emph {et~al.}(2020)\citenamefont {Ivanov}, \citenamefont {Ivanova}, \citenamefont {Caballero-Benitez},\ and\ \citenamefont {Mekhov}}]{ivanov20}%
  \BibitemOpen
  \bibfield  {author} {\bibinfo {author} {\bibfnamefont {D.~A.}\ \bibnamefont {Ivanov}}, \bibinfo {author} {\bibfnamefont {T.~Y.}\ \bibnamefont {Ivanova}}, \bibinfo {author} {\bibfnamefont {S.~F.}\ \bibnamefont {Caballero-Benitez}},\ and\ \bibinfo {author} {\bibfnamefont {I.~B.}\ \bibnamefont {Mekhov}},\ }\bibfield  {title} {\bibinfo {title} {Feedback-induced quantum phase transitions using weak measurements},\ }\href {https://doi.org/10.1103/PhysRevLett.124.010603} {\bibfield  {journal} {\bibinfo  {journal} {Phys. Rev. Lett.}\ }\textbf {\bibinfo {volume} {124}},\ \bibinfo {pages} {010603} (\bibinfo {year} {2020})}\BibitemShut {NoStop}%
\bibitem [{\citenamefont {Bao}\ \emph {et~al.}(2021)\citenamefont {Bao}, \citenamefont {Choi},\ and\ \citenamefont {Altman}}]{bao21}%
  \BibitemOpen
  \bibfield  {author} {\bibinfo {author} {\bibfnamefont {Y.}~\bibnamefont {Bao}}, \bibinfo {author} {\bibfnamefont {S.}~\bibnamefont {Choi}},\ and\ \bibinfo {author} {\bibfnamefont {E.}~\bibnamefont {Altman}},\ }\bibfield  {title} {\bibinfo {title} {Symmetry enriched phases of quantum circuits},\ }\href {https://doi.org/https://doi.org/10.1016/j.aop.2021.168618} {\bibfield  {journal} {\bibinfo  {journal} {Ann. Phys. (N.Y.)}\ }\textbf {\bibinfo {volume} {435}},\ \bibinfo {pages} {168618} (\bibinfo {year} {2021})}\BibitemShut {NoStop}%
\bibitem [{\citenamefont {Sang}\ and\ \citenamefont {Hsieh}(2021)}]{sang21}%
  \BibitemOpen
  \bibfield  {author} {\bibinfo {author} {\bibfnamefont {S.}~\bibnamefont {Sang}}\ and\ \bibinfo {author} {\bibfnamefont {T.~H.}\ \bibnamefont {Hsieh}},\ }\bibfield  {title} {\bibinfo {title} {Measurement-protected quantum phases},\ }\href {https://doi.org/10.1103/PhysRevResearch.3.023200} {\bibfield  {journal} {\bibinfo  {journal} {Phys. Rev. Research}\ }\textbf {\bibinfo {volume} {3}},\ \bibinfo {pages} {023200} (\bibinfo {year} {2021})}\BibitemShut {NoStop}%
\bibitem [{\citenamefont {Iaconis}\ \emph {et~al.}(2020)\citenamefont {Iaconis}, \citenamefont {Lucas},\ and\ \citenamefont {Chen}}]{Iaconis21}%
  \BibitemOpen
  \bibfield  {author} {\bibinfo {author} {\bibfnamefont {J.}~\bibnamefont {Iaconis}}, \bibinfo {author} {\bibfnamefont {A.}~\bibnamefont {Lucas}},\ and\ \bibinfo {author} {\bibfnamefont {X.}~\bibnamefont {Chen}},\ }\bibfield  {title} {\bibinfo {title} {Measurement-induced phase transitions in quantum automaton circuits},\ }\href {https://doi.org/10.1103/PhysRevB.102.224311} {\bibfield  {journal} {\bibinfo  {journal} {Phys. Rev. B}\ }\textbf {\bibinfo {volume} {102}},\ \bibinfo {pages} {224311} (\bibinfo {year} {2020})}\BibitemShut {NoStop}%
\bibitem [{\citenamefont {Lavasani}\ \emph {et~al.}(2021)\citenamefont {Lavasani}, \citenamefont {Alavirad},\ and\ \citenamefont {Barkeshli}}]{lavasani21}%
  \BibitemOpen
  \bibfield  {author} {\bibinfo {author} {\bibfnamefont {A.}~\bibnamefont {Lavasani}}, \bibinfo {author} {\bibfnamefont {Y.}~\bibnamefont {Alavirad}},\ and\ \bibinfo {author} {\bibfnamefont {M.}~\bibnamefont {Barkeshli}},\ }\bibfield  {title} {\bibinfo {title} {Measurement-induced topological entanglement transitions in symmetric random quantum circuits},\ }\href {https://doi.org/10.1038/s41567-020-01112-z} {\bibfield  {journal} {\bibinfo  {journal} {Nat. Phys}\ }\textbf {\bibinfo {volume} {17}},\ \bibinfo {pages} {342} (\bibinfo {year} {2021})}\BibitemShut {NoStop}%
\bibitem [{\citenamefont {Ippoliti}\ \emph {et~al.}(2021)\citenamefont {Ippoliti}, \citenamefont {Gullans}, \citenamefont {Gopalakrishnan}, \citenamefont {Huse},\ and\ \citenamefont {Khemani}}]{ippoliti21b}%
  \BibitemOpen
  \bibfield  {author} {\bibinfo {author} {\bibfnamefont {M.}~\bibnamefont {Ippoliti}}, \bibinfo {author} {\bibfnamefont {M.~J.}\ \bibnamefont {Gullans}}, \bibinfo {author} {\bibfnamefont {S.}~\bibnamefont {Gopalakrishnan}}, \bibinfo {author} {\bibfnamefont {D.~A.}\ \bibnamefont {Huse}},\ and\ \bibinfo {author} {\bibfnamefont {V.}~\bibnamefont {Khemani}},\ }\bibfield  {title} {\bibinfo {title} {Entanglement phase transitions in measurement-only dynamics},\ }\href {https://doi.org/10.1103/PhysRevX.11.011030} {\bibfield  {journal} {\bibinfo  {journal} {Phys. Rev. X}\ }\textbf {\bibinfo {volume} {11}},\ \bibinfo {pages} {011030} (\bibinfo {year} {2021})}\BibitemShut {NoStop}%
\bibitem [{\citenamefont {Ippoliti}\ and\ \citenamefont {Khemani}(2021)}]{ippoliti21c}%
  \BibitemOpen
  \bibfield  {author} {\bibinfo {author} {\bibfnamefont {M.}~\bibnamefont {Ippoliti}}\ and\ \bibinfo {author} {\bibfnamefont {V.}~\bibnamefont {Khemani}},\ }\bibfield  {title} {\bibinfo {title} {Postselection-free entanglement dynamics via spacetime duality},\ }\href {https://doi.org/10.1103/PhysRevLett.126.060501} {\bibfield  {journal} {\bibinfo  {journal} {Phys. Rev. Lett.}\ }\textbf {\bibinfo {volume} {126}},\ \bibinfo {pages} {060501} (\bibinfo {year} {2021})}\BibitemShut {NoStop}%
\bibitem [{\citenamefont {Zabalo}\ \emph {et~al.}(2022)\citenamefont {Zabalo}, \citenamefont {Gullans}, \citenamefont {Wilson}, \citenamefont {Vasseur}, \citenamefont {Ludwig}, \citenamefont {Gopalakrishnan}, \citenamefont {Huse},\ and\ \citenamefont {Pixley}}]{zabalo22}%
  \BibitemOpen
  \bibfield  {author} {\bibinfo {author} {\bibfnamefont {A.}~\bibnamefont {Zabalo}}, \bibinfo {author} {\bibfnamefont {M.~J.}\ \bibnamefont {Gullans}}, \bibinfo {author} {\bibfnamefont {J.~H.}\ \bibnamefont {Wilson}}, \bibinfo {author} {\bibfnamefont {R.}~\bibnamefont {Vasseur}}, \bibinfo {author} {\bibfnamefont {A.~W.~W.}\ \bibnamefont {Ludwig}}, \bibinfo {author} {\bibfnamefont {S.}~\bibnamefont {Gopalakrishnan}}, \bibinfo {author} {\bibfnamefont {D.~A.}\ \bibnamefont {Huse}},\ and\ \bibinfo {author} {\bibfnamefont {J.~H.}\ \bibnamefont {Pixley}},\ }\bibfield  {title} {\bibinfo {title} {Operator scaling dimensions and multifractality at measurement-induced transitions},\ }\href {https://doi.org/10.1103/PhysRevLett.128.050602} {\bibfield  {journal} {\bibinfo  {journal} {Phys. Rev. Lett.}\ }\textbf {\bibinfo {volume} {128}},\ \bibinfo {pages} {050602} (\bibinfo {year} {2022})}\BibitemShut {NoStop}%
\bibitem [{\citenamefont {Jian}\ \emph {et~al.}(2021)\citenamefont {Jian}, \citenamefont {Liu}, \citenamefont {Chen}, \citenamefont {Swingle},\ and\ \citenamefont {Zhang}}]{jian21b}%
  \BibitemOpen
  \bibfield  {author} {\bibinfo {author} {\bibfnamefont {S.-K.}\ \bibnamefont {Jian}}, \bibinfo {author} {\bibfnamefont {C.}~\bibnamefont {Liu}}, \bibinfo {author} {\bibfnamefont {X.}~\bibnamefont {Chen}}, \bibinfo {author} {\bibfnamefont {B.}~\bibnamefont {Swingle}},\ and\ \bibinfo {author} {\bibfnamefont {P.}~\bibnamefont {Zhang}},\ }\bibfield  {title} {\bibinfo {title} {Measurement-induced phase transition in the monitored {Sachdev-Ye-Kitaev} model},\ }\href {https://doi.org/10.1103/PhysRevLett.127.140601} {\bibfield  {journal} {\bibinfo  {journal} {Phys. Rev. Lett.}\ }\textbf {\bibinfo {volume} {127}},\ \bibinfo {pages} {140601} (\bibinfo {year} {2021})}\BibitemShut {NoStop}%
\bibitem [{\citenamefont {Sierant}\ and\ \citenamefont {Turkeshi}(2022)}]{sierant21}%
  \BibitemOpen
  \bibfield  {author} {\bibinfo {author} {\bibfnamefont {P.}~\bibnamefont {Sierant}}\ and\ \bibinfo {author} {\bibfnamefont {X.}~\bibnamefont {Turkeshi}},\ }\bibfield  {title} {\bibinfo {title} {Universal behavior beyond multifractality of wave-functions at measurement--induced phase transitions},\ }\href {https://doi.org/10.1103/PhysRevLett.128.130605} {\bibfield  {journal} {\bibinfo  {journal} {Phys. Rev. Lett.}\ }\textbf {\bibinfo {volume} {128}},\ \bibinfo {pages} {130605} (\bibinfo {year} {2022})}\BibitemShut {NoStop}%
\bibitem [{\citenamefont {Lu}\ and\ \citenamefont {Grover}(2021)}]{lu21}%
  \BibitemOpen
  \bibfield  {author} {\bibinfo {author} {\bibfnamefont {T.-C.}\ \bibnamefont {Lu}}\ and\ \bibinfo {author} {\bibfnamefont {T.}~\bibnamefont {Grover}},\ }\bibfield  {title} {\bibinfo {title} {Spacetime duality between localization transitions and measurement-induced transitions},\ }\href {https://doi.org/10.1103/PRXQuantum.2.040319} {\bibfield  {journal} {\bibinfo  {journal} {PRX Quantum}\ }\textbf {\bibinfo {volume} {2}},\ \bibinfo {pages} {040319} (\bibinfo {year} {2021})}\BibitemShut {NoStop}%
\bibitem [{\citenamefont {C\^ot\'e}\ and\ \citenamefont {Kourtis}(2022)}]{cote21}%
  \BibitemOpen
  \bibfield  {author} {\bibinfo {author} {\bibfnamefont {J.}~\bibnamefont {C\^ot\'e}}\ and\ \bibinfo {author} {\bibfnamefont {S.}~\bibnamefont {Kourtis}},\ }\bibfield  {title} {\bibinfo {title} {Entanglement phase transition with spin glass criticality},\ }\href {https://doi.org/10.1103/PhysRevLett.128.240601} {\bibfield  {journal} {\bibinfo  {journal} {Phys. Rev. Lett.}\ }\textbf {\bibinfo {volume} {128}},\ \bibinfo {pages} {240601} (\bibinfo {year} {2022})}\BibitemShut {NoStop}%
\bibitem [{\citenamefont {M\"uller}\ \emph {et~al.}(2022)\citenamefont {M\"uller}, \citenamefont {Diehl},\ and\ \citenamefont {Buchhold}}]{muller22}%
  \BibitemOpen
  \bibfield  {author} {\bibinfo {author} {\bibfnamefont {T.}~\bibnamefont {M\"uller}}, \bibinfo {author} {\bibfnamefont {S.}~\bibnamefont {Diehl}},\ and\ \bibinfo {author} {\bibfnamefont {M.}~\bibnamefont {Buchhold}},\ }\bibfield  {title} {\bibinfo {title} {Measurement-induced dark state phase transitions in long-ranged fermion systems},\ }\href {https://doi.org/10.1103/PhysRevLett.128.010605} {\bibfield  {journal} {\bibinfo  {journal} {Phys. Rev. Lett.}\ }\textbf {\bibinfo {volume} {128}},\ \bibinfo {pages} {010605} (\bibinfo {year} {2022})}\BibitemShut {NoStop}%
\bibitem [{\citenamefont {Sharma}\ \emph {et~al.}(2022)\citenamefont {Sharma}, \citenamefont {Turkeshi}, \citenamefont {Fazio},\ and\ \citenamefont {Dalmonte}}]{sharma22}%
  \BibitemOpen
  \bibfield  {author} {\bibinfo {author} {\bibfnamefont {S.}~\bibnamefont {Sharma}}, \bibinfo {author} {\bibfnamefont {X.}~\bibnamefont {Turkeshi}}, \bibinfo {author} {\bibfnamefont {R.}~\bibnamefont {Fazio}},\ and\ \bibinfo {author} {\bibfnamefont {M.}~\bibnamefont {Dalmonte}},\ }\bibfield  {title} {\bibinfo {title} {Measurement-induced criticality in extended and long-range unitary circuits},\ }\href {https://doi.org/10.21468/SciPostPhysCore.5.2.023} {\bibfield  {journal} {\bibinfo  {journal} {SciPost Phys. Core}\ }\textbf {\bibinfo {volume} {5}},\ \bibinfo {pages} {023} (\bibinfo {year} {2022})}\BibitemShut {NoStop}%
\bibitem [{\citenamefont {Liu}\ \emph {et~al.}(2023)\citenamefont {Liu}, \citenamefont {Li}, \citenamefont {Zhang}, \citenamefont {Jian},\ and\ \citenamefont {Yao}}]{liu2023}%
  \BibitemOpen
  \bibfield  {author} {\bibinfo {author} {\bibfnamefont {S.}~\bibnamefont {Liu}}, \bibinfo {author} {\bibfnamefont {M.-R.}\ \bibnamefont {Li}}, \bibinfo {author} {\bibfnamefont {S.-X.}\ \bibnamefont {Zhang}}, \bibinfo {author} {\bibfnamefont {S.-K.}\ \bibnamefont {Jian}},\ and\ \bibinfo {author} {\bibfnamefont {H.}~\bibnamefont {Yao}},\ }\bibfield  {title} {\bibinfo {title} {Universal kardar-parisi-zhang scaling in noisy hybrid quantum circuits},\ }\href {https://doi.org/10.1103/PhysRevB.107.L201113} {\bibfield  {journal} {\bibinfo  {journal} {Phys. Rev. B}\ }\textbf {\bibinfo {volume} {107}},\ \bibinfo {pages} {L201113} (\bibinfo {year} {2023})}\BibitemShut {NoStop}%
\bibitem [{\citenamefont {Buchhold}\ \emph {et~al.}(2021)\citenamefont {Buchhold}, \citenamefont {Minoguchi}, \citenamefont {Altland},\ and\ \citenamefont {Diehl}}]{buchhold21}%
  \BibitemOpen
  \bibfield  {author} {\bibinfo {author} {\bibfnamefont {M.}~\bibnamefont {Buchhold}}, \bibinfo {author} {\bibfnamefont {Y.}~\bibnamefont {Minoguchi}}, \bibinfo {author} {\bibfnamefont {A.}~\bibnamefont {Altland}},\ and\ \bibinfo {author} {\bibfnamefont {S.}~\bibnamefont {Diehl}},\ }\bibfield  {title} {\bibinfo {title} {Effective theory for the measurement-induced phase transition of {Dirac} fermions},\ }\href {https://doi.org/10.1103/PhysRevX.11.041004} {\bibfield  {journal} {\bibinfo  {journal} {Phys. Rev. X}\ }\textbf {\bibinfo {volume} {11}},\ \bibinfo {pages} {041004} (\bibinfo {year} {2021})}\BibitemShut {NoStop}%
\bibitem [{\citenamefont {Doggen}\ \emph {et~al.}(2022)\citenamefont {Doggen}, \citenamefont {Gefen}, \citenamefont {Gornyi}, \citenamefont {Mirlin},\ and\ \citenamefont {Polyakov}}]{doggen22}%
  \BibitemOpen
  \bibfield  {author} {\bibinfo {author} {\bibfnamefont {E.~V.~H.}\ \bibnamefont {Doggen}}, \bibinfo {author} {\bibfnamefont {Y.}~\bibnamefont {Gefen}}, \bibinfo {author} {\bibfnamefont {I.~V.}\ \bibnamefont {Gornyi}}, \bibinfo {author} {\bibfnamefont {A.~D.}\ \bibnamefont {Mirlin}},\ and\ \bibinfo {author} {\bibfnamefont {D.~G.}\ \bibnamefont {Polyakov}},\ }\bibfield  {title} {\bibinfo {title} {Generalized quantum measurements with matrix product states: Entanglement phase transition and clusterization},\ }\href {https://doi.org/10.1103/PhysRevResearch.4.023146} {\bibfield  {journal} {\bibinfo  {journal} {Phys. Rev. Research}\ }\textbf {\bibinfo {volume} {4}},\ \bibinfo {pages} {023146} (\bibinfo {year} {2022})}\BibitemShut {NoStop}%
\bibitem [{\citenamefont {Li}\ \emph {et~al.}(2021)\citenamefont {Li}, \citenamefont {Chen}, \citenamefont {Ludwig},\ and\ \citenamefont {Fisher}}]{li21}%
  \BibitemOpen
  \bibfield  {author} {\bibinfo {author} {\bibfnamefont {Y.}~\bibnamefont {Li}}, \bibinfo {author} {\bibfnamefont {X.}~\bibnamefont {Chen}}, \bibinfo {author} {\bibfnamefont {A.~W.~W.}\ \bibnamefont {Ludwig}},\ and\ \bibinfo {author} {\bibfnamefont {M.~P.~A.}\ \bibnamefont {Fisher}},\ }\bibfield  {title} {\bibinfo {title} {Conformal invariance and quantum nonlocality in critical hybrid circuits},\ }\href {https://doi.org/10.1103/PhysRevB.104.104305} {\bibfield  {journal} {\bibinfo  {journal} {Phys. Rev. B}\ }\textbf {\bibinfo {volume} {104}},\ \bibinfo {pages} {104305} (\bibinfo {year} {2021})}\BibitemShut {NoStop}%
\bibitem [{\citenamefont {Zabalo}\ \emph {et~al.}(2020)\citenamefont {Zabalo}, \citenamefont {Gullans}, \citenamefont {Wilson}, \citenamefont {Gopalakrishnan}, \citenamefont {Huse},\ and\ \citenamefont {Pixley}}]{zabalo20}%
  \BibitemOpen
  \bibfield  {author} {\bibinfo {author} {\bibfnamefont {A.}~\bibnamefont {Zabalo}}, \bibinfo {author} {\bibfnamefont {M.~J.}\ \bibnamefont {Gullans}}, \bibinfo {author} {\bibfnamefont {J.~H.}\ \bibnamefont {Wilson}}, \bibinfo {author} {\bibfnamefont {S.}~\bibnamefont {Gopalakrishnan}}, \bibinfo {author} {\bibfnamefont {D.~A.}\ \bibnamefont {Huse}},\ and\ \bibinfo {author} {\bibfnamefont {J.~H.}\ \bibnamefont {Pixley}},\ }\bibfield  {title} {\bibinfo {title} {Critical properties of the measurement-induced transition in random quantum circuits},\ }\href {https://doi.org/10.1103/PhysRevB.101.060301} {\bibfield  {journal} {\bibinfo  {journal} {Phys. Rev. B}\ }\textbf {\bibinfo {volume} {101}},\ \bibinfo {pages} {060301(R)} (\bibinfo {year} {2020})}\BibitemShut {NoStop}%
\bibitem [{\citenamefont {Preskill}(2018)}]{preskill18}%
  \BibitemOpen
  \bibfield  {author} {\bibinfo {author} {\bibfnamefont {J.}~\bibnamefont {Preskill}},\ }\bibfield  {title} {\bibinfo {title} {Quantum {C}omputing in the {NISQ} era and beyond},\ }\href {https://doi.org/10.22331/q-2018-08-06-79} {\bibfield  {journal} {\bibinfo  {journal} {{Quantum}}\ }\textbf {\bibinfo {volume} {2}},\ \bibinfo {pages} {79} (\bibinfo {year} {2018})}\BibitemShut {NoStop}%
\bibitem [{\citenamefont {Gullans}\ \emph {et~al.}(2021)\citenamefont {Gullans}, \citenamefont {Krastanov}, \citenamefont {Huse}, \citenamefont {Jiang},\ and\ \citenamefont {Flammia}}]{gullans21}%
  \BibitemOpen
  \bibfield  {author} {\bibinfo {author} {\bibfnamefont {M.~J.}\ \bibnamefont {Gullans}}, \bibinfo {author} {\bibfnamefont {S.}~\bibnamefont {Krastanov}}, \bibinfo {author} {\bibfnamefont {D.~A.}\ \bibnamefont {Huse}}, \bibinfo {author} {\bibfnamefont {L.}~\bibnamefont {Jiang}},\ and\ \bibinfo {author} {\bibfnamefont {S.~T.}\ \bibnamefont {Flammia}},\ }\bibfield  {title} {\bibinfo {title} {Quantum coding with low-depth random circuits},\ }\href {https://doi.org/10.1103/PhysRevX.11.031066} {\bibfield  {journal} {\bibinfo  {journal} {Phys. Rev. X}\ }\textbf {\bibinfo {volume} {11}},\ \bibinfo {pages} {031066} (\bibinfo {year} {2021})}\BibitemShut {NoStop}%
\bibitem [{\citenamefont {Kelly}\ \emph {et~al.}(2023)\citenamefont {Kelly}, \citenamefont {Poschinger}, \citenamefont {Schmidt-Kaler}, \citenamefont {Fisher},\ and\ \citenamefont {Marino}}]{kelly22}%
  \BibitemOpen
  \bibfield  {author} {\bibinfo {author} {\bibfnamefont {S.~P.}\ \bibnamefont {Kelly}}, \bibinfo {author} {\bibfnamefont {U.}~\bibnamefont {Poschinger}}, \bibinfo {author} {\bibfnamefont {F.}~\bibnamefont {Schmidt-Kaler}}, \bibinfo {author} {\bibfnamefont {M.~P.~A.}\ \bibnamefont {Fisher}},\ and\ \bibinfo {author} {\bibfnamefont {J.}~\bibnamefont {Marino}},\ }\bibfield  {title} {\bibinfo {title} {{Coherence requirements for quantum communication from hybrid circuit dynamics}},\ }\href {https://doi.org/10.21468/SciPostPhys.15.6.250} {\bibfield  {journal} {\bibinfo  {journal} {SciPost Phys.}\ }\textbf {\bibinfo {volume} {15}},\ \bibinfo {pages} {250} (\bibinfo {year} {2023})}\BibitemShut {NoStop}%
\bibitem [{\citenamefont {Dehghani}\ \emph {et~al.}(2023)\citenamefont {Dehghani}, \citenamefont {Lavasani}, \citenamefont {Hafezi},\ and\ \citenamefont {Gullans}}]{dehghani2023}%
  \BibitemOpen
  \bibfield  {author} {\bibinfo {author} {\bibfnamefont {H.}~\bibnamefont {Dehghani}}, \bibinfo {author} {\bibfnamefont {A.}~\bibnamefont {Lavasani}}, \bibinfo {author} {\bibfnamefont {M.}~\bibnamefont {Hafezi}},\ and\ \bibinfo {author} {\bibfnamefont {M.~J.}\ \bibnamefont {Gullans}},\ }\bibfield  {title} {\bibinfo {title} {Neural-network decoders for measurement induced phase transitions},\ }\href {https://doi.org/10.1038/s41467-023-37902-1} {\bibfield  {journal} {\bibinfo  {journal} {Nat. Commun.}\ }\textbf {\bibinfo {volume} {14}},\ \bibinfo {pages} {2918} (\bibinfo {year} {2023})}\BibitemShut {NoStop}%
\bibitem [{\citenamefont {Weinstein}\ \emph {et~al.}(2023)\citenamefont {Weinstein}, \citenamefont {Kelly}, \citenamefont {Marino},\ and\ \citenamefont {Altman}}]{weinstein22}%
  \BibitemOpen
  \bibfield  {author} {\bibinfo {author} {\bibfnamefont {Z.}~\bibnamefont {Weinstein}}, \bibinfo {author} {\bibfnamefont {S.~P.}\ \bibnamefont {Kelly}}, \bibinfo {author} {\bibfnamefont {J.}~\bibnamefont {Marino}},\ and\ \bibinfo {author} {\bibfnamefont {E.}~\bibnamefont {Altman}},\ }\bibfield  {title} {\bibinfo {title} {Scrambling transition in a radiative random unitary circuit},\ }\href {https://doi.org/10.1103/PhysRevLett.131.220404} {\bibfield  {journal} {\bibinfo  {journal} {Phys. Rev. Lett.}\ }\textbf {\bibinfo {volume} {131}},\ \bibinfo {pages} {220404} (\bibinfo {year} {2023})}\BibitemShut {NoStop}%
\bibitem [{\citenamefont {Noel}\ \emph {et~al.}(2022)\citenamefont {Noel}, \citenamefont {Niroula}, \citenamefont {Zhu}, \citenamefont {Risinger}, \citenamefont {Egan}, \citenamefont {Biswas}, \citenamefont {Cetina}, \citenamefont {Gorshkov}, \citenamefont {Gullans}, \citenamefont {Huse},\ and\ \citenamefont {Monroe}}]{Noel2022}%
  \BibitemOpen
  \bibfield  {author} {\bibinfo {author} {\bibfnamefont {C.}~\bibnamefont {Noel}}, \bibinfo {author} {\bibfnamefont {P.}~\bibnamefont {Niroula}}, \bibinfo {author} {\bibfnamefont {D.}~\bibnamefont {Zhu}}, \bibinfo {author} {\bibfnamefont {A.}~\bibnamefont {Risinger}}, \bibinfo {author} {\bibfnamefont {L.}~\bibnamefont {Egan}}, \bibinfo {author} {\bibfnamefont {D.}~\bibnamefont {Biswas}}, \bibinfo {author} {\bibfnamefont {M.}~\bibnamefont {Cetina}}, \bibinfo {author} {\bibfnamefont {A.~V.}\ \bibnamefont {Gorshkov}}, \bibinfo {author} {\bibfnamefont {M.~J.}\ \bibnamefont {Gullans}}, \bibinfo {author} {\bibfnamefont {D.~A.}\ \bibnamefont {Huse}},\ and\ \bibinfo {author} {\bibfnamefont {C.}~\bibnamefont {Monroe}},\ }\bibfield  {title} {\bibinfo {title} {Measurement-induced quantum phases realized in a trapped-ion quantum computer},\ }\href {https://doi.org/10.1038/s41567-022-01619-7} {\bibfield  {journal} {\bibinfo  {journal} {Nat. Phys.}\ }\textbf {\bibinfo {volume} {18}},\ \bibinfo {pages} {760} (\bibinfo
  {year} {2022})}\BibitemShut {NoStop}%
\bibitem [{\citenamefont {Yoshida}(2021)}]{yoshida21}%
  \BibitemOpen
  \bibfield  {author} {\bibinfo {author} {\bibfnamefont {B.}~\bibnamefont {Yoshida}},\ }\bibfield  {title} {\bibinfo {title} {Decoding the entanglement structure of monitored quantum circuits},\ }\href {https://doi.org/10.48550/ARXIV.2109.08691} {\bibfield  {journal} {\bibinfo  {journal} {arXiv:2109.09691}\ } (\bibinfo {year} {2021})}\BibitemShut {NoStop}%
\bibitem [{\citenamefont {Cazalilla}\ \emph {et~al.}(2011)\citenamefont {Cazalilla}, \citenamefont {Citro}, \citenamefont {Giamarchi}, \citenamefont {Orignac},\ and\ \citenamefont {Rigol}}]{cazalilla11}%
  \BibitemOpen
  \bibfield  {author} {\bibinfo {author} {\bibfnamefont {M.~A.}\ \bibnamefont {Cazalilla}}, \bibinfo {author} {\bibfnamefont {R.}~\bibnamefont {Citro}}, \bibinfo {author} {\bibfnamefont {T.}~\bibnamefont {Giamarchi}}, \bibinfo {author} {\bibfnamefont {E.}~\bibnamefont {Orignac}},\ and\ \bibinfo {author} {\bibfnamefont {M.}~\bibnamefont {Rigol}},\ }\bibfield  {title} {\bibinfo {title} {One dimensional bosons: From condensed matter systems to ultracold gases},\ }\href {https://doi.org/10.1103/RevModPhys.83.1405} {\bibfield  {journal} {\bibinfo  {journal} {Rev. Mod. Phys.}\ }\textbf {\bibinfo {volume} {83}},\ \bibinfo {pages} {1405} (\bibinfo {year} {2011})}\BibitemShut {NoStop}%
\bibitem [{\citenamefont {Barends}\ \emph {et~al.}(2013)\citenamefont {Barends}, \citenamefont {Kelly}, \citenamefont {Megrant}, \citenamefont {Sank}, \citenamefont {Jeffrey}, \citenamefont {Chen}, \citenamefont {Yin}, \citenamefont {Chiaro}, \citenamefont {Mutus}, \citenamefont {Neill}, \citenamefont {O'Malley}, \citenamefont {Roushan}, \citenamefont {Wenner}, \citenamefont {White}, \citenamefont {Cleland},\ and\ \citenamefont {Martinis}}]{barends13}%
  \BibitemOpen
  \bibfield  {author} {\bibinfo {author} {\bibfnamefont {R.}~\bibnamefont {Barends}}, \bibinfo {author} {\bibfnamefont {J.}~\bibnamefont {Kelly}}, \bibinfo {author} {\bibfnamefont {A.}~\bibnamefont {Megrant}}, \bibinfo {author} {\bibfnamefont {D.}~\bibnamefont {Sank}}, \bibinfo {author} {\bibfnamefont {E.}~\bibnamefont {Jeffrey}}, \bibinfo {author} {\bibfnamefont {Y.}~\bibnamefont {Chen}}, \bibinfo {author} {\bibfnamefont {Y.}~\bibnamefont {Yin}}, \bibinfo {author} {\bibfnamefont {B.}~\bibnamefont {Chiaro}}, \bibinfo {author} {\bibfnamefont {J.}~\bibnamefont {Mutus}}, \bibinfo {author} {\bibfnamefont {C.}~\bibnamefont {Neill}}, \bibinfo {author} {\bibfnamefont {P.}~\bibnamefont {O'Malley}}, \bibinfo {author} {\bibfnamefont {P.}~\bibnamefont {Roushan}}, \bibinfo {author} {\bibfnamefont {J.}~\bibnamefont {Wenner}}, \bibinfo {author} {\bibfnamefont {T.~C.}\ \bibnamefont {White}}, \bibinfo {author} {\bibfnamefont {A.~N.}\ \bibnamefont {Cleland}},\ and\ \bibinfo {author} {\bibfnamefont {J.~M.}\ \bibnamefont
  {Martinis}},\ }\bibfield  {title} {\bibinfo {title} {Coherent josephson qubit suitable for scalable quantum integrated circuits},\ }\href {https://doi.org/10.1103/PhysRevLett.111.080502} {\bibfield  {journal} {\bibinfo  {journal} {Phys. Rev. Lett.}\ }\textbf {\bibinfo {volume} {111}},\ \bibinfo {pages} {080502} (\bibinfo {year} {2013})}\BibitemShut {NoStop}%
\bibitem [{\citenamefont {Hacohen-Gourgy}\ \emph {et~al.}(2015)\citenamefont {Hacohen-Gourgy}, \citenamefont {Ramasesh}, \citenamefont {De~Grandi}, \citenamefont {Siddiqi},\ and\ \citenamefont {Girvin}}]{hacohen-gourgy15}%
  \BibitemOpen
  \bibfield  {author} {\bibinfo {author} {\bibfnamefont {S.}~\bibnamefont {Hacohen-Gourgy}}, \bibinfo {author} {\bibfnamefont {V.~V.}\ \bibnamefont {Ramasesh}}, \bibinfo {author} {\bibfnamefont {C.}~\bibnamefont {De~Grandi}}, \bibinfo {author} {\bibfnamefont {I.}~\bibnamefont {Siddiqi}},\ and\ \bibinfo {author} {\bibfnamefont {S.~M.}\ \bibnamefont {Girvin}},\ }\bibfield  {title} {\bibinfo {title} {Cooling and autonomous feedback in a {Bose-Hubbard} chain with attractive interactions},\ }\href {https://doi.org/10.1103/PhysRevLett.115.240501} {\bibfield  {journal} {\bibinfo  {journal} {Phys. Rev. Lett.}\ }\textbf {\bibinfo {volume} {115}},\ \bibinfo {pages} {240501} (\bibinfo {year} {2015})}\BibitemShut {NoStop}%
\bibitem [{\citenamefont {Roushan}\ \emph {et~al.}(2017)\citenamefont {Roushan} \emph {et~al.}}]{roushan17}%
  \BibitemOpen
  \bibfield  {author} {\bibinfo {author} {\bibfnamefont {P.}~\bibnamefont {Roushan}} \emph {et~al.},\ }\bibfield  {title} {\bibinfo {title} {Spectroscopic signatures of localization with interacting photons in superconducting qubits},\ }\href {https://doi.org/10.1126/science.aao1401} {\bibfield  {journal} {\bibinfo  {journal} {Science}\ }\textbf {\bibinfo {volume} {358}},\ \bibinfo {pages} {1175} (\bibinfo {year} {2017})}\BibitemShut {NoStop}%
\bibitem [{\citenamefont {Ma}\ \emph {et~al.}(2019)\citenamefont {Ma}, \citenamefont {Saxberg}, \citenamefont {Owens}, \citenamefont {Leung}, \citenamefont {Lu}, \citenamefont {Simon},\ and\ \citenamefont {Schuster}}]{ma19}%
  \BibitemOpen
  \bibfield  {author} {\bibinfo {author} {\bibfnamefont {R.}~\bibnamefont {Ma}}, \bibinfo {author} {\bibfnamefont {B.}~\bibnamefont {Saxberg}}, \bibinfo {author} {\bibfnamefont {C.}~\bibnamefont {Owens}}, \bibinfo {author} {\bibfnamefont {N.}~\bibnamefont {Leung}}, \bibinfo {author} {\bibfnamefont {Y.}~\bibnamefont {Lu}}, \bibinfo {author} {\bibfnamefont {J.}~\bibnamefont {Simon}},\ and\ \bibinfo {author} {\bibfnamefont {D.~I.}\ \bibnamefont {Schuster}},\ }\bibfield  {title} {\bibinfo {title} {A dissipatively stabilized {Mott} insulator of photons},\ }\href {https://doi.org/10.1038/s41586-019-0897-9} {\bibfield  {journal} {\bibinfo  {journal} {Nature}\ }\textbf {\bibinfo {volume} {566}},\ \bibinfo {pages} {51} (\bibinfo {year} {2019})}\BibitemShut {NoStop}%
\bibitem [{\citenamefont {Yan}\ \emph {et~al.}(2019)\citenamefont {Yan} \emph {et~al.}}]{yan19}%
  \BibitemOpen
  \bibfield  {author} {\bibinfo {author} {\bibfnamefont {Z.}~\bibnamefont {Yan}} \emph {et~al.},\ }\bibfield  {title} {\bibinfo {title} {Strongly correlated quantum walks with a 12-qubit superconducting processor},\ }\href {https://doi.org/10.1126/science.aaw1611} {\bibfield  {journal} {\bibinfo  {journal} {Science}\ }\textbf {\bibinfo {volume} {364}},\ \bibinfo {pages} {753} (\bibinfo {year} {2019})}\BibitemShut {NoStop}%
\bibitem [{\citenamefont {Kjaergaard}\ \emph {et~al.}(2020)\citenamefont {Kjaergaard}, \citenamefont {Schwartz}, \citenamefont {Braumüller}, \citenamefont {Krantz}, \citenamefont {Wang}, \citenamefont {Gustavsson},\ and\ \citenamefont {Oliver}}]{kjaergaard20}%
  \BibitemOpen
  \bibfield  {author} {\bibinfo {author} {\bibfnamefont {M.}~\bibnamefont {Kjaergaard}}, \bibinfo {author} {\bibfnamefont {M.~E.}\ \bibnamefont {Schwartz}}, \bibinfo {author} {\bibfnamefont {J.}~\bibnamefont {Braumüller}}, \bibinfo {author} {\bibfnamefont {P.}~\bibnamefont {Krantz}}, \bibinfo {author} {\bibfnamefont {J.~I.-J.}\ \bibnamefont {Wang}}, \bibinfo {author} {\bibfnamefont {S.}~\bibnamefont {Gustavsson}},\ and\ \bibinfo {author} {\bibfnamefont {W.~D.}\ \bibnamefont {Oliver}},\ }\bibfield  {title} {\bibinfo {title} {Superconducting qubits: Current state of play},\ }\href {https://doi.org/10.1146/annurev-conmatphys-031119-050605} {\bibfield  {journal} {\bibinfo  {journal} {Annu. Rev. Condens. Matter Phys.}\ }\textbf {\bibinfo {volume} {11}},\ \bibinfo {pages} {369} (\bibinfo {year} {2020})}\BibitemShut {NoStop}%
\bibitem [{\citenamefont {Mart{\'{\i}}n-V{\'{a}}zquez}\ and\ \citenamefont {Sab{\'{\i}}n}(2020)}]{martinvazquez20}%
  \BibitemOpen
  \bibfield  {author} {\bibinfo {author} {\bibfnamefont {G.}~\bibnamefont {Mart{\'{\i}}n-V{\'{a}}zquez}}\ and\ \bibinfo {author} {\bibfnamefont {C.}~\bibnamefont {Sab{\'{\i}}n}},\ }\bibfield  {title} {\bibinfo {title} {Closed timelike curves and chronology protection in quantum and classical simulators},\ }\href {https://doi.org/10.1088/1361-6382/ab5f3f} {\bibfield  {journal} {\bibinfo  {journal} {Class. Quantum Gravity}\ }\textbf {\bibinfo {volume} {37}},\ \bibinfo {pages} {045013} (\bibinfo {year} {2020})}\BibitemShut {NoStop}%
\bibitem [{\citenamefont {Koh}\ \emph {et~al.}(2023)\citenamefont {Koh}, \citenamefont {Sun}, \citenamefont {Motta},\ and\ \citenamefont {Minnich}}]{Koh2023}%
  \BibitemOpen
  \bibfield  {author} {\bibinfo {author} {\bibfnamefont {J.~M.}\ \bibnamefont {Koh}}, \bibinfo {author} {\bibfnamefont {S.-N.}\ \bibnamefont {Sun}}, \bibinfo {author} {\bibfnamefont {M.}~\bibnamefont {Motta}},\ and\ \bibinfo {author} {\bibfnamefont {A.~J.}\ \bibnamefont {Minnich}},\ }\bibfield  {title} {\bibinfo {title} {Measurement-induced entanglement phase transition on a superconducting quantum processor with mid-circuit readout},\ }\href {https://doi.org/10.1038/s41567-023-02076-6} {\bibfield  {journal} {\bibinfo  {journal} {Nat. Phys.}\ } (\bibinfo {year} {2023})}\BibitemShut {NoStop}%
\bibitem [{\citenamefont {Czischek}\ \emph {et~al.}(2021)\citenamefont {Czischek}, \citenamefont {Torlai}, \citenamefont {Ray}, \citenamefont {Islam},\ and\ \citenamefont {Melko}}]{Czischek2021}%
  \BibitemOpen
  \bibfield  {author} {\bibinfo {author} {\bibfnamefont {S.}~\bibnamefont {Czischek}}, \bibinfo {author} {\bibfnamefont {G.}~\bibnamefont {Torlai}}, \bibinfo {author} {\bibfnamefont {S.}~\bibnamefont {Ray}}, \bibinfo {author} {\bibfnamefont {R.}~\bibnamefont {Islam}},\ and\ \bibinfo {author} {\bibfnamefont {R.~G.}\ \bibnamefont {Melko}},\ }\bibfield  {title} {\bibinfo {title} {Simulating a measurement-induced phase transition for trapped-ion circuits},\ }\href {https://doi.org/10.1103/PhysRevA.104.062405} {\bibfield  {journal} {\bibinfo  {journal} {Phys. Rev. A}\ }\textbf {\bibinfo {volume} {104}},\ \bibinfo {pages} {062405} (\bibinfo {year} {2021})}\BibitemShut {NoStop}%
\bibitem [{\citenamefont {Gopalakrishnan}\ and\ \citenamefont {Gullans}(2021)}]{gopalakrishnan21}%
  \BibitemOpen
  \bibfield  {author} {\bibinfo {author} {\bibfnamefont {S.}~\bibnamefont {Gopalakrishnan}}\ and\ \bibinfo {author} {\bibfnamefont {M.~J.}\ \bibnamefont {Gullans}},\ }\bibfield  {title} {\bibinfo {title} {Entanglement and purification transitions in non-{{Hermitian}} quantum mechanics},\ }\href {https://doi.org/10.1103/PhysRevLett.126.170503} {\bibfield  {journal} {\bibinfo  {journal} {Phys. Rev. Lett.}\ }\textbf {\bibinfo {volume} {126}},\ \bibinfo {pages} {170503} (\bibinfo {year} {2021})}\BibitemShut {NoStop}%
\bibitem [{\citenamefont {Nahum}\ \emph {et~al.}(2021)\citenamefont {Nahum}, \citenamefont {Roy}, \citenamefont {Skinner},\ and\ \citenamefont {Ruhman}}]{nahum21}%
  \BibitemOpen
  \bibfield  {author} {\bibinfo {author} {\bibfnamefont {A.}~\bibnamefont {Nahum}}, \bibinfo {author} {\bibfnamefont {S.}~\bibnamefont {Roy}}, \bibinfo {author} {\bibfnamefont {B.}~\bibnamefont {Skinner}},\ and\ \bibinfo {author} {\bibfnamefont {J.}~\bibnamefont {Ruhman}},\ }\bibfield  {title} {\bibinfo {title} {Measurement and entanglement phase transitions in all-to-all quantum circuits, on quantum trees, and in {Landau-Ginsburg} theory},\ }\href {https://doi.org/10.1103/PRXQuantum.2.010352} {\bibfield  {journal} {\bibinfo  {journal} {PRX Quantum}\ }\textbf {\bibinfo {volume} {2}},\ \bibinfo {pages} {010352} (\bibinfo {year} {2021})}\BibitemShut {NoStop}%
\bibitem [{\citenamefont {Buchhold}\ \emph {et~al.}(2022)\citenamefont {Buchhold}, \citenamefont {Müller},\ and\ \citenamefont {Diehl}}]{buchhold22}%
  \BibitemOpen
  \bibfield  {author} {\bibinfo {author} {\bibfnamefont {M.}~\bibnamefont {Buchhold}}, \bibinfo {author} {\bibfnamefont {T.}~\bibnamefont {Müller}},\ and\ \bibinfo {author} {\bibfnamefont {S.}~\bibnamefont {Diehl}},\ }\bibfield  {title} {\bibinfo {title} {Revealing measurement-induced phase transitions by pre-selection},\ }\href {https://doi.org/10.48550/ARXIV.2208.10506} {\bibfield  {journal} {\bibinfo  {journal} {arXiv:2208.10506}\ } (\bibinfo {year} {2022})}\BibitemShut {NoStop}%
\bibitem [{\citenamefont {Iadecola}\ \emph {et~al.}(2023)\citenamefont {Iadecola}, \citenamefont {Ganeshan}, \citenamefont {Pixley},\ and\ \citenamefont {Wilson}}]{iadecola22}%
  \BibitemOpen
  \bibfield  {author} {\bibinfo {author} {\bibfnamefont {T.}~\bibnamefont {Iadecola}}, \bibinfo {author} {\bibfnamefont {S.}~\bibnamefont {Ganeshan}}, \bibinfo {author} {\bibfnamefont {J.~H.}\ \bibnamefont {Pixley}},\ and\ \bibinfo {author} {\bibfnamefont {J.~H.}\ \bibnamefont {Wilson}},\ }\bibfield  {title} {\bibinfo {title} {Measurement and feedback driven entanglement transition in the probabilistic control of chaos},\ }\href {https://doi.org/10.1103/PhysRevLett.131.060403} {\bibfield  {journal} {\bibinfo  {journal} {Phys. Rev. Lett.}\ }\textbf {\bibinfo {volume} {131}},\ \bibinfo {pages} {060403} (\bibinfo {year} {2023})}\BibitemShut {NoStop}%
\bibitem [{\citenamefont {O'Dea}\ \emph {et~al.}(2024)\citenamefont {O'Dea}, \citenamefont {Morningstar}, \citenamefont {Gopalakrishnan},\ and\ \citenamefont {Khemani}}]{odea22}%
  \BibitemOpen
  \bibfield  {author} {\bibinfo {author} {\bibfnamefont {N.}~\bibnamefont {O'Dea}}, \bibinfo {author} {\bibfnamefont {A.}~\bibnamefont {Morningstar}}, \bibinfo {author} {\bibfnamefont {S.}~\bibnamefont {Gopalakrishnan}},\ and\ \bibinfo {author} {\bibfnamefont {V.}~\bibnamefont {Khemani}},\ }\bibfield  {title} {\bibinfo {title} {Entanglement and absorbing-state transitions in interactive quantum dynamics},\ }\href {https://doi.org/10.1103/PhysRevB.109.L020304} {\bibfield  {journal} {\bibinfo  {journal} {Phys. Rev. B}\ }\textbf {\bibinfo {volume} {109}},\ \bibinfo {pages} {L020304} (\bibinfo {year} {2024})}\BibitemShut {NoStop}%
\bibitem [{\citenamefont {Ravindranath}\ \emph {et~al.}(2023)\citenamefont {Ravindranath}, \citenamefont {Han}, \citenamefont {Yang},\ and\ \citenamefont {Chen}}]{ravindranath22}%
  \BibitemOpen
  \bibfield  {author} {\bibinfo {author} {\bibfnamefont {V.}~\bibnamefont {Ravindranath}}, \bibinfo {author} {\bibfnamefont {Y.}~\bibnamefont {Han}}, \bibinfo {author} {\bibfnamefont {Z.-C.}\ \bibnamefont {Yang}},\ and\ \bibinfo {author} {\bibfnamefont {X.}~\bibnamefont {Chen}},\ }\bibfield  {title} {\bibinfo {title} {Entanglement steering in adaptive circuits with feedback},\ }\href {https://doi.org/10.1103/PhysRevB.108.L041103} {\bibfield  {journal} {\bibinfo  {journal} {Phys. Rev. B}\ }\textbf {\bibinfo {volume} {108}},\ \bibinfo {pages} {L041103} (\bibinfo {year} {2023})}\BibitemShut {NoStop}%
\bibitem [{\citenamefont {Sierant}\ and\ \citenamefont {Turkeshi}(2023)}]{piotr2023}%
  \BibitemOpen
  \bibfield  {author} {\bibinfo {author} {\bibfnamefont {P.}~\bibnamefont {Sierant}}\ and\ \bibinfo {author} {\bibfnamefont {X.}~\bibnamefont {Turkeshi}},\ }\bibfield  {title} {\bibinfo {title} {Controlling entanglement at absorbing state phase transitions in random circuits},\ }\href {https://doi.org/10.1103/PhysRevLett.130.120402} {\bibfield  {journal} {\bibinfo  {journal} {Phys. Rev. Lett.}\ }\textbf {\bibinfo {volume} {130}},\ \bibinfo {pages} {120402} (\bibinfo {year} {2023})}\BibitemShut {NoStop}%
\bibitem [{\citenamefont {Piroli}\ \emph {et~al.}(2023)\citenamefont {Piroli}, \citenamefont {Li}, \citenamefont {Vasseur},\ and\ \citenamefont {Nahum}}]{piroli2023}%
  \BibitemOpen
  \bibfield  {author} {\bibinfo {author} {\bibfnamefont {L.}~\bibnamefont {Piroli}}, \bibinfo {author} {\bibfnamefont {Y.}~\bibnamefont {Li}}, \bibinfo {author} {\bibfnamefont {R.}~\bibnamefont {Vasseur}},\ and\ \bibinfo {author} {\bibfnamefont {A.}~\bibnamefont {Nahum}},\ }\bibfield  {title} {\bibinfo {title} {Triviality of quantum trajectories close to a directed percolation transition},\ }\href {https://doi.org/10.1103/PhysRevB.107.224303} {\bibfield  {journal} {\bibinfo  {journal} {Phys. Rev. B}\ }\textbf {\bibinfo {volume} {107}},\ \bibinfo {pages} {224303} (\bibinfo {year} {2023})}\BibitemShut {NoStop}%
\bibitem [{\citenamefont {Koch}\ \emph {et~al.}(2007)\citenamefont {Koch} \emph {et~al.}}]{koch2007}%
  \BibitemOpen
  \bibfield  {author} {\bibinfo {author} {\bibfnamefont {J.}~\bibnamefont {Koch}} \emph {et~al.},\ }\bibfield  {title} {\bibinfo {title} {Charge-insensitive qubit design derived from the {Cooper} pair box},\ }\href {https://doi.org/10.1103/PhysRevA.76.042319} {\bibfield  {journal} {\bibinfo  {journal} {Phys. Rev. A}\ }\textbf {\bibinfo {volume} {76}},\ \bibinfo {pages} {042319} (\bibinfo {year} {2007})}\BibitemShut {NoStop}%
\bibitem [{\citenamefont {Paik}\ \emph {et~al.}(2011)\citenamefont {Paik} \emph {et~al.}}]{paik2011}%
  \BibitemOpen
  \bibfield  {author} {\bibinfo {author} {\bibfnamefont {H.}~\bibnamefont {Paik}} \emph {et~al.},\ }\bibfield  {title} {\bibinfo {title} {Observation of high coherence in {Josephson} junction qubits measured in a three-dimensional circuit {QED} architecture},\ }\href {https://doi.org/10.1103/PhysRevLett.107.240501} {\bibfield  {journal} {\bibinfo  {journal} {Phys. Rev. Lett.}\ }\textbf {\bibinfo {volume} {107}},\ \bibinfo {pages} {240501} (\bibinfo {year} {2011})}\BibitemShut {NoStop}%
\bibitem [{\citenamefont {Arute}\ \emph {et~al.}(2019)\citenamefont {Arute} \emph {et~al.}}]{arute2019}%
  \BibitemOpen
  \bibfield  {author} {\bibinfo {author} {\bibfnamefont {F.}~\bibnamefont {Arute}} \emph {et~al.},\ }\bibfield  {title} {\bibinfo {title} {Quantum supremacy using a programmable superconducting processor},\ }\href {https://doi.org/10.1038/s41586-019-1666-5} {\bibfield  {journal} {\bibinfo  {journal} {Nature}\ }\textbf {\bibinfo {volume} {574}},\ \bibinfo {pages} {505} (\bibinfo {year} {2019})}\BibitemShut {NoStop}%
\bibitem [{\citenamefont {Orell}\ \emph {et~al.}(2019)\citenamefont {Orell}, \citenamefont {Michailidis}, \citenamefont {Serbyn},\ and\ \citenamefont {Silveri}}]{Orell2019}%
  \BibitemOpen
  \bibfield  {author} {\bibinfo {author} {\bibfnamefont {T.}~\bibnamefont {Orell}}, \bibinfo {author} {\bibfnamefont {A.~A.}\ \bibnamefont {Michailidis}}, \bibinfo {author} {\bibfnamefont {M.}~\bibnamefont {Serbyn}},\ and\ \bibinfo {author} {\bibfnamefont {M.}~\bibnamefont {Silveri}},\ }\bibfield  {title} {\bibinfo {title} {Probing the many-body localization phase transition with superconducting circuits},\ }\href {https://doi.org/10.1103/PhysRevB.100.134504} {\bibfield  {journal} {\bibinfo  {journal} {Phys. Rev. B}\ }\textbf {\bibinfo {volume} {100}},\ \bibinfo {pages} {134504} (\bibinfo {year} {2019})}\BibitemShut {NoStop}%
\bibitem [{\citenamefont {Piekarska}\ and\ \citenamefont {Kope\ifmmode~\acute{c}\else \'{c}\fi{}}(2018)}]{Pierkarska2018}%
  \BibitemOpen
  \bibfield  {author} {\bibinfo {author} {\bibfnamefont {A.~M.}\ \bibnamefont {Piekarska}}\ and\ \bibinfo {author} {\bibfnamefont {T.~K.}\ \bibnamefont {Kope\ifmmode~\acute{c}\else \'{c}\fi{}}},\ }\bibfield  {title} {\bibinfo {title} {Quantum glass of interacting bosons with off-diagonal disorder},\ }\href {https://doi.org/10.1103/PhysRevLett.120.160401} {\bibfield  {journal} {\bibinfo  {journal} {Phys. Rev. Lett.}\ }\textbf {\bibinfo {volume} {120}},\ \bibinfo {pages} {160401} (\bibinfo {year} {2018})}\BibitemShut {NoStop}%
\bibitem [{\citenamefont {Jaschke}\ and\ \citenamefont {Carr}(2018)}]{Jaschke2018}%
  \BibitemOpen
  \bibfield  {author} {\bibinfo {author} {\bibfnamefont {D.}~\bibnamefont {Jaschke}}\ and\ \bibinfo {author} {\bibfnamefont {L.~D.}\ \bibnamefont {Carr}},\ }\bibfield  {title} {\bibinfo {title} {Open source matrix product states: {Exact} diagonalization and other entanglement-accurate methods revisited in quantum systems},\ }\href {https://doi.org/10.1088/1751-8121/aae4d1} {\bibfield  {journal} {\bibinfo  {journal} {J. Phys. A Math. Theor.}\ }\textbf {\bibinfo {volume} {51}},\ \bibinfo {pages} {465302} (\bibinfo {year} {2018})}\BibitemShut {NoStop}%
\bibitem [{\citenamefont {Sieberer}\ \emph {et~al.}(2019)\citenamefont {Sieberer}, \citenamefont {Olsacher}, \citenamefont {Elben}, \citenamefont {Heyl}, \citenamefont {Hauke}, \citenamefont {Haake},\ and\ \citenamefont {Zoller}}]{Sieberer19}%
  \BibitemOpen
  \bibfield  {author} {\bibinfo {author} {\bibfnamefont {L.~M.}\ \bibnamefont {Sieberer}}, \bibinfo {author} {\bibfnamefont {T.}~\bibnamefont {Olsacher}}, \bibinfo {author} {\bibfnamefont {A.}~\bibnamefont {Elben}}, \bibinfo {author} {\bibfnamefont {M.}~\bibnamefont {Heyl}}, \bibinfo {author} {\bibfnamefont {P.}~\bibnamefont {Hauke}}, \bibinfo {author} {\bibfnamefont {F.}~\bibnamefont {Haake}},\ and\ \bibinfo {author} {\bibfnamefont {P.}~\bibnamefont {Zoller}},\ }\bibfield  {title} {\bibinfo {title} {Digital quantum simulation, trotter errors, and quantum chaos of the kicked top},\ }\href {https://doi.org/10.1038/s41534-019-0192-5} {\bibfield  {journal} {\bibinfo  {journal} {npj Quantum Inf.}\ }\textbf {\bibinfo {volume} {5}},\ \bibinfo {pages} {78} (\bibinfo {year} {2019})}\BibitemShut {NoStop}%
\bibitem [{\citenamefont {Barbiero}\ \emph {et~al.}(2020)\citenamefont {Barbiero}, \citenamefont {Chomaz}, \citenamefont {Nascimbene},\ and\ \citenamefont {Goldman}}]{Barbiero2020}%
  \BibitemOpen
  \bibfield  {author} {\bibinfo {author} {\bibfnamefont {L.}~\bibnamefont {Barbiero}}, \bibinfo {author} {\bibfnamefont {L.}~\bibnamefont {Chomaz}}, \bibinfo {author} {\bibfnamefont {S.}~\bibnamefont {Nascimbene}},\ and\ \bibinfo {author} {\bibfnamefont {N.}~\bibnamefont {Goldman}},\ }\bibfield  {title} {\bibinfo {title} {{Bose-Hubbard} physics in synthetic dimensions from interaction trotterization},\ }\href {https://doi.org/10.1103/PhysRevResearch.2.043340} {\bibfield  {journal} {\bibinfo  {journal} {Phys. Rev. Research}\ }\textbf {\bibinfo {volume} {2}},\ \bibinfo {pages} {043340} (\bibinfo {year} {2020})}\BibitemShut {NoStop}%
\bibitem [{\citenamefont {Kargi}\ \emph {et~al.}(2021)\citenamefont {Kargi}, \citenamefont {Dehollain}, \citenamefont {Henriques}, \citenamefont {Sieberer}, \citenamefont {Olsacher}, \citenamefont {Hauke}, \citenamefont {Heyl}, \citenamefont {Zoller},\ and\ \citenamefont {Langford}}]{kargi21}%
  \BibitemOpen
  \bibfield  {author} {\bibinfo {author} {\bibfnamefont {C.}~\bibnamefont {Kargi}}, \bibinfo {author} {\bibfnamefont {J.~P.}\ \bibnamefont {Dehollain}}, \bibinfo {author} {\bibfnamefont {F.}~\bibnamefont {Henriques}}, \bibinfo {author} {\bibfnamefont {L.~M.}\ \bibnamefont {Sieberer}}, \bibinfo {author} {\bibfnamefont {T.}~\bibnamefont {Olsacher}}, \bibinfo {author} {\bibfnamefont {P.}~\bibnamefont {Hauke}}, \bibinfo {author} {\bibfnamefont {M.}~\bibnamefont {Heyl}}, \bibinfo {author} {\bibfnamefont {P.}~\bibnamefont {Zoller}},\ and\ \bibinfo {author} {\bibfnamefont {N.~K.}\ \bibnamefont {Langford}},\ }\bibfield  {title} {\bibinfo {title} {Quantum chaos and universal trotterisation performance behaviours in digital quantum simulation},\ }in\ \href {https://doi.org/10.1364/QIM.2021.W3A.1} {\emph {\bibinfo {booktitle} {Quantum Information and Measurement VI 2021}}}\ (\bibinfo  {publisher} {Optica Publishing Group},\ \bibinfo {year} {2021})\ p.\ \bibinfo {pages} {W3A.1}\BibitemShut {NoStop}%
\bibitem [{\citenamefont {Zhou}\ and\ \citenamefont {Nahum}(2019)}]{zhou19}%
  \BibitemOpen
  \bibfield  {author} {\bibinfo {author} {\bibfnamefont {T.}~\bibnamefont {Zhou}}\ and\ \bibinfo {author} {\bibfnamefont {A.}~\bibnamefont {Nahum}},\ }\bibfield  {title} {\bibinfo {title} {Emergent statistical mechanics of entanglement in random unitary circuits},\ }\href {https://doi.org/10.1103/PhysRevB.99.174205} {\bibfield  {journal} {\bibinfo  {journal} {Phys. Rev. B}\ }\textbf {\bibinfo {volume} {99}},\ \bibinfo {pages} {174205} (\bibinfo {year} {2019})}\BibitemShut {NoStop}%
\bibitem [{\citenamefont {Vasseur}\ \emph {et~al.}(2019)\citenamefont {Vasseur}, \citenamefont {Potter}, \citenamefont {You},\ and\ \citenamefont {Ludwig}}]{vasseur19}%
  \BibitemOpen
  \bibfield  {author} {\bibinfo {author} {\bibfnamefont {R.}~\bibnamefont {Vasseur}}, \bibinfo {author} {\bibfnamefont {A.~C.}\ \bibnamefont {Potter}}, \bibinfo {author} {\bibfnamefont {Y.-Z.}\ \bibnamefont {You}},\ and\ \bibinfo {author} {\bibfnamefont {A.~W.~W.}\ \bibnamefont {Ludwig}},\ }\bibfield  {title} {\bibinfo {title} {Entanglement transitions from holographic random tensor networks},\ }\href {https://doi.org/10.1103/PhysRevB.100.134203} {\bibfield  {journal} {\bibinfo  {journal} {Phys. Rev. B}\ }\textbf {\bibinfo {volume} {100}},\ \bibinfo {pages} {134203} (\bibinfo {year} {2019})}\BibitemShut {NoStop}%
\bibitem [{\citenamefont {Barbier}\ and\ \citenamefont {Macris}(2022)}]{barbier21}%
  \BibitemOpen
  \bibfield  {author} {\bibinfo {author} {\bibfnamefont {J.}~\bibnamefont {Barbier}}\ and\ \bibinfo {author} {\bibfnamefont {N.}~\bibnamefont {Macris}},\ }\bibfield  {title} {\bibinfo {title} {Statistical limits of dictionary learning: Random matrix theory and the spectral replica method},\ }\href {https://doi.org/10.1103/PhysRevE.106.024136} {\bibfield  {journal} {\bibinfo  {journal} {Phys. Rev. E}\ }\textbf {\bibinfo {volume} {106}},\ \bibinfo {pages} {024136} (\bibinfo {year} {2022})}\BibitemShut {NoStop}%
\bibitem [{\citenamefont {Jian}\ and\ \citenamefont {Swingle}(2021)}]{jian21}%
  \BibitemOpen
  \bibfield  {author} {\bibinfo {author} {\bibfnamefont {S.-K.}\ \bibnamefont {Jian}}\ and\ \bibinfo {author} {\bibfnamefont {B.}~\bibnamefont {Swingle}},\ }\bibfield  {title} {\bibinfo {title} {Phase transition in von {Neumann} entanglement entropy from replica symmetry breaking},\ }\href {https://doi.org/10.48550/arXiv.2108.11973} {\bibfield  {journal} {\bibinfo  {journal} {arXiv:2108.11973}\ } (\bibinfo {year} {2021})}\BibitemShut {NoStop}%
\bibitem [{\citenamefont {Oshima}\ and\ \citenamefont {Fuji}(2023)}]{oshima2023}%
  \BibitemOpen
  \bibfield  {author} {\bibinfo {author} {\bibfnamefont {H.}~\bibnamefont {Oshima}}\ and\ \bibinfo {author} {\bibfnamefont {Y.}~\bibnamefont {Fuji}},\ }\bibfield  {title} {\bibinfo {title} {Charge fluctuation and charge-resolved entanglement in a monitored quantum circuit with {U(1)} symmetry},\ }\href {https://doi.org/10.1103/PhysRevB.107.014308} {\bibfield  {journal} {\bibinfo  {journal} {Phys. Rev. B}\ }\textbf {\bibinfo {volume} {107}},\ \bibinfo {pages} {014308} (\bibinfo {year} {2023})}\BibitemShut {NoStop}%
\bibitem [{\citenamefont {Agrawal}\ \emph {et~al.}(2022)\citenamefont {Agrawal}, \citenamefont {Zabalo}, \citenamefont {Chen}, \citenamefont {Wilson}, \citenamefont {Potter}, \citenamefont {Pixley}, \citenamefont {Gopalakrishnan},\ and\ \citenamefont {Vasseur}}]{agrawal2022}%
  \BibitemOpen
  \bibfield  {author} {\bibinfo {author} {\bibfnamefont {U.}~\bibnamefont {Agrawal}}, \bibinfo {author} {\bibfnamefont {A.}~\bibnamefont {Zabalo}}, \bibinfo {author} {\bibfnamefont {K.}~\bibnamefont {Chen}}, \bibinfo {author} {\bibfnamefont {J.~H.}\ \bibnamefont {Wilson}}, \bibinfo {author} {\bibfnamefont {A.~C.}\ \bibnamefont {Potter}}, \bibinfo {author} {\bibfnamefont {J.~H.}\ \bibnamefont {Pixley}}, \bibinfo {author} {\bibfnamefont {S.}~\bibnamefont {Gopalakrishnan}},\ and\ \bibinfo {author} {\bibfnamefont {R.}~\bibnamefont {Vasseur}},\ }\bibfield  {title} {\bibinfo {title} {Entanglement and charge-sharpening transitions in {U(1)} symmetric monitored quantum circuits},\ }\href {https://doi.org/10.1103/PhysRevX.12.041002} {\bibfield  {journal} {\bibinfo  {journal} {Phys. Rev. X}\ }\textbf {\bibinfo {volume} {12}},\ \bibinfo {pages} {041002} (\bibinfo {year} {2022})}\BibitemShut {NoStop}%
\bibitem [{\citenamefont {Thiel}\ and\ \citenamefont {Kessler}(2020)}]{thiel20}%
  \BibitemOpen
  \bibfield  {author} {\bibinfo {author} {\bibfnamefont {F.}~\bibnamefont {Thiel}}\ and\ \bibinfo {author} {\bibfnamefont {D.~A.}\ \bibnamefont {Kessler}},\ }\bibfield  {title} {\bibinfo {title} {Non-{Hermitian} and {Zeno} limit of quantum systems under rapid measurements},\ }\href {https://doi.org/10.1103/PhysRevA.102.012218} {\bibfield  {journal} {\bibinfo  {journal} {Phys. Rev. A}\ }\textbf {\bibinfo {volume} {102}},\ \bibinfo {pages} {012218} (\bibinfo {year} {2020})}\BibitemShut {NoStop}%
\bibitem [{\citenamefont {Dubey}\ \emph {et~al.}(2021)\citenamefont {Dubey}, \citenamefont {Bernardin},\ and\ \citenamefont {Dhar}}]{dubey21}%
  \BibitemOpen
  \bibfield  {author} {\bibinfo {author} {\bibfnamefont {V.}~\bibnamefont {Dubey}}, \bibinfo {author} {\bibfnamefont {C.}~\bibnamefont {Bernardin}},\ and\ \bibinfo {author} {\bibfnamefont {A.}~\bibnamefont {Dhar}},\ }\bibfield  {title} {\bibinfo {title} {Quantum dynamics under continuous projective measurements: Non-{Hermitian} description and the continuum-space limit},\ }\href {https://doi.org/10.1103/PhysRevA.103.032221} {\bibfield  {journal} {\bibinfo  {journal} {Phys. Rev. A}\ }\textbf {\bibinfo {volume} {103}},\ \bibinfo {pages} {032221} (\bibinfo {year} {2021})}\BibitemShut {NoStop}%
\bibitem [{\citenamefont {Altland}\ \emph {et~al.}(2022)\citenamefont {Altland}, \citenamefont {Buchhold}, \citenamefont {Diehl},\ and\ \citenamefont {Micklitz}}]{altland22}%
  \BibitemOpen
  \bibfield  {author} {\bibinfo {author} {\bibfnamefont {A.}~\bibnamefont {Altland}}, \bibinfo {author} {\bibfnamefont {M.}~\bibnamefont {Buchhold}}, \bibinfo {author} {\bibfnamefont {S.}~\bibnamefont {Diehl}},\ and\ \bibinfo {author} {\bibfnamefont {T.}~\bibnamefont {Micklitz}},\ }\bibfield  {title} {\bibinfo {title} {Dynamics of measured many-body quantum chaotic systems},\ }\href {https://doi.org/10.1103/PhysRevResearch.4.L022066} {\bibfield  {journal} {\bibinfo  {journal} {Phys. Rev. Research}\ }\textbf {\bibinfo {volume} {4}},\ \bibinfo {pages} {L022066} (\bibinfo {year} {2022})}\BibitemShut {NoStop}%
\bibitem [{\citenamefont {Sternheim}\ and\ \citenamefont {Walker}(1972)}]{sterheim1972}%
  \BibitemOpen
  \bibfield  {author} {\bibinfo {author} {\bibfnamefont {M.~M.}\ \bibnamefont {Sternheim}}\ and\ \bibinfo {author} {\bibfnamefont {J.~F.}\ \bibnamefont {Walker}},\ }\bibfield  {title} {\bibinfo {title} {Non-{Hermitian} {Hamiltonians}, decaying states, and perturbation theory},\ }\href {https://doi.org/10.1103/PhysRevC.6.114} {\bibfield  {journal} {\bibinfo  {journal} {Phys. Rev. C}\ }\textbf {\bibinfo {volume} {6}},\ \bibinfo {pages} {114} (\bibinfo {year} {1972})}\BibitemShut {NoStop}%
\bibitem [{\citenamefont {Brody}(2013)}]{Brody_2013}%
  \BibitemOpen
  \bibfield  {author} {\bibinfo {author} {\bibfnamefont {D.~C.}\ \bibnamefont {Brody}},\ }\bibfield  {title} {\bibinfo {title} {Biorthogonal quantum mechanics},\ }\href {https://doi.org/10.1088/1751-8113/47/3/035305} {\bibfield  {journal} {\bibinfo  {journal} {J. Phys. A: Math. Theor.}\ }\textbf {\bibinfo {volume} {47}},\ \bibinfo {pages} {035305} (\bibinfo {year} {2013})}\BibitemShut {NoStop}%
\bibitem [{\citenamefont {Houdayer}\ and\ \citenamefont {Hartmann}(2004)}]{houdayer04}%
  \BibitemOpen
  \bibfield  {author} {\bibinfo {author} {\bibfnamefont {J.}~\bibnamefont {Houdayer}}\ and\ \bibinfo {author} {\bibfnamefont {A.~K.}\ \bibnamefont {Hartmann}},\ }\bibfield  {title} {\bibinfo {title} {Low-temperature behavior of two-dimensional gaussian {Ising} spin glasses},\ }\href {https://doi.org/10.1103/PhysRevB.70.014418} {\bibfield  {journal} {\bibinfo  {journal} {Phys. Rev. B}\ }\textbf {\bibinfo {volume} {70}},\ \bibinfo {pages} {014418} (\bibinfo {year} {2004})}\BibitemShut {NoStop}%
\bibitem [{\citenamefont {Alberton}\ \emph {et~al.}(2021)\citenamefont {Alberton}, \citenamefont {Buchhold},\ and\ \citenamefont {Diehl}}]{alberton21}%
  \BibitemOpen
  \bibfield  {author} {\bibinfo {author} {\bibfnamefont {O.}~\bibnamefont {Alberton}}, \bibinfo {author} {\bibfnamefont {M.}~\bibnamefont {Buchhold}},\ and\ \bibinfo {author} {\bibfnamefont {S.}~\bibnamefont {Diehl}},\ }\bibfield  {title} {\bibinfo {title} {Entanglement transition in a monitored free-fermion chain: From extended criticality to area law},\ }\href {https://doi.org/10.1103/PhysRevLett.126.170602} {\bibfield  {journal} {\bibinfo  {journal} {Phys. Rev. Lett.}\ }\textbf {\bibinfo {volume} {126}},\ \bibinfo {pages} {170602} (\bibinfo {year} {2021})}\BibitemShut {NoStop}%
\bibitem [{\citenamefont {Rigol}\ and\ \citenamefont {Muramatsu}(2005)}]{Rigol2005}%
  \BibitemOpen
  \bibfield  {author} {\bibinfo {author} {\bibfnamefont {M.}~\bibnamefont {Rigol}}\ and\ \bibinfo {author} {\bibfnamefont {A.}~\bibnamefont {Muramatsu}},\ }\bibfield  {title} {\bibinfo {title} {Hard-core bosons and spinless fermions trapped on 1d lattices},\ }\href {https://doi.org/10.1007/s10909-005-2274-3} {\bibfield  {journal} {\bibinfo  {journal} {J. Low Temp. Phys.}\ }\textbf {\bibinfo {volume} {138}},\ \bibinfo {pages} {645} (\bibinfo {year} {2005})}\BibitemShut {NoStop}%
\bibitem [{\citenamefont {Karamlou}\ \emph {et~al.}(2022)\citenamefont {Karamlou} \emph {et~al.}}]{Karamlou22}%
  \BibitemOpen
  \bibfield  {author} {\bibinfo {author} {\bibfnamefont {A.~H.}\ \bibnamefont {Karamlou}} \emph {et~al.},\ }\bibfield  {title} {\bibinfo {title} {Quantum transport and localization in {1D} and {2D} tight-binding lattices},\ }\href {https://doi.org/10.1038/s41534-022-00528-0} {\bibfield  {journal} {\bibinfo  {journal} {npj Quantum Inf.}\ }\textbf {\bibinfo {volume} {8}},\ \bibinfo {pages} {1} (\bibinfo {year} {2022})}\BibitemShut {NoStop}%
\bibitem [{\citenamefont {Moghaddam}\ \emph {et~al.}(2023)\citenamefont {Moghaddam}, \citenamefont {P\"oyh\"onen},\ and\ \citenamefont {Ojanen}}]{moghaddam2023}%
  \BibitemOpen
  \bibfield  {author} {\bibinfo {author} {\bibfnamefont {A.~G.}\ \bibnamefont {Moghaddam}}, \bibinfo {author} {\bibfnamefont {K.}~\bibnamefont {P\"oyh\"onen}},\ and\ \bibinfo {author} {\bibfnamefont {T.}~\bibnamefont {Ojanen}},\ }\bibfield  {title} {\bibinfo {title} {Exponential shortcut to measurement-induced entanglement phase transitions},\ }\href {https://doi.org/10.1103/PhysRevLett.131.020401} {\bibfield  {journal} {\bibinfo  {journal} {Phys. Rev. Lett.}\ }\textbf {\bibinfo {volume} {131}},\ \bibinfo {pages} {020401} (\bibinfo {year} {2023})}\BibitemShut {NoStop}%
\bibitem [{\citenamefont {Mansikkam\"aki}\ \emph {et~al.}(2021)\citenamefont {Mansikkam\"aki}, \citenamefont {Laine},\ and\ \citenamefont {Silveri}}]{mansikkamaki21}%
  \BibitemOpen
  \bibfield  {author} {\bibinfo {author} {\bibfnamefont {O.}~\bibnamefont {Mansikkam\"aki}}, \bibinfo {author} {\bibfnamefont {S.}~\bibnamefont {Laine}},\ and\ \bibinfo {author} {\bibfnamefont {M.}~\bibnamefont {Silveri}},\ }\bibfield  {title} {\bibinfo {title} {Phases of the disordered {Bose-Hubbard} model with attractive interactions},\ }\href {https://doi.org/10.1103/PhysRevB.103.L220202} {\bibfield  {journal} {\bibinfo  {journal} {Phys. Rev. B}\ }\textbf {\bibinfo {volume} {103}},\ \bibinfo {pages} {L220202} (\bibinfo {year} {2021})}\BibitemShut {NoStop}%
\bibitem [{\citenamefont {Yamamoto}\ and\ \citenamefont {Hamazaki}(2023)}]{yamamoto2023}%
  \BibitemOpen
  \bibfield  {author} {\bibinfo {author} {\bibfnamefont {K.}~\bibnamefont {Yamamoto}}\ and\ \bibinfo {author} {\bibfnamefont {R.}~\bibnamefont {Hamazaki}},\ }\bibfield  {title} {\bibinfo {title} {Localization properties in disordered quantum many-body dynamics under continuous measurement},\ }\href {https://doi.org/10.1103/PhysRevB.107.L220201} {\bibfield  {journal} {\bibinfo  {journal} {Phys. Rev. B}\ }\textbf {\bibinfo {volume} {107}},\ \bibinfo {pages} {L220201} (\bibinfo {year} {2023})}\BibitemShut {NoStop}%
\bibitem [{\citenamefont {Vidal}(2003)}]{vidal03}%
  \BibitemOpen
  \bibfield  {author} {\bibinfo {author} {\bibfnamefont {G.}~\bibnamefont {Vidal}},\ }\bibfield  {title} {\bibinfo {title} {Efficient classical simulation of slightly entangled quantum computations},\ }\href {https://doi.org/10.1103/PhysRevLett.91.147902} {\bibfield  {journal} {\bibinfo  {journal} {Phys. Rev. Lett.}\ }\textbf {\bibinfo {volume} {91}},\ \bibinfo {pages} {147902} (\bibinfo {year} {2003})}\BibitemShut {NoStop}%
\bibitem [{\citenamefont {Vidal}(2004)}]{vidal04}%
  \BibitemOpen
  \bibfield  {author} {\bibinfo {author} {\bibfnamefont {G.}~\bibnamefont {Vidal}},\ }\bibfield  {title} {\bibinfo {title} {Efficient simulation of one-dimensional quantum many-body systems},\ }\href {https://doi.org/10.1103/PhysRevLett.93.040502} {\bibfield  {journal} {\bibinfo  {journal} {Phys. Rev. Lett.}\ }\textbf {\bibinfo {volume} {93}},\ \bibinfo {pages} {040502} (\bibinfo {year} {2004})}\BibitemShut {NoStop}%
\bibitem [{\citenamefont {Suzuki}(1990)}]{suzuki90}%
  \BibitemOpen
  \bibfield  {author} {\bibinfo {author} {\bibfnamefont {M.}~\bibnamefont {Suzuki}},\ }\bibfield  {title} {\bibinfo {title} {Fractal decomposition of exponential operators with applications to many-body theories and {Monte Carlo} simulations},\ }\href {https://doi.org/https://doi.org/10.1016/0375-9601(90)90962-N} {\bibfield  {journal} {\bibinfo  {journal} {Phys. Lett. A}\ }\textbf {\bibinfo {volume} {146}},\ \bibinfo {pages} {319} (\bibinfo {year} {1990})}\BibitemShut {NoStop}%
\bibitem [{\citenamefont {Kamakari}\ \emph {et~al.}(2022)\citenamefont {Kamakari}, \citenamefont {Sun}, \citenamefont {Motta},\ and\ \citenamefont {Minnich}}]{kamakari2022}%
  \BibitemOpen
  \bibfield  {author} {\bibinfo {author} {\bibfnamefont {H.}~\bibnamefont {Kamakari}}, \bibinfo {author} {\bibfnamefont {S.-N.}\ \bibnamefont {Sun}}, \bibinfo {author} {\bibfnamefont {M.}~\bibnamefont {Motta}},\ and\ \bibinfo {author} {\bibfnamefont {A.~J.}\ \bibnamefont {Minnich}},\ }\bibfield  {title} {\bibinfo {title} {Digital quantum simulation of open quantum systems using quantum imaginary--time evolution},\ }\href {https://doi.org/10.1103/PRXQuantum.3.010320} {\bibfield  {journal} {\bibinfo  {journal} {PRX Quantum}\ }\textbf {\bibinfo {volume} {3}},\ \bibinfo {pages} {010320} (\bibinfo {year} {2022})}\BibitemShut {NoStop}%
\bibitem [{\citenamefont {Bezanson}\ \emph {et~al.}(2017)\citenamefont {Bezanson}, \citenamefont {Edelman}, \citenamefont {Karpinski},\ and\ \citenamefont {Shah}}]{bezanson17}%
  \BibitemOpen
  \bibfield  {author} {\bibinfo {author} {\bibfnamefont {J.}~\bibnamefont {Bezanson}}, \bibinfo {author} {\bibfnamefont {A.}~\bibnamefont {Edelman}}, \bibinfo {author} {\bibfnamefont {S.}~\bibnamefont {Karpinski}},\ and\ \bibinfo {author} {\bibfnamefont {V.~B.}\ \bibnamefont {Shah}},\ }\bibfield  {title} {\bibinfo {title} {Julia: A fresh approach to numerical computing},\ }\href {https://doi.org/10.1137/141000671} {\bibfield  {journal} {\bibinfo  {journal} {SIAM Rev.}\ }\textbf {\bibinfo {volume} {59}},\ \bibinfo {pages} {65} (\bibinfo {year} {2017})}\BibitemShut {NoStop}%
\bibitem [{\citenamefont {Saad}(1992)}]{saad92}%
  \BibitemOpen
  \bibfield  {author} {\bibinfo {author} {\bibfnamefont {Y.}~\bibnamefont {Saad}},\ }\bibfield  {title} {\bibinfo {title} {Analysis of some {Krylov} subspace approximations to the matrix exponential operator},\ }\href {https://doi.org/10.1137/0729014} {\bibfield  {journal} {\bibinfo  {journal} {SIAM J. Numer. Anal.}\ }\textbf {\bibinfo {volume} {29}},\ \bibinfo {pages} {209} (\bibinfo {year} {1992})}\BibitemShut {NoStop}%
\bibitem [{\citenamefont {Kawashima}\ and\ \citenamefont {Ito}(1993)}]{kawashima93}%
  \BibitemOpen
  \bibfield  {author} {\bibinfo {author} {\bibfnamefont {N.}~\bibnamefont {Kawashima}}\ and\ \bibinfo {author} {\bibfnamefont {N.}~\bibnamefont {Ito}},\ }\bibfield  {title} {\bibinfo {title} {Critical behavior of the three-dimensional {±J} model in a magnetic field},\ }\href {https://doi.org/10.1143/JPSJ.62.435} {\bibfield  {journal} {\bibinfo  {journal} {J. Phys. Soc. Jpn.}\ }\textbf {\bibinfo {volume} {62}},\ \bibinfo {pages} {435} (\bibinfo {year} {1993})}\BibitemShut {NoStop}%
\bibitem [{\citenamefont {Mansikkam\"aki}\ \emph {et~al.}(2022)\citenamefont {Mansikkam\"aki}, \citenamefont {Laine}, \citenamefont {Piltonen},\ and\ \citenamefont {Silveri}}]{mansikkamaki22}%
  \BibitemOpen
  \bibfield  {author} {\bibinfo {author} {\bibfnamefont {O.}~\bibnamefont {Mansikkam\"aki}}, \bibinfo {author} {\bibfnamefont {S.}~\bibnamefont {Laine}}, \bibinfo {author} {\bibfnamefont {A.}~\bibnamefont {Piltonen}},\ and\ \bibinfo {author} {\bibfnamefont {M.}~\bibnamefont {Silveri}},\ }\bibfield  {title} {\bibinfo {title} {Beyond hard-core bosons in transmon arrays},\ }\href {https://doi.org/10.1103/PRXQuantum.3.040314} {\bibfield  {journal} {\bibinfo  {journal} {PRX Quantum}\ }\textbf {\bibinfo {volume} {3}},\ \bibinfo {pages} {040314} (\bibinfo {year} {2022})}\BibitemShut {NoStop}%
\end{thebibliography}%

\end{document}